\documentclass{ws-rv9x6}  

\usepackage{graphicx}% Include figure files
\usepackage{epsfig}

\newcommand{\rem}[1]{}
\newcommand{\scite}[1]{${}^{\mbox{\protect{\scriptsize \cite{#1}}}}$}
\newcommand{\bcite}[1]{[\cite{#1}]}
\newcommand{\bigcite}[1]{[\cite{#1}]}
\newcommand{\figcite}[1]{\protect\bcite{#1}}

\newcommand{\Tab}[1]{Table \ref{tab:#1}}
\newcommand{\Fig}[1]{Fig.\ref{fig:#1}}
\newcommand{\Eq}[1]{(\ref{eq:#1})}

\newcommand{\vect}[1]{{\mbox{\boldmath $#1$}}}

\newcommand{\Ham}{{\cal H}}
\newcommand{\Sec}[1]{section \ref{sec:#1}}
\newcommand{\SecNum}[1]{\ref{sec:#1}}
\newcommand{\Ssc}[1]{subsection \ref{ssc:#1}}
\newcommand{\ssclabel}[1]{\label{ssc:#1}}
\newcommand{\dontshow}[1]{}
\newcommand{\sast}{${}^{\ast}$}

\newcommand{\bc}{\begin{center}}
\newcommand{\ec}{\end{center}}
\newcommand{\be}{\begin{equation}}
\newcommand{\ee}{\end{equation}}
\newcommand{\ba}{\begin{array}}
\newcommand{\ea}{\end{array}}
\newcommand{\beqn}{\begin{eqnarray}}
\newcommand{\eeqn}{\end{eqnarray}}
\newcommand{\na}{\quad ---}

\newcommand{\deffig}[4]{
\begin{figure}[th]
\centerline{\psfig{figure=#2,width=#3 \textwidth,angle=0}}
\vspace*{8pt}
\caption{
#4
\label{fig:#1}
}
\end{figure}
}

\makeatletter
\def\@cite#1#2{#1\if@tempswa , #2\fi}
\makeatother

\begin{document}

\setcounter{chapter}{5}

\chapter{Recent Progress in Spin Glasses}

\markboth{N. Kawashima and H. Rieger}
{Recent Progress in Spin Glass Problem and Related Issues}

\author{N. Kawashima$^1$ and H. Rieger$^2$}
\medskip

\address{Department of Physics, Tokyo Metropolitan University,\\
$^1$ Hachioji, Tokyo 192-0397, Japan\\
E-mail: nao@phys.metro-u.ac.jp\\
\smallskip
$^2$ Theoretische Physik, Universit\"at des Saarlandes,\\
66041 Saarbr\"ucken, Germany\\
E-mail: H.Rieger@mx.uni-saarland.de}

\begin{abstract}
We review recent findings on spin glass models.
Both the equilibrium properties and the dynamic properties are covered.
We focus on progress in theoretical, in particular numerical, studies,
while its relationship to real magnetic materials is also mentioned.
\end{abstract}
\ \\
\ \\
\rem{The upper critical dimension is 6 even for the Heisenberg model? \\}
\rem{Note to myself: There are a few places where the bond distribution,
$\pm J$ or Gaussian, is not specified whereas the distinction is important.
In particular, it must be clarified to what extent we can rely on
the asymptotic form \Eq{DropletScalingForCorrelationFunction}
for the $\pm J$ model. \\}
\rem{Note to myself: Another figure by Marinari suggesting the
RSB picture might be included. \\}
\rem{Heiko, don't you think we should write a summary? \\}

The motivation that pulls researchers
toward spin glasses is possibly not the potential future
use of spin glass materials in ``practical'' applications.
It is rather the expectation that there must be something 
very fundamental in systems with randomness and frustration.
It seems that this expectation has been acquiring a firmer ground as
the difficulty of the problem is appreciated more clearly.
For example, a close relationship has been established between
spin glass problems and a class of optimization problems 
known to be $NP$-hard.
It is now widely accepted that a good (i.e. polynomial)
computational algorithm for solving any one of $NP$-hard problems 
most probably does not exist.
Therefore, it is quite natural to expect that the Ising spin glass 
problem, as one of the problem in the class,
would be very hard to solve,
which indeed turns out to be the case
in a vast number of numerical studies.
None the less, it is often the case that only numerical studies can 
decide whether a certain hypothetical picture applies to a given instance.
Consequently, the major part of what we present below is inevitably a 
description of the current status of spin glass studies,
what is being said and done on the issue, 
rather than an account of established facts.
In what follows, we put an emphasis on the theoretical and the numerical
progress achieved in the field in the last decade.
We mention older theoretical and experimental works
only when it is necessary for readers to be able to follow 
current topics without reading through many other articles.
Many of the important older works and experimental results, therefore,
have been left out.
For a review on these, the readers are referred to 
\bcite{HemmenM1983},
\bcite{HemmenM1986},
\bcite{FischerH1991},
\bcite{Mydosh1993},
and 
\bcite{Young1998}.

The present paper is organized as follows.
In \Sec{TwoPictures}, we first give a brief overview of 
two well-known paradigms on spin glasses and
summarize the predictions derived from them.
Then, in the subsequent sections \SecNum{Ising2D},
\SecNum{Ising3D} and \SecNum{Ising4D}, 
we discuss equilibrium properties of Ising spin glass models.
In particular, in \Sec{Ising3D}, a recent active debate,
as to which paradigm is appropriate for realistic short-range 
spin glass models in three dimensions, is presented.
Then, we proceed with the dynamical properties in \Sec{Aging} to 
examine these paradigms on a different ground.
In this section, an emphasis is put on aging phenomena
while some other non-equilibrium properties are also discussed.
Models with continuous degrees of freedom as well as
the Potts spin glass models are mentioned in \Sec{ContinuousSpinModels}.
The effects of weak disorder discussed in \Sec{WeakDisorder}
provide us with a set of interesting issues quite different 
from those arising from strong disorder dealt with 
in the preceding sections.
Several exact relations can be found and play an important role
in shedding light on the issue under debate.
Finally, we see in \Sec{QuantumSpinModels}
what results from the interplay between quantum fluctuations 
and randomness in spin glasses.

%%%%%%%%%%%%%%%%%%%%%%%%%%%%%%%%%%%%%%%%%%%%%%%%%%%%%%%%%%%%%%%%%%%%%%%%%%%%%%%
\section{Two Pictures}
%%%%%%%%%%%%%%%%%%%%%%%%%%%%%%%%%%%%%%%%%%%%%%%%%%%%%%%%%%%%%%%%%%%%%%%%%%%%%%%
\label{sec:TwoPictures}
The equilibrium properties of Ising spin-glass models 
in finite dimensions have been investigated mainly through
developments of working hypotheses and numerical techniques.
In principle, we can judge which working hypothesis is correct
by numerical computations.
The working hypotheses are often called ``pictures''.
Among various pictures, unarguably the most frequently mentioned
are the replica-symmetry-breaking (RSB) and the droplet picture.
In fact, the dichotomy between these two has been the
central issue in the spin glass research for more than a decade.
However, it is rather difficult to perform a conclusive numerical
test in the three dimensional case.
One thing that renders the test so difficult is that
the lower critical dimension appears to be close to three.
Another factor is, as mentioned in the introduction, that
there is no polynomial algorithm for obtaining the ground state
of a given instance in three (and higher) dimension.
Therefore, the best we can do at present is to summarize on-going
arguments, discriminating what has been established from what has not,
and presenting the most important numerical evidences.
We try to do these in what follows.

It should also be pointed out that most of the
``pictures'' are asymptotic theories which are supposed to be correct
only for large systems. In contrast to most regular systems the utility of an
asymptotic theory for glassy systems cannot be taken for granted. It
was pointed out\scite{JoenssonYN2002} that in most real spin glass
materials the typical length scale of an equilibrated domain within the
experimental time scale may not exceed a small number of lattice spacings.  
If this is the case, considering the numerical evidence of
the existence of large corrections to scaling, 
it can happen that the asymptotically correct theory
fails to explain experimental results.
The fact that many experiments on spin glasses, 
those at temperatures below the transition in particular,
are not performed in equilibrium
implies that an equilibrium theory may not be appropriate. 
Some attempts to construct a theory for a realistic time 
and length scales are reviewed in \Sec{Aging}.

%==============================================================================
\subsection{Mean-Field Picture}
%==============================================================================
We start with a brief description of the two pictures mentioned above
in the present and the next subsection.  
We do not, however, intend to
repeat the whole history of the development of each picture or discuss
every technical detail.  The reader may find a complete description of
these pictures in a number of review 
articles.\scite{HemmenM1983,HemmenM1986,BinderY1986,MezardPV1987,MarinariPRRZ2000,NewmanS2003}

The RSB picture or the mean-field picture is based on Parisi's 
solution\scite{Parisi1979,Parisi1980a,Parisi1980b,Parisi1983}
of the Sherrington-Kirkpatric (SK) model\scite{SherringtonK1975} and
its interpretation in terms of a multitude of thermodynamic states
\bcite{MezardPSTV1984a,MezardPSTV1984b}.
The SK model of $N$ Ising spins, $S_i$, is defined by the Hamiltonian
$$ 
  H = - \sum_{(i,j)} J_{ij} S_i S_j 
$$ 
where the summation is taken
over all $N(N-1)/2$ pairs of spins $(i,j)$ with $i\ne j$.  The
coupling constants $J_{ij}$ are quenched Gaussian random variables
with zero mean and variance $J/N$ where $J$ is a positive constant of
${\cal O}(1)$ that sets the energy scale. In \bcite{SherringtonK1975},
the replica method was used in which $n$
identical copies (replicas) of the system are introduced to perform
the disorder average of the free energy: $-\beta F=[\ln Z]_{\rm
av}=\lim_{n\to0} ([Z^n]_{\rm av}-1)/n$.  They found a solution of the
mean-field equations using a replica symmetric Ansatz (see
below). This solution has a phase transition and the transition point
is located at the temperature $T_{\rm c} = J$. Below this temperature,
the order parameter
\begin{equation}
  q_{\rm EA} =
  q^{\alpha\beta} \equiv \frac1{N} \sum_{i} S_i^{\alpha} S_i^{\beta}
  \label{eq:EAOrderParameter}
\end{equation}
has a non-zero value where indices $\alpha$ and $\beta$ specify
replicas.
The replica symmetric Ansatz means that 
only solutions to the mean-field equations with
$q^{\alpha\beta}$ being independent of $\alpha$ and $\beta$ 
(before taking the limit $n\to0$) are allowed. 
However, this solution did
not describe the low temperature phase correctly since the entropy
predicted by this solution became negative at low temperatures.

In \bcite{ThoulessAP1977} a different approach was taken for the SK model. 
They constructed a set of
equations (called the TAP equations) that expresses the equation of
state for each bond-realization of the SK model in terms of the local
magnetization, $m_i = \langle S_i \rangle$.  They studied the
eigen-modes of the stability matrix of stable solutions of the TAP
equations and found that the spectrum of the eigen-values extends down
to zero, suggesting that the solutions are only marginally stable in
the thermodynamic limit.  Furthermore, 
the stability matrix in the replica space, i.e., 
the second derivatives of the free energy with respect 
to the replica order parameters, $q^{\alpha\beta}$,
was examined.\scite{AlmeidaT1978}
It turned out that the replica symmetric solution is stable only in the
paramagnetic phase and becomes unstable below the phase boundary
called the AT line in the $T$-$H$ plane.  Therefore, the SK solution
is correct above the AT line whereas it is not below the AT line.

To find the correct solution of the SK model below the AT line
the replica symmetry had to be broken.
The way the symmetry is to be broken is, however, highly non-trivial.
A solution, called the Parisi solution, was 
proposed\scite{Parisi1979,Parisi1980a,Parisi1980b,Parisi1983}
based on a novel metric structure of the replica-index space.
It has now been proved with acceptable mathematical rigor that the
solution is exact.\scite{Guerra1}
In the Parisi solution, a metric structure is introduced
in the replica-index space and it is assumed that the overlap $q$ between
two replicas depends only on the distance between the two in this
index space.  As a result, the order parameter is
actually not a single number, in contrast to the EA order parameter $q_{\rm EA}$,
but a function $q(x)$ $(0\le x\le 1)$, where $x$ is a
suitable parameterization of the distance between two replica indices.
A remarkable feature of Parisi's solution below the AT line was that
$q(x)$ varies continuously as a function of $x$.

The order-parameter function $q(x)$ first appeared to be a mathematical artifact,
but its physical meaning was later clarified\scite{Parisi1983}
by relating $q(x)$ to the probability distribution $P(q)$ of the
overlap $q$ between {\it pure states}, rather than the replicas.
A pure state is defined as an extremal equilibrium distribution 
in the thermodynamic limit that cannot be expressed as a
linear combination of any other distributions.
Denoting the inverse function of $q(x)$ as $x(q)$,
it was found that
\begin{equation}
  x(q) = \int_{-\infty}^q dq' P(q').
  \label{eq:Relation_between_P_and_x}
\end{equation}
Here, the overlap distribution $P(q)$ is formally (and somewhat 
symbolically) defined as
\begin{equation}
  P(q) = \sum_{\alpha,\beta} 
  P_{\alpha}P_{\beta}\delta(q-q^{\alpha\beta}).
  \label{eq:Symbolic_definition_of_P_of_Q}
\end{equation}
Here the indices $\alpha$ and $\beta$ specify pure states, not
replicas, $q_{\alpha\beta}$ is the overlap between the two pure
states $\alpha$ and $\beta$, and $P_{\alpha}$ is the probability
for the system to be in the pure state $\alpha$.

Equation \Eq{Relation_between_P_and_x} may provide a way to
investigate the structure of the space of the equilibrium
distributions with the replica method.  
We can conclude, for instance, that more than one pure state exists
and the overlap between two pure states varies depending on 
the particular pair of states if $q(x)$ is not constant.
If $q(x)$ changes continuously in a finite range of $x$, 
as in Parisi's solution for the SK model, 
there must even be an infinite number of pure states.
Simply stated, the RSB picture for systems in finite dimensions with
short range interactions consists of the hypothesis that there are
infinitely many pure states with varying overlaps and
various predictions derived from the hypothesis.

The definition \Eq{Symbolic_definition_of_P_of_Q} of $P(q)$, 
however, must be taken with caution since 
the existence of pure states as a well-defined thermodynamic limit
is not clear for disordered systems.
In fact, it was shown\scite{MezardPSTV1984a,MezardPSTV1984b}
that $P_J(q)$ for a bond realization $J$ 
depends on $J$ even in the thermodynamic limit.
As discussed below, this means\scite{NewmanS1996} 
that a unique thermodynamic limit of $P_J(q)$ with
fixed bond realization $J$ does not exist.
Therefore, the order parameter, $q(x)$, is not self-averaging for 
the SK model, in striking contrast to homogeneous systems.

Because of this subtle nature of the order parameter,
the exact implication of the RSB picture for more realistic 
spin glass models in finite dimensions is somewhat ambiguous 
and has been a subject of debates.\scite{MarinariPRRZ2000,NewmanS2003}
One focus is the interpretation of the numerical result
for finite systems.
The definition of $q$ and $P(q)$ often used in numerical simulations is
based on the overlap between two spin configurations, $S$ and $S'$,
independently chosen from the equilibrium distribution of 
the same bond realization. Namely,
\begin{eqnarray}
  q(S,S') 
  & \equiv & 
  \frac1N\sum_{i=1}^N S_i S'_i
  \;,\qquad  
  P(q;L) 
  \equiv 
  [P_J(q;L)]\;,
  \label{eq:RealReplicaPofQ}
%  \label{eq:Overlap}
  \\
  P_J(q;L) & \equiv & \sum_{S}\sum_{S'} P_J(S;L)P_J(S';L)
  \delta\left(q - q(S,S')\right) \nonumber 
\end{eqnarray}
where $P_J(S;L)$ is the normalized Boltzmann weight of the spin
configuration $S$ in a system of the linear size $L$ with the
bond configuration $J$.
Throughout this article, unless stated otherwise,
the angular bracket $[\cdots]$ denotes the bond-configuration 
average with the distribution
$
  P(J) = \prod_{(ij)} P_1(J_{ij}),
$
where the single-bond distribution $P_1$ is independent of $(ij)$.
The following link-overlap and its distribution are also often discussed
recently:
\begin{eqnarray}
  q_l(S,S') & \equiv & \frac1{N_l}\sum_{(ij)} S_i S_j S'_i S'_j
  \;,\qquad
  P(q_l;L) \equiv [P_J(q_l;L)]\;,
%  \label{eq:LinkOverlap}
  \\
  P_J(q_l;L) & \equiv & \sum_{S}\sum_{S'}P_J(S;L)P_J(S';L)
  \delta\left(q_l - q_l(S,S')\right)\;,\nonumber 
%  \label{eq:RealReplicaOverlapDistribution}
\end{eqnarray}
where the summation concerning $(ij)$ is over all 
the pairs of nearest neighbor spins, and 
$N_l$ is the total number of the pairs in the system.

The following is the list of features that follows from
the RSB picture.
\begin{enumerate}
\item The overlap distribution, $P(q;L)$, converges to a function $P(q)$
      that has a continuous part in the limit of $L\to\infty$.
      In particular, $P(0;L)$ converges to a finite value $P(0) > 0$.
\item Some quantities, such as $P_J(q;L)$, are non-self-averaging, i.e.,
      a unique thermodynamic limit does not exist for a fixed bond
      realization.
\item The bond-averaged link-overlap distribution $P(q_l;L)$ 
      is not $\delta(q_l)$ in the thermodynamic limit, i.e., 
      its width converges to a finite value.
\item Global excitations with the energy cost of ${\cal O}(1)$ exist.
\item A change in the boundary conditions generally affects 
      spins located far away from the boundary.
\end{enumerate}

%==============================================================================
\subsection{Droplet Picture}
%==============================================================================
Another well-known 
picture\scite{MooreB1985,BrayM1986,FisherH1986,FisherH1988a,FisherH1988b}
is based on a scaling hypothesis on local excitations and 
produces markedly different conclusion for various quantities,
in particular, $P(q)$.
The basic assumption in this picture is that even below the
transition temperature in zero magnetic field
there are only two pure states that are mapped to each other by the
total inversion of spins as in the homogeneous ferromagnets.
Then, several scaling properties are assumed for compact excitations
with varying size, which are called {\it droplets}.
To be specific, a droplet of the scale $l$ has a typical
excitation free energy $\epsilon_l \sim \Upsilon l^{\theta}$,
where $\theta$ is called the droplet exponent.
Since this exponent relates the energy scale to the length,
similar to the stiffness exponent $\theta_{\rm S}$ that relates 
the domain wall excitation energy to the system size,
the simplest assumption is to identify $\theta$ with the
stiffness exponent $\theta_{\rm S}$.

Among various results derived from this picture, of particular
importance are the scaling forms of the correlation 
functions.\scite{FisherH1986,BrayM1986}
For example, it is predicted for models with a continuous bond distribution
and with no magnetic field that the following asymptotic form should apply:
\begin{equation}
  [\langle S_i S_j \rangle^2] 
  - [\langle S_i \rangle^2] [\langle S_j \rangle^2]
  \sim 
  T/\Upsilon R^{\theta}
  \label{eq:DropletScalingForCorrelationFunction}
\end{equation}
near zero temperature,
where $R$ is the distance between site $i$ and site $j$ and
the thermal average $\langle \cdots\rangle$ is taken in
a single pure state (if there are two).
It follows that the variance of the distribution function $P(q;L)$
defined in \Eq{RealReplicaPofQ}
(more strictly, the variance of $P(|q|;L)$)
decreases as the system size increases:
$$
  (\Delta q)^2 \equiv [\langle q^2 \rangle] - [\langle |q| \rangle]^2
  \propto L^{-\theta} \to 0.
$$
Therefore, when $P(q)$ is defined as the thermodynamic limit of
$P(q;L)$, it consists of a pair of delta peaks at $q = \pm q_{\rm EA}$.

For the models with discrete energy levels, such as the $\pm J$
model, the scaling form of the correlation function near $T=0$
must be modified.
However, it is widely believed that the 
discretized nature of the energy level is not relevant
at finite temperatures and the behavior
is qualitatively the same as the one 
of the models with continuous energy levels.

Below, we list the defining properties of the droplet picture
together with some predictions derived from it.
\begin{enumerate}
\item The distribution of the excitation free energy of droplets
      of the scale $l$, $P(\epsilon_l)$,
      is continuous down to zero energy.
\item $P(\epsilon_l)$ has the typical energy scale
      $\Upsilon l^{\theta}$ with $\Upsilon$ being an ${\cal O}(1)$ constant.
      Specifically, $P(\epsilon_l)$ has the scaling form
      $$
        P(\epsilon_l) \sim \frac{1}{\Upsilon l^{\theta}}
        \tilde P\left(\frac{\epsilon_l}{\Upsilon l^{\theta}}\right).
      $$
\item $P(q)$ is self-averaging and it consists of
      a pair of delta peaks below the transition temperature.
\item Since $P(0;L)$ is proportional to the excitation probability of 
      a droplet that contains approximately one half of all the spins,
      for a system with a continuous bond distribution,
      $P(0;L)$ has the following temperature and size dependence near $T=0$;
      \begin{equation}
         P(0;L) \sim [{\rm Prob}\ (\epsilon_{L} < T)]
         \sim \frac{T}{\overline{\epsilon_{L}}} 
         \propto T L^{-\theta}.
         \label{eq:TemperatureSizeDependenceOfPzero}
      \end{equation}
\item For systems with a continuous bond distribution
      the bond-averaged link-overlap distribution $P(q_l;L)$ 
      consists of a single delta function in the thermodynamic limit.
      At low temperature, its width depends on the size and 
      the temperature as\scite{DrosselBMB2000}
      \begin{equation}
        (\Delta q_l)^2 \propto T L^{-\mu}
        \qquad
        (\mu_l \equiv 2d_{\rm S}-2d-\theta)
        \label{eq:TemperatureSizeDependenceOfDeltaQl}
      \end{equation}
      where $d_{\rm S}$ is the fractal dimension of the surface
      of droplets.
\item For systems with finite temperature spin glass transition,
      $\theta$ is positive. Therefore, a global excitation costs 
      an infinite energy.
\item A change in the boundary conditions does not affect 
      spins located far away from the boundary,
      except for a potential simultaneous inversion of all spins.
\end{enumerate}

%==============================================================================
\section{Equilibrium Properties of Two-Dimensional Ising Spin Glasses}
%==============================================================================
\label{sec:Ising2D}
Two-dimensional spin glass models are easier to study than
three-dimensional models from a technical point of view.
However, there are still some open questions.
Among others, the possibility of the existence of a phase
transition at non-zero temperatures 
in the discrete distribution models is important.

%------------------------------------------------------------------------------
\subsection{Zero-Temperature Transition?}
\label{ssc:ZTT}
%------------------------------------------------------------------------------
One of the arguments that supports the zero-temperature phase
transition scenario is based on the numerical estimates of the
stiffness exponent that characterizes the size dependence of the
domain-wall free-energy. The stiffness is simply the difference in the
free energy caused by a particular change in the boundary conditions.
The most frequent choice in numerical computations is the application of
periodic versus anti-periodic boundary condition. The stiffness exponent 
$\theta_S$ is then determined by the asymptotic size-dependence of
the stiffness, $\Delta F$:
$$ 
\Delta F \propto L^{\theta_S}.  
$$ 
According to \bcite{BrayM1986}, when
the distribution of coupling constants is continuous and has a
non-zero weight at $J=0$, whether a phase transition takes place at a
finite temperature or not is determined by the sign of the stiffness
exponent. If it is positive, the zero temperature phase is
strongly ordered, leading to a finite-temperature phase transition,
whereas otherwise the order is fragile and infinitesimal thermal
fluctuations would destroy it, leading to criticality at zero
temperature.  Numerical computations of the ground states of finite
systems in two dimensions were carried
out\scite{BrayM1986,McMillan1984a,McMillan1984b} for estimating the
stiffness exponent at $T=0$.  For the Gaussian distribution of the
bonds, it turned out that $\theta_S\sim -0.3$.  At present, one of the
most accurate and reliable estimates of the stiffness exponent is
given in\bcite{HartmannY2001} as $ \theta_{\rm S} =
-0.282(2) $ for the Gaussian bond distribution with 
periodic/anti-periodic boundary condition. Due to the negative value
of the stiffness exponent, it is widely believed that there is no
finite temperature transition in this model and that the system is
critical at zero-temperature.

The stiffness exponent of $\pm J$ model, on the other hand, 
is much closer to the marginal value, 0.
It was estimated,\scite{KawashimaR1997}
with the assumption of the power-low dependence of the stiffness
on the system size, as
$$
  \theta_S = -0.060(4),
$$
while the possibility of $\theta_S = 0$ was not ruled out.
The possibility of the stiffness exponent being zero
was strongly suggested by another ground
state computation.\scite{HartmannY2001}

Recently, it was pointed out that the stiffness exponent have to be
interpreted differently for $\pm J$ models.  Amoruso {\it et
al.}\scite{AmorusoMMP2003} carried out a renormalization calculation
using the Migdal-Kadanoff method.  They found that the stiffness
exponent, calculated within this approximation, is zero in any
dimension lower than the lower critical dimension.  This finding
suggests that the stiffness exponent being zero does not necessarily
imply that the system is exactly at the lower critical dimension.
This type of dimensional dependence of the stiffness exponent
was observed only for the class of the bond distribution for which
the discretized nature of the coupling constant is not smeared out by
the renormalization.
Since the $\pm J$ bond distribution falls into this class,
the above-mentioned numerical estimates, $\theta_{\rm S} \sim 0$,
may only mean that the lower-critical dimension is two {\it or above}.
For the $\pm J$ model, however, even the possibility of 
a finite temperature phase transition was 
suggested,\scite{ShirakuraM1996} as we discuss in greater 
detail further below.

%------------------------------------------------------------------------------
\subsection{Droplet Argument for Gaussian-Coupling Models}
%------------------------------------------------------------------------------
For two dimensional systems 
with a symmetric $(P(J) = P(-J))$ and continuous bond distribution,
it was argued\scite{BrayM1986} that 
there is only one independent critical exponent.
In other words, all critical indices are related to 
the stiffness exponent via scaling laws.
We consider the following finite size scaling form
of the singular part of the free energy $F_{\rm s}$,
$$
  \Phi(T,H,L) \equiv \beta F_{\rm s} = -[\log Z] \sim \phi(TL^y,HL^{y_h}).
$$
The condition that the total magnetization 
$[M] \equiv \lim_{H\to 0} \partial F_{\rm s}/\partial H$
is proportional to  $L^{d/2}$ at $T=0$
relates $y$ to $y_h$ via
$$
  y_h = y + \frac{d}{2}.
$$
Then, by differentiating $\Phi$ with respect to $T$ and $H$,
we can express any critical index by $y$.
For example, the magnetization per spin depends on the magnetic field as
$m = L^{-d} M \sim = L^{-\frac{d}{2}}
\sim H^{\frac{d}{2y_h}}$, which means that the exponent $\delta$ appearing
in the scaling of the magnetization with the magnetic field is 
$\delta=1-2y/d$.
%$$
%  \delta = \frac{2y_h}{d} = 1 + \frac{2y}{d}.
%$$
The non-linear susceptibility
$\chi_2 \equiv L^{-d} \partial^3 M / \partial H^3$ 
at $H=0$ depends on the temperature as
$\chi_2 \sim T^{-\gamma_2}$ with $\gamma_2 = 3 + \frac{d}{y}$.
Since the spin glass susceptibility $\chi_{\rm SG}$ at $H=0$
is related to $\chi_2$ as
$\chi_{\rm SG} \equiv 
L^{-d} \sum_{i,j} [\langle S_i S_j \rangle^2]
= T^3 \chi_2$, the corresponding exponent is 
$\gamma_{\rm SG} = d/y$.
Similarly, the specific heat exponent can be expressed as
$\alpha = -d/y$.
As for the exponent $\eta_{\rm SG}$, that characterizes the
asymptotic form of the two-point spin glass correlation function
$[\langle S_i S_j \rangle^2] \sim R_{ij}^{d - 2 + \eta_{\rm SG}}$,
we have $2-\eta_{\rm SG} = y \gamma_{\rm SG}$, which yields
$\eta_{\rm SG} = 2-d$.
This is consistent with the fact that the correlation function
does not decay at zero-temperature.

\subsection{Droplets in Gaussian-Coupling Models: Numerics}
\label{ssc:TwoDimensionalGaussian}
%------------------------------------------------------------------------------
As mentioned above, one naturally expects\scite{FisherH1986}
that the exponent $-y$ is identical to the stiffness
exponent $\theta_S$ since both the exponents relate the energy
scale to the length scale.
However, the estimates of various critical 
exponents\scite{KawashimaS1992,KawashimaHS1992,Liang1992,Rieger1997,chaos-young}
based on the system size up to $L \sim 50$
seem to satisfy the scaling relation with $y \sim 0.5$ rather than 
$y = -\theta_S \sim 0.3$.

In order to check if $y$ coincides with $-\theta$,
a direct numerical estimate of the droplet excitation exponent $\theta$
was carried out using a heuristic optimization 
procedure\scite{KawashimaA1999,Kawashima2000} applied to
the EA model with a Gaussian bond-distribution in two dimensions. For
each realization of the model, the ground state spin-configuration was
computed with free boundary condition. Then, the spins on the
boundary were fixed as they were in the original ground state whereas
the spin at the center was fixed in the opposite direction. These
constraints lead to a new ground state that is identical to the old
one on the boundary but differs from it in the vicinity of the
central spin by a droplet of flipped spins around the center.
The droplet is typically as large as the system itself.
The system-size dependence of the droplet volume $V$ 
and the droplet excitation energy $E$ could be described well by the scaling law
$$
  V(L) \propto L^{d_D},\quad\mbox{and}\quad 
  E(L) \propto (V(L))^{\theta/d_D}
$$
with 
$$
  d_D = 1.80(2),\quad\mbox{and}\quad
  -\theta = 0.47(5)
$$
for the range of the system size $5 \le L \le 49$.
In particular, the value of $-\theta$ agreed with
most of the previous estimates of $y$.
This can be understood also within the droplet theory
because all scaling forms derived by droplet arguments
are identical to what one can get via the
ordinary finite size scaling by identifying $\theta$ and $-y$.
%since one can derive from the argument the scaling forms 
%of various quantities in which $-\theta$ appears in place of $y$.

However, a recent computation\scite{HartmannM2003a}
demonstrates the presence of a cross-over in the
droplet excitation energy. They performed essentially the same 
calculations as the one described above for larger systems (up to $L = 160$).
and argued that there may be a correction term
due to the self-interaction of the surface of the droplets
and that the droplet excitation energy has the form
$$
  \Delta E(L) \sim A L^{\theta} + B L^{-\omega}
$$
with $\omega > |\theta_D|$.
The numerical data could be well fitted by 
$$
  -\theta_D =0.29,\quad\mbox{and}\quad \omega = 0.97(5).
$$
as is shown in \Fig{HartmannM2003a_fig1and2}(a).
%%%%%%%%%%%%%%%%%%%%%%%%
\deffig{HartmannM2003a_fig1and2}{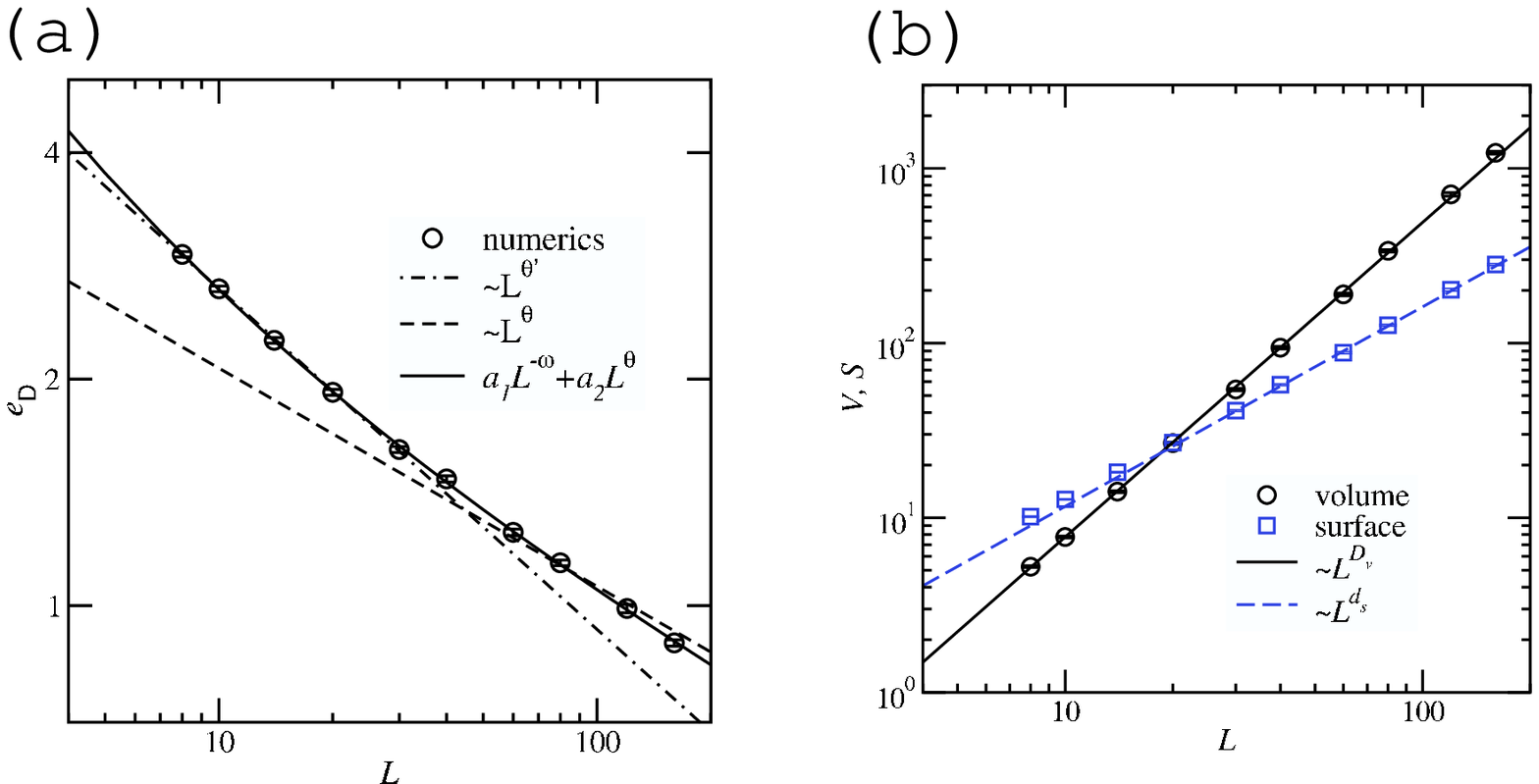}{0.8}
{
(a) The droplet energy as a function of the system size.
The two dashed straight lines represent the algebraic
dependences $L^{-0.47}$ for smaller systems and $L^{-0.29}$
for larger ones, respectively.
The solid curve is a fitting function (see text).
(b) The volume and surface of the droplets.
(From \figcite{HartmannM2003a}.)
}
%%%%%%%%%%%%%%%%%%%%%%%%
The fact that the estimate of the droplet exponent
now becomes close to the stiffness exponent indicates
the validity of the simplest scenario.
However, the result on the fractal dimension of droplets
seems to suggest the contorary;
their result appears to confirm the previous result
$d_D \sim 1.8$ (\Fig{HartmannM2003a_fig1and2}(b)).
While this may be another transient behavior,
the problem of compactness of the droplets
remains open in the two-dimensional system.

The same model was studied in \bcite{Middleton1999} and then in 
\bcite{PalassiniY1999a}
at zero-temperature using a different approach.
For each realization of the model,
the ground state for periodic boundary condition was compared
with the ground state for another set of boundary conditions, such as
anti-periodic and random boundary conditions.
(In the latter, half of the bonds across the boundary 
are chosen at random and inverted while the other half are kept
unchanged.)
%In the RSB picture, such a perturbation would change a ground state into 
%a different one that has nearly zero overlap with the first,
%whereas in the droplet-picture the only possible change that
%a such a perturbation may cause is the creation of a domain wall.
The authors of \bcite{PalassiniY1999a} argued that 
in the RSB picture the change in the boundary condition would induce 
an excitation whose surface-to-volume ratio should not diminish as the
system size increases,
whereas in the droplet picture
it should decrease down to zero.
Therefore, we can differentiate between the two pictures
by applying a boundary condition that induces a domain wall 
in the system and focusing on a small box located at the 
center of the system.
If the probability of having the domain wall crossing this part 
is finite even in the thermodynamic limit,
it would indicate the validity of the RSB picture.
On the other hand, it should decrease to zero 
if the droplet picture is valid.

They performed a calculation of exact ground states of
many instances of the EA model in two dimensions with the
Gaussian bond distribution.
The system size explored was up to $L=30$.
They considered a box of the linear size of $L_{\rm box}=2$
located at the center of the system,
and measured the distribution of the overlap $q_{\rm box}$
on this box:
$$
  q_{\rm box}^{\alpha\beta} \equiv \frac{1}{L_{\rm box}^d}
  \sum_{i\in {\rm ``box''}}
  S_i^{\alpha} S_i^{\beta}
$$
where $\alpha$ and $\beta$ are the indices specifying real replicas.
If the boundary passes the box with a finite probability,
the distribution of $q_{\rm box}^{\alpha\beta}$ should have
a finite weight around $q_{\rm box}^{\alpha\beta} = 0$.
This weight turned out to be decreasing 
with the power of $-0.7$ as a function of the system size.
Based on this result, they concluded that the zero-temperature 
structure of the two-dimensional EA model is trivial, i.e., 
only two pure states exist. (See \Fig{BoxOverlap}.)

%------------------------------------------------------------------------------
\subsection{Finite-Temperature Transition?}
%------------------------------------------------------------------------------
There are some numerical results\scite{ShirakuraM1996} 
of Monte Carlo simulation on the $\pm J$ model
that suggest the existence of a phase transition at a finite temperature.
Specifically, the overlap distribution $P(q;L)$ 
and the binder parameter $g$ were computed.
The estimates of the Binder parameter
\begin{equation}
  g_q \equiv \frac12\left( 
  3 - \frac{[\langle q^4 \rangle]}{[\langle q^2 \rangle]^2}
  \right)
  \label{eq:BinderParameter}
\end{equation}
as a function of the temperature for various system sizes appeared to
have a common crossing point.  The finite-size-scaling plot 
seems better when a finite transition temperature $T_c \sim 0.24J$ 
was assumed in stead of zero transition temperature. 
Their estimate for the critical indices of this finite
temperature transition are $\nu \sim 1.8$ and $\eta \sim 0.2$ The
overlap distribution $P(q;L)$ could be nicely scaled with the same
$\eta$ but with a slightly higher temperature $T_c \approx 0.29J$. 
Similar results were obtained\scite{ShirakuraM1997} for an asymmetric bond
distribution.  

Numerical evidences suggest that the stiffness exponent
of the two-dimensional $\pm J$ model is non-positive,
as discussed in \Ssc{ZTT},
which indicates that the lower critical dimension of the
$\pm J$ model is equal to or greater than two.
Now, if a spin-glass transition takes place at a finite
temperature as suggested, it means that the lower critical dimension
is not greater than two.
Therefore the only scenario consistent with all the
available numerical and analytical results is that the lower-critical
dimension is exactly two and that the stiffness
exponent is exactly zero in two dimensions. 
More evidences, however, appear to be necessary to settle this issue
beyond reasonable doubts.

%==============================================================================
\section{Equilibrium Properties of Three-Dimensional Models}
%==============================================================================
\label{sec:Ising3D}
\rem{Have to make sure all the entries in Table 1 are correct.}
The problem of spin glasses in three dimensions is the central topic 
of the field.
Whereas the RSB nature of the low-temperature phase in
four dimensions is much less controversial, 
the nature of the low-temperature phase in three dimensions 
still remains the subject of an active debate.

Below we present a number of numerical results.
Because of the severe technical limitations for three dimensional cases,
all the results obtained are for small systems, typically up to the linear
size of about 10 lattice spacings or slightly more.
Therefore, many important issues are left open,
in particular the question as to what picture yields 
the correct description of the low-temperature phase of the EA model,
which is the main subject of
\Ssc{LowTemperaturePhasePMJ} through \Ssc{Arguments}.

%------------------------------------------------------------------------------
\subsection{Finite Temperature Transition?}
%------------------------------------------------------------------------------
Even the very existence of a phase transition at finite temperature
was not easy to establish.
An evidence of the existence was obtained through the calculation of
the domain wall energy at zero-temperature,\scite{McMillan1984b}
in which a positive estimate for the stiffness exponent was obtained.
However, the system size was rather limited ($L\le 8$)
and the estimated value of the stiffness exponent was small
($\theta \sim 0.2$).
Therefore, this finding about the stiffness exponent
alone was not sufficient to establish the existence of 
a finite-temperature phase transition.
Finite temperature approaches could not settle the issue, either.
For example, while 
Monte Carlo simulations\scite{BhattY1985,Ogielski1985} strongly 
suggested the existence of a transition at a finite temperature, 
they could not rule out the possibility of zero-temperature singularity
with an exponentially diverging correlation length;
it was suggested\scite{MarinariPR1994} 
that all existing data at that time (namely 1994) were 
consistent with both hypotheses: $T_c>0$ and $T_c = 0$.
In particular, the simulation results of \bcite{MarinariPR1994} 
for the spin-glass
susceptibility could be fitted by a functional form consistent with
a zero-temperature singularity,
$
  \chi_{\rm SG} = 1 + A/(T-T_c)^{\gamma},
$
just as well as the one consistent with a
finite-temperature transition
$
  \chi_{\rm SG} = A\left(e^{(B/T)^p}-1\right) + C.
$

However, another set of Monte-Carlo results\scite{KawashimaY1996}
showed beyond reasonable doubts the existence of a finite
temperature transition in the $\pm J$ model in three dimensions.
Specifically, it demonstrated that the Binder parameter curves 
for different system sizes cross at the same point 
(\Fig{pmJ_3D_BinderParameter}) near
$$
  T \sim 1.1,\quad \mbox{and}\quad g_{\rm sg} \sim 0.75.
$$
With the assumption of the algebraic singularity at the critical point,
the critical temperature and the critical indices were estimated as
$$
  T_c = 1.11(4),\quad \nu = 1.7(3),\quad\mbox{and}\quad \eta = -0.35(5).
$$
These estimates were consistent with previous 
ones\scite{BhattY1985} and confirmed by other simulations as presented in
\Tab{ExponentsInThreeDimensions}.
\deffig{pmJ_3D_BinderParameter}{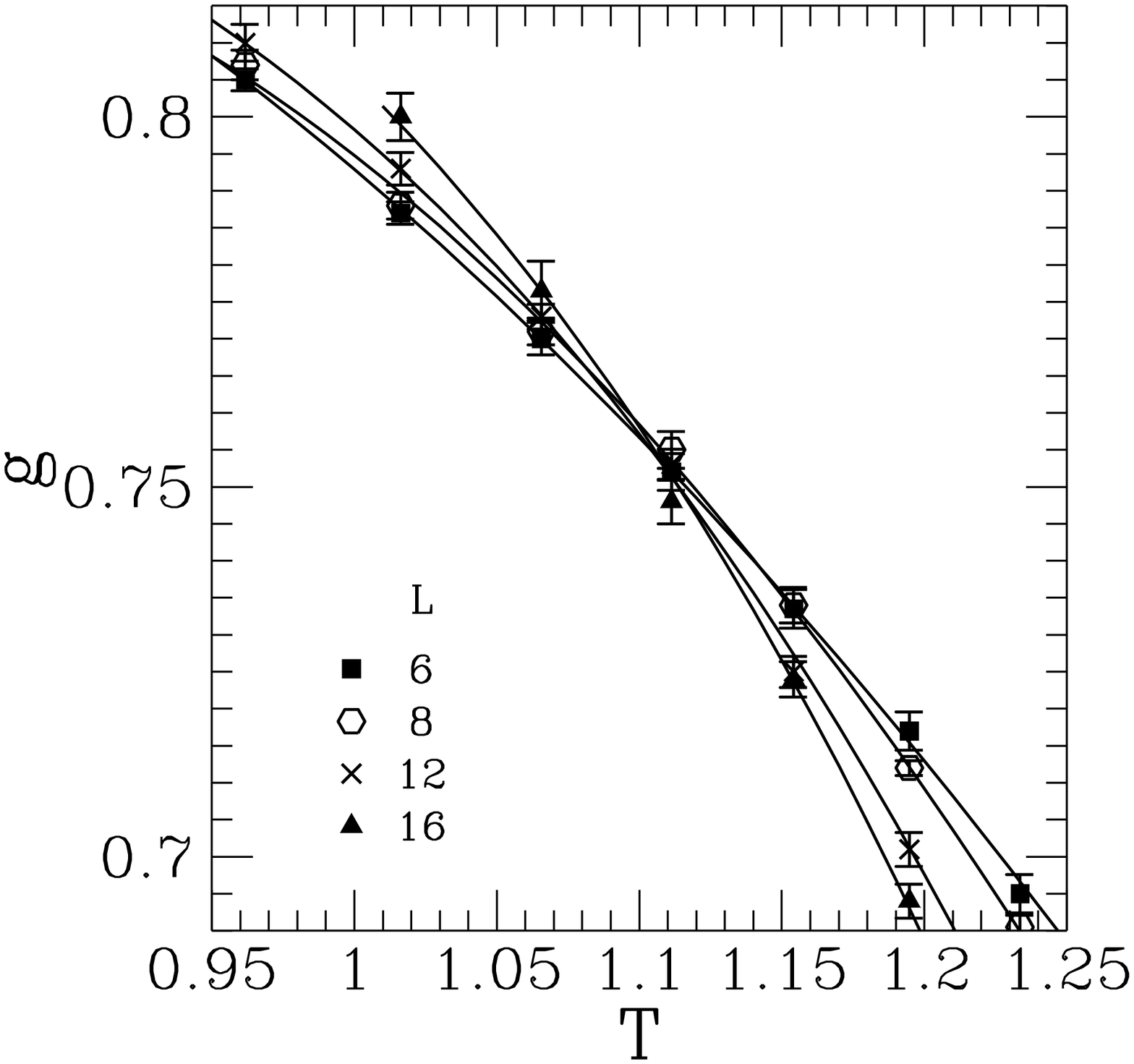}{0.5}
{The Binder parameter vs temperature for the $\pm J$
model in three dimensions.
(From \figcite{KawashimaY1996}.)}

A Monte Carlo simulation for larger system sizes\scite{PalassiniC1999}
clarified the issue further.  The results of the correlation length
and the spin-glass susceptibility were extrapolated to the infinite
system-size limit via finite size scaling, and were well fitted by a
curve representing an algebraic divergence at a finite temperature.
While the existence of a finite temperature phase
transition could be concluded from this result, 
the nature of the singularities at the
critical point could not unambiguously be settled.
(The singularity could be an essential rather than 
an algebraic singularity.)
In the next subsection, we present other numerical results
concerning the nature of the critical point.

%------------------------------------------------------------------------------
\subsection{Universality Class}
%------------------------------------------------------------------------------
Early Monte Carlo simulations such as \bcite{BhattY1985} 
indicated a scenario
that the system stays critical in the whole low temperature region.
Namely, these data could be explained by
a line of critical points terminating at $T_c = 1.2$, 
similar to the Berezinskii-Kosterlitz-Thouless
transition in the two-dimensional XY ferromagnet.
The conclusion of \bcite{PalassiniC1999} on this issue was similar,
while it was suggested that a finite-temperature transition 
with a algebraic divergence is most likely.
With the assumption of the algebraic divergence,
the critical parameters were estimated to be
consistent with previous ones mentioned above.
%$$
%  Tc = 1.156(15),\quad
%  \nu = 1.8(2),\quad \mbox{and}
%  \eta = -0.26(4).
%$$
However, the data were also consistent with an exponential 
divergence as in the BKT transition.

A large-scale Monte Carlo simulation\scite{BallesterosETAL2000}
clarified the issue of the nature of the phase transition.
Using special purpose machines and parallel tempering 
method\scite{TesiROW1996,HukushimaN1996}
the authors succeeded to equilibrate the $\pm J$ model of sizes up to
$L=20$ down to a temperature low enough to cover a sufficiently large
region around the transition point.  They obtained various results
consistent with an ordinary second order phase transition. For
example, they observed a clear crossing in the effective correlation
length\scite{CooperFP1989} $\xi(L)$ divided by the system size $L$.
(They defined the effective 
correlation length as $\xi^2(L)	\equiv \tilde k_m^{-2} 
(C_q(0)/C_q(\vect{k}_m) - 1)$,
where $C_q(\vect{k})$ is the static structure factor,
$\vect{k}_m$ is the smallest non-zero wave number compatible
to the boundary condition, and 
$\tilde k^2 \equiv 4 ( \sin^2(k^x/2) + \sin^2(k^y/2) + \sin^2(k^z/2) )$.)

%%%%%%%%%%%%%%%%%%%%%%%%
\deffig{BallesterosETAL2000_fig3}{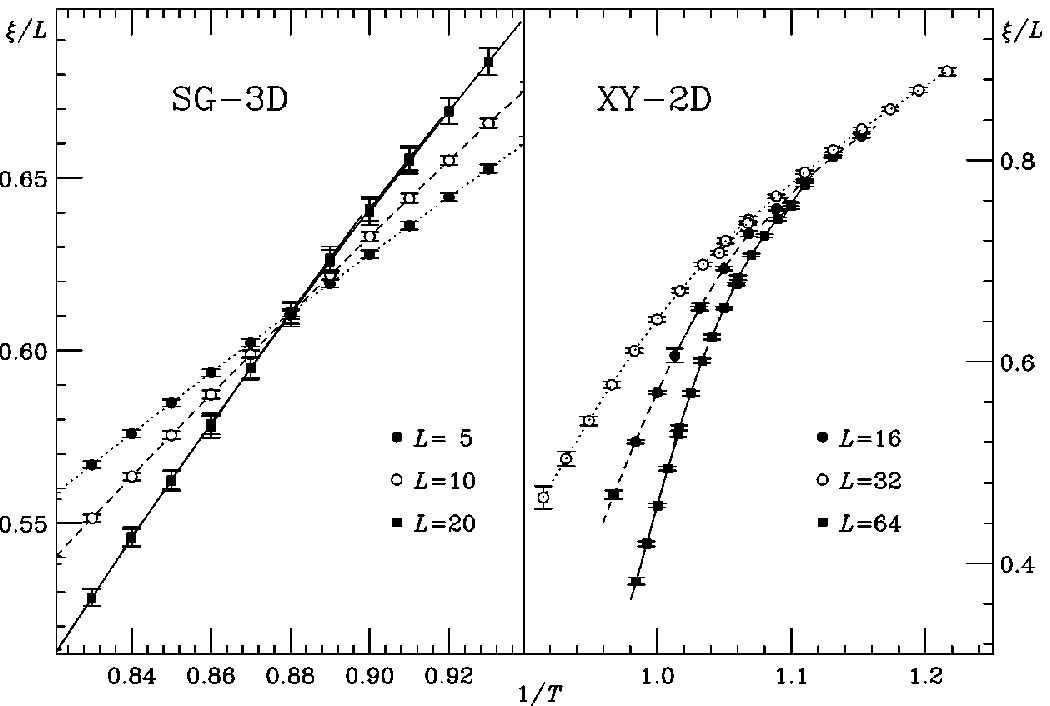}{0.8}
{
The effective correlation length divided by the system size.
The left panel is for the $\pm J$ model in three dimensions
whereas the right panel is for the $XY$ model in two dimensions.
(From \figcite{BallesterosETAL2000}.)
}
%%%%%%%%%%%%%%%%%%%%%%%%
%
The scaling dimension of the effective correlation length is zero;
the curves for various system sizes should intersects at the same point
similar to the Binder parameter.
Indeed, it shows a common intersection point
as in the right panel of \Fig{BallesterosETAL2000_fig3}.
This is in marked contrast to what we typically see 
in a BKT type transition, which is shown in 
the right panel of \Fig{BallesterosETAL2000_fig3} for comparison.
From this result one concludes that the phase transition is
an ordinary second-order one.
For the critical temperature and the indices,
again, see \Tab{ExponentsInThreeDimensions}.

The results of \bcite{BallesterosETAL2000}
provided also indications in favor of an 
RSB nature of the system below the
critical temperature.
Namely, the estimates of a quantity called the $G$-parameter 
seems to have a finite value even below the critical point.
The $G$-parameter is defined\scite{MarinariETAL1998} as
$$
  G \equiv 
  \frac{[\langle q^2 \rangle^2]-[\langle q^2\rangle]^2}
       {3[\langle q^2 \rangle]^2 - [\langle q^4 \rangle]}.
$$
The numerator is finite only if the sample-to-sample dependence of
$P(q)$ does not vanish, while the denominator can be finite 
even if there is no sample dependence.
Therefore, a observed non-vanishing value of $G$ 
implies that $P(q)$ is not self-averaging.

However, there are a number of evidences by other groups that
can be interpreted otherwise, namely, in favor of the absence of an RSB 
nature of the low temperature phase, as we discuss below.

\begin{sidewaystable}
\tbl{
The estimates of the critical temperature and the exponents for the three
dimensional spin glass models. 
The entries are categorized into three groups according to
the anisotropy; easy-axis (top), isotropic (middle), and easy-plane (bottom).
The entries with ``${}^*$'' are not quoted in the original paper
but estimated through the scaling relations by the present authors.
The entries with ``${}^{\rm c}$'' are for the chiral glass transition.
\label{tab:ExponentsInThreeDimensions}.
}
{\begin{tabular}{llllllll}
\hline
Authors\                 & Material/Model   & $T_c/J$ & $\nu$      & $\eta$       & $\beta$    & $\gamma$  & z \\
%%%%%%%%%%%%%%%%%%%%%%%%%%%%%%%%%%%%%%%%%%%%%%%%%%%%%%%%%%%%%%%%%%%%%%%%%%%%%%%%%%%%%%%%%%%%%%%%%%%%%%%%%%%%%%%
\hline
Gunnarsson {\it et al.}\scite{GunnarssonETAL1991} &
             FeMnTiO                     & \na        & 1.7        & $-$0.35      & 0.54       & 4.0(3)     & 6.2 \\
Ogielski\scite{Ogielski1985} &
             Ising, $\pm J$              & 1.175(25)  & 1.3(1)     & $-$0.22(5)   & 0.5        & 2.9(3)     & 6.0(5) \\
Bhatt-Young\scite{BhattY1985} &
             Ising, $\pm J$        & 1.2${+0.1\atop -0.2}$ & 1.3(3)  & $-$0.3(2)  & 0.46\sast  & 3.2        & \na \\
Singh-Chakravarty\scite{SinghC1987} &
             Ising, $\pm J$              & 1.2(1)     & \na        & \na          & \na        & 2.9(5)     & \na \\
Bhatt-Young\scite{BhattY1988} &
             Ising, Gaussian             & 0.9        & 1.6(4)     & $-$0.4(2)    & \na        & \na        & \na \\
Kawashima-Young\scite{KawashimaY1996} &
             Ising, $\pm J$              & 1.11(4)    & 1.7(3)     & $-$0.35(5)   & 0.55\sast  & 4.0\sast   & \na \\
I\~nigues {\it et al.}\scite{IniguesPR1996} &
             Ising, Gaussian             & 1.02(5)    & 1.5(3)     & \na          & \na        & \na        & \na \\
Marinari {\it et al.}\scite{MarinariPR1998} &
             Ising, Gaussian             & 0.98(5)    & 2.00(15)   & $-$0.36(6)   & 0.64\sast  & 4.72\sast  & \na \\
Berg-Janke\scite{BergJ1998} &
             Ising, $\pm J$              & 0.88       & \na        & $-$0.37(4)   & \na        & \na        & \na \\
Palassini-Caracciolo\scite{PalassiniC1999} &
             Ising, $\pm J$              & 1.156(15)  & 1.8(2)     & $-$0.26(4)   & 0.65\sast  & 4.1(5)     & \na \\
Mari-Campbell\scite{MariC1999} &
             Ising, $\pm J$              & 1.19(1)    & 1.33(5)    & $-$0.22(2)   & 0.52\sast  & 2.95(15)   & \na \\
Ballesteros {\it et al.}\scite{BallesterosETAL2000} &
             Ising, $\pm J$              & 1.138(10)  & 2.15(15)   & $-$0.337(15) & 0.73\sast  & 5.0\sast   & \na \\
Mari-Campbell\scite{MariC2002} &
             Ising, $\pm J$              & 1.195(15)  & 1.35(10)   & $-$0.225(25) & 0.55\sast  & 2.95(30)   & 5.65(15) \\
Nakamura {\it et al.}\scite{NakamuraEY2003} &
             Ising, $\pm J$              & 1.17(4)    & 1.5(3)     & $-$0.4(1)    & 0.45\sast  & 3.6(6)     & 6.2(2) \\
%%%%%%%%%%%%%%%%%%%%%%%%%%%%%%%%%%%%%%%%%%%%%%%%%%%%%%%%%%%%%%%%%%%%%%%%%%%%%%%%%%%%%%%%%%%%%%%%%%%%%%%%%%%%%%%
\hline
de Courtenary {\it et al.}\scite{CourtenaryBHF1986} &
             CuMn,AgMn                   & \na        & 1.4\sast   & 0.4\sast     & 1.0(1)     & 2.2(1)     & \na \\
Bouchiat\scite{Bouchiat1986} &
             AgMn                        & \na        & 1.4\sast   & 0.4\sast     & 1.0(1)     & 2.2(2)     & \na \\
Levy-Ogielski\scite{LevyO1986} &
             AgMn                        & \na        & 1.3(2)     & 0.4\sast     & 0.9(2)     & 2.1(1)     & 5.5 \\
Simpson\scite{Simpson1979} &
             CuAlMn                      & \na        & 1.3\sast   & 0.5\sast     & 1.0\sast   & 1.9\sast   & \na \\
Coles-Williams\scite{ColesW1988} &
             PdMn                        & \na        & 1.3\sast   & 0.4\sast     & 0.90(15)   & 2.0(2)     & \na \\
Vincent-Hamman\scite{VincentH1987} &
             CdCrInS                     & \na        & 1.25(25)   & 0.2\sast     & 0.75(10)   & 2.3(4)     & 5.5 \\
Kawamura\scite{Kawamura1998} &
             Heisenberg, Gaussian        & 0.157(10)$^{\rm c}$ & \na & \na          & 1.1(1)$^{\rm c}$ & \na    & \na \\
Hukushima-Kawamura\scite{HukushimaK2000a} &
             Heisenberg, Gaussian        & 0.160(5)$^{\rm c}$ & 1.2$^{\rm c}$ & 0.8$^{\rm c}$ & 1.1(1)$^{\rm c}$ & 1.5(3)$^{\rm c}$ & \na \\
Matsubara {\it et al.}\scite{MatsubaraSE2001} &
             Heisenberg, $\pm J$         & 0.18       & \na        & \na          & \na        & \na        & \na \\
Nakamura-Endoh\scite{NakamuraE2002} &
             Heisenberg, $\pm J$         & 0.21${+0.01\atop -0.03}$ & 1.1(2) & 0.3\sast & 0.72(6)    & 1.9(4) & 4.5 \\
Lee-Young\scite{LeeY2003} &
             Heisenberg, Gaussian        & 0.16(2)    & 1.1(2)     & \na          & \na        & \na        & \na \\
Nakamura {\it et al.}\scite{NakamuraEY2003} &
             Heisenberg, $\pm J$         & 0.20(2)    & 0.8(2)     & $-$0.3(3)    & \na        & 1.9(5)     & 6.2(5) \\
%%%%%%%%%%%%%%%%%%%%%%%%%%%%%%%%%%%%%%%%%%%%%%%%%%%%%%%%%%%%%%%%%%%%%%%%%%%%%%%%%%%%%%%%%%%%%%%%%%%%%%%%%%%%%%%
\hline
Kawamura-Li\scite{KawamuraL2001} &
             $XY$, $\pm J$               & 0.39(3)$^{\rm c}$ & 1.2(2)$^{\rm c}$ & 0.15(20)$^{\rm c}$ & \na & \na & 7.4(10)$^{\rm c}$  \\
Lee-Young\scite{LeeY2003} &
             $XY$, Gaussian              & 0.34(2)    & 1.2(2)     & \na          & \na        & \na        & \na \\
Nakamura {\it et al.}\scite{NakamuraEY2003} &
             $XY$, $\pm J$               & 0.43(3)    & \na        & $-$0.4(2)    & \na        & \na        & 6.8(5) \\
%%%%%%%%%%%%%%%%%%%%%%%%%%%%%%%%%%%%%%%%%%%%%%%%%%%%%%%%%%%%%%%%%%%%%%%%%%%%%%%%%%%%%%%%%%%%%%%%%%%%%%%%%%%%%%%
\hline
\end{tabular}}
\end{sidewaystable}

%------------------------------------------------------------------------------
\subsection{Low-Temperature Phase of the $\pm J$ Model}
\ssclabel{LowTemperaturePhasePMJ}
%------------------------------------------------------------------------------
There is a very active debate on the nature of the low-temperature
phase of three dimensional spin glass models.
In the following few subsections, we review various theories,
arguments and numerical calculations that were made or done
with the ultimate aim to clarify the issue. But none
of the pictures, RSB, droplet, or others,
could be established so far.
Nonetheless, the debate itself is interesting and each picture
is worth being scrutinized with detailed numerical calculations. 
While the common belief is that the $\pm J$ model and the 
Gaussian-coupling model show essentially the same physics 
at finite temperature,
we present them separately since many of the computations are done
at zero temperature where the two may differ.
The $\pm J$ model is discussed in the present subsection 
and the Gaussian-coupling model in the next.

A finite temperature simulation on the $\pm J$ model
was performed\scite{PalassiniY2001}
using the replica exchange Monte Carlo simulation\scite{HukushimaN1996}
for $\pm J$ model reaching down to $T=0.2$ for the system size 
up to $L=10$.
The integrated overlap probability distribution 
$$
  x(1/2) \equiv \int_0^{1/2} dq P(q)
$$
was computed as a function of 
the temperature and the system size.
Inspired by \bcite{KrzakalaM2001a},
the scaling --- $x(1/2)$ being proportional to $L^{\lambda}$
for small $L$ and to $L^{\theta}$ for larger $L$ --- was assumed.
In other words, 
$$
  x(1/2) = T L^{-\theta} f(T L^{\lambda})
$$
was assumed.
The fitting to the numerical results worked nicely with
$\theta = 0$ yielding
$$
  \lambda = 0.9(1),
$$
whereas a fitting with $\theta = 0.2$, the droplet prediction, 
turned out to be significantly worse.
The estimate of $\lambda$
is considerably smaller than a preceding estimate\scite{Hartmann1999c} 
while consistent with the value
$\lambda = 0.72(12)$ in \bcite{BergHC1994}.
%\deffig{IntegratedProbability}{PalassiniY2001_fig5.eps}{0.5}
%{The scaling plot of the integrated probability distribution
%of $q$ in the interval $-0.5 \le q \le 0.5$ for
%the $\pm J$ model in three dimensions.
%(Adopted from \figcite{PalassiniY2001}.)}

Zero-temperature calculations offer a powerful alternative 
to the finite temperature approach with Monte Carlo simulations,
because one might expect that the differences between 
the two scenarios should be more prominent at
lower temperature, and much better methods are available for
systems at zero-temperature
than for a very low but finite temperature.
For three dimensional models, most methods for solving 
zero-temperature problems are based on
heuristic optimization since no good exact method 
is available due to the NP-hardness of the problem.
A computation of the ground states of the $\pm J$ Ising model up to $L=14$
was done\scite{Hartmann1997,Hartmann1998} with a heuristic algorithm
called the cluster-exact approximation method.\scite{Hartmann-Rieger}
Later the computation was redone,\scite{Hartmann1999c,Hartmann2000a}
in order to fix the problem of the biased sampling.\scite{Sandvik1999}
It was found that the width of the overlap distribution, $P(q)$,
decreases as the system becomes larger, 
indicating the triviality of $P(q)$.
The validity of the ultra-metric relation,
 $q_{12} = q_{23} < q_{31}$, was also examined,
where $q_{ij}$'s are the overlaps among three randomly 
chosen ground states.
% $S_1^0$, $S_2^0$ and $S_3^0$.
The numerical results indicated that the ultra-metric relation holds 
for a typical triplet of ground states with relatively large mutual distances.
However, it was also found that the contribution from these triplets 
to $P(q)$, which constitutes the ``non-trivial'' (i.e., continuous)
component of $P(q)$, decreases as the system becomes larger.
In fact, the integrated weight of $P(q)$ systematically decreased
toward zero as
$$
  x(q) \equiv \int_{-q}^q dq P(q) \propto L^{-\lambda}
$$
with $\lambda = 1.25(5)$ for $q=0.5$.
This indicated a trivial structure of $P(q)$ at zero temperature.
As for the exponent $\lambda$, a different estimate was 
obtained\scite{PalassiniY2001} as mentioned below.

The authors of \bcite{KrzakalaM2001a} considered $P(q)$
for the $\pm J$ model at $T=0$, and pointed out that
the ground state space may be dominated by a single valley or a single
pure state no matter which picture is valid.
They argued that even if there are
multiple valleys in the energy landscape of the
discrete energy model, 
the distribution of the overlap $P(q)$ may still be trivial,
indicating that it is impossible to discriminate
between the two scenarios by a computation of $P(q)$
such as the one mentioned above.

The underlying assumption of the argument
is the many-valley structure in the phase space and 
the sponge-like structure of low-lying excitations in the real space.
(The latter is further discussed below in \Ssc{Sponge} and \Ssc{TNT}.)
A droplet typically has a finite volume
and when flipped it takes a state in a valley to another state in
the same valley.
In contrast, a typical sponge-like cluster is supposed to occupy 
a finite fraction of the whole system and
flipping it generally causes a transition from one valley to another.

This argument is based on an estimate of the entropy by counting 
zero-energy excitations from a particular ground state.
In general, the continuous part of $P(q)$, if any,
is caused by excitations of various scales whose excitation free
energies are smaller than the temperature.
In particular, at zero temperature, 
it is caused by zero-energy ``excitations''.
They argued that $P(q)$ has 
a trivial structure even if there are multiple valleys.
To see this, it suffices to consider the case of two valleys.
Each valley contributes to $P(q)$ according to its weight
that is the number of distinct spin configurations in the valley.
This number is roughly the same as the number of zero-energy droplets.
We can choose two representative configurations, one from each
valley, that can be transformed to each other by flipping a
sponge-like cluster.
Then, the zero-energy droplet excitations in one configuration 
differ from those in the other configuration
only on (or near) the boundary of the sponge-like cluster.
%This is because the relative orientations of nearest neighbor spins
%in one representative configuration differ from those 
%in the other configuration only on the surface of the sponge-like cluster,
%the zero-energy droplet excitations around one of the two configurations 
%can differ from those around the other configuration
Therefore, if $A$ is the area of the surface of the sponge-like cluster,
the droplets differ only at $A$ positions, 
This difference results in the entropy difference of ${\cal O}(A^{1/2})$.
(The power $1/2$ comes from the assumption that the number of
the droplet-like excitation located on the surface is a
random variable.)
It follows that the contribution (to the continuous part of $P(q)$)
from one valley differs from that from the other typically by a factor 
$e^{\pm \mbox{(const)}\times A^{1/2}}$.
We now see that almost certainly $P(q)$ is dominated by
a contribution from a single valley in the thermodynamic limit
where $A\to\infty$,
leading to a trivial structure in $P(q)$.
It is clear that the presence of more than two valleys 
does not affect the result
as long as the number of valleys does not grow too fast as
the system size increases.
Based on this result, they argued that a trivial $P(q)$ for the
$\pm J$ model in three dimensions at $T=0$\scite{Hartmann2000a}
does not necessarily indicate the absence of the RSB.

However, it is not known 
how the number of valleys depends on the system size.
It is not too unrealistic to assume that the 
number of valleys grows faster than or proportional to
the number of spins, $N$.
If so, a typical minimum entropy difference between two valleys
would be less than $A^{1/2}/N < N^{-1/2} \to 0$.
Therefore, it is not clear if this argument really invalidates
the numerical evidence in \bcite{Hartmann2000a}
for triviality of the energy landscape structure.

%------------------------------------------------------------------------------
\subsection{Low-temperature Phase of the Gaussian-Coupling Model}
\ssclabel{LowTemperaturePhaseGCM}
%------------------------------------------------------------------------------
The ground state of models with the Gaussian bond distribution,
is unique up to a trivial degeneracy 
due to the $Z_2$ symmetry.
However, it is still possible to extract useful information 
about low-lying excitations from zero-temperature computations.
The authors of \bcite{PalassiniY1999b}
attempted to use the same strategy 
that they used for two dimensional models with
the Gaussian bond distribution\scite{PalassiniY1999a} 
(see \Ssc{TwoDimensionalGaussian})
to discriminate between the two scenarios
in three dimensions.
They estimated the probability of a domain-wall passing through
a small imaginary box placed inside the system.
A clear decreasing behavior as a function of the system size
was found, in favor of the droplet picture.
However, the amount of the total decrease that they could observe 
by changing the system size was only of a factor of 1.3 or 1.4, 
due to a severe system size limitation.
(In the case of two dimensions, the same quantity varies by almost 
an order of magnitude as can be seen in \Fig{BoxOverlap}.)
\deffig{BoxOverlap}{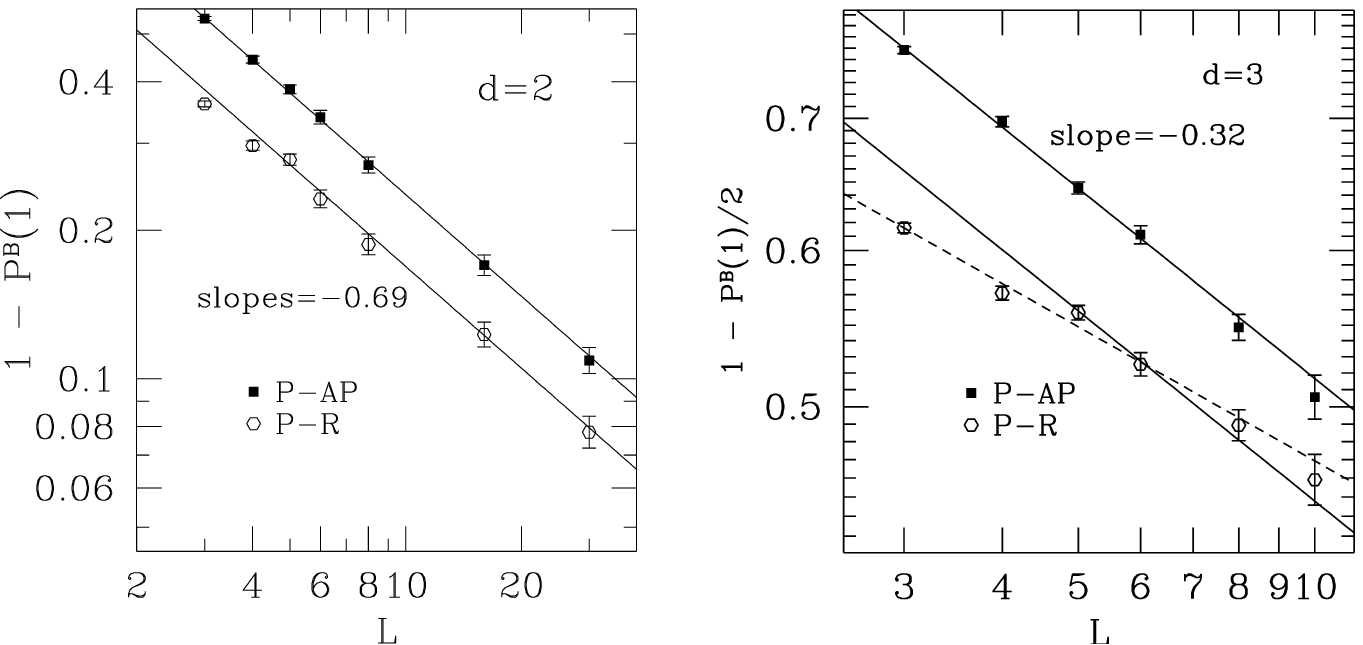}{0.9}
{The probability of a domain wall passing through the box.
%($P^{\rm B}(1)$ is the probability of the perfect overlap.)
The left panel is for the EA model with the Gaussian bond
distribution in two dimensions at $T=0$ and $H=0$,
whereas the right panel is for the same model in three dimensions.
(From \figcite{PalassiniY1999b}.)}

In addition to the response to the boundary condition,
the effect of a weak perturbation applied to the whole system
was also examined.\scite{PalassiniY2000a}
The perturbation was chosen in favor of the excited states
against the ground state.
For each bond realization, they first obtained the ground state,
which we denote $\{S_i^{(0)}\}$.
Then to each coupling constant $J_{ij}$ the following perturbation
was added
$$
  \Delta J_{ij} \equiv - \Delta J S_i^{(0)} S_j^{(0)}.
$$
where $\Delta J \equiv \epsilon / N_{\rm bond}$ 
was chosen to be of ${\cal O}(L^{-d})$.
With this perturbation the ground state energy is increased,
exactly, by $\epsilon$, whereas excited state energies are changed
according to its similarity to the ground state.
Therefore by adjusting the amplitude of the perturbation and
examining the overlap between the original ground state
and the ground state of the perturbed system,
one may obtain some information about the excitations.

The RSB picture predicts the existence of many spin states that are
local minima of the Hamiltonian.
These states differ from the ground state by only an infinitesimal amount
of energy per spin but differ by a macroscopic number of spins
such that the average Hamming distance 
between them (normalized by the system size) converges to some
finite value in the large system size limit.  
This means for the bulk perturbation considered in
\bigcite{PalassiniY2000a} that
the overlap between the two such minima, 
the one with the perturbation and the other without,
should be truly smaller than unity in the thermodynamic limit,
no matter how small the perturbation may be.
Therefore in the RSB picture, $1-[q]$ should be finite where
$q$ is the overlap between the two minima.
On the other hand, in the droplet picture, 
$1-[q]$ should depend on the system size by a power-law
being characterized by the droplet energy exponent $\theta$
and an exponent related to the geometrical properties of the excitations.
To be specific, since $1-[q]$ is dominated by the contribution from 
droplets whose linear size is comparable to the system size $L$,
it is roughly equal to the volume fraction of a typical droplet
of size $L$ multiplied by the probability of such a droplet
being excited.
Namely, 
\begin{equation}
  1-[q] \propto 
  \frac{L^{d_{\rm v}}}{L^{d}} \times \frac{\Delta_L}{\epsilon_L}
\end{equation}
where $\Delta_L$ is the energy gain by the droplet excitation
due to the perturbation 
and $\epsilon_L$ is the droplet excitation energy without the perturbation.
If the droplets are compact, as assumed in \bigcite{PalassiniY2000a},
only the second factor matters since $d_{\rm v} = d$.
For $\Delta_L$, the authors of \bcite{PalassiniY2000a}
assumed that $\epsilon \sim {\cal O}(1)$ when
integrated over the whole system, which is
equivalent to setting $\Delta_L \propto L^{-d+d_{\rm s}}$ where
$d_{\rm s}$ is the fractal dimension of the boundary of droplets.
Based on this, the following scaling form for the overlap was proposed
\bigcite{PalassiniY2000a}:
\begin{equation}
  v(\epsilon, L) \equiv 1-[q]
  \sim \tilde V(\epsilon / L^{d-d_{\rm s}-\theta}).
  \label{eq:BulkPerturbationOverlap}
\end{equation}
For the link overlap $q_l$, a similar argument yields
\begin{equation}
  s(\epsilon, L) \equiv 1-[q_l]
  \sim L^{-(d-d_{\rm s})} \tilde S(\epsilon / L^{d-d_{\rm s}-\theta}).
  \label{eq:SVRatio}
\end{equation}

In the system size range that could be explored, (namely, $L\le 8$), 
the numerical results for $[q]$ and $[q_l]$ shows
that $v(\epsilon,L)$ and $s(\epsilon,L)$ 
obeys the power law and decreases to zero as the
system size increases, contradicting the RSB picture.
However, fitting the numerical results
to the forms \Eq{BulkPerturbationOverlap} and \Eq{SVRatio} yielded 
$$
  d-d_{\rm s} = 0.42(2),\quad \mbox{and}\quad
  d-d_{\rm s}+\theta = 0.44(2).
$$
This implies
$$
  \theta = 0.02(3).
$$
The difference between the value and the stiffness exponent
($\sim 0.2$) obtained from the domain wall calculation 
is statistically significant,
i.e., the result contradicts not only to the RSB picture
but also to the droplet picture unless this seeming
contradiction is caused by a large correction to scaling.

Concerning the possible source of the correction,
two arguments\scite{Middleton2001,Moore2002,HartmannM2003a}
were presented to explain the inconsistency observed in numerical results 
within the framework of the droplet argument.
It was pointed out\scite{Middleton2001} that if one assumes 
a ``clean'' scaling with no correction for the droplet excitation energy,
the contribution from small droplets gives rise to a correction
to scaling in quantities such as the spin-glass susceptibility and 
the magnetic susceptibility.
The droplet excitation energy computed with the condition used in
\bigcite{KawashimaA1999,Kawashima2000}
and the link overlap computed in \bigcite{PalassiniY2000a}
may have been affected by such a correction to scaling.
Another possible source of a correction to scaling is
the interaction between domain walls.
It was argued\scite{Moore2002,HartmannM2003a}
that the energy of a domain wall may be increased 
by the presence of another domain wall and that 
the energy shift due to this interaction may have the form
$
  l^{-\omega'}
$
where $l$ is the distance between two domain walls.
Similarly, a ``self-interaction'' of the domain wall may give rise
to a correction to scaling in the droplet excitations as
$
  E = A l^{\theta} + B l^{-\omega}.
$
In this case $l$ stands for the size of the droplets.
As the correction-to-scaling exponent due to this mechanism,
the authors of \bcite{Moore2002} quoted
the value $\omega \sim 0.13(2)$\scite{LamarcqBMM2002}.

In response to some reports (see the following paragraphs in the
present section) contradicting to their conclusion,
the authors of \bcite{PalassiniLJY2002} further pursued the ground state 
nature of the three dimensional system with the Gaussian bond-distribution, 
along the same line as their own preceding calculation.\scite{PalassiniY2000a}
This time, however, larger systems ($L=12$) were 
dealt with the branch-and-cut algorithm.\scite{JuengerRT1995}
This algorithm guarantees that the states found are the true ground states.
In addition, a greater care was taken for various possibilities
of fitting functions and different sources of corrections to scaling.
Specifically, they considered a few different fitting functions
for size dependent quantities such as the link overlap and the box overlap.
Fitting functions consistent with the droplet, the TNT, and the
RSB scenario, respectively, were considered.
(For the TNT scenario, see \Ssc{Sponge} and \Ssc{TNT} below.)
It was found that the size dependence of the surface-to-volume ratio, i.e.,
$1 - q_{\rm l}$, could be explained by any one of three pictures,
and also that the size dependence of the box overlap could be 
explained by any one of three pictures. 
The estimates of the exponent $\mu \equiv d - d_{\rm s} + \theta$ 
turned out to depend on the boundary condition.
With the free boundary condition, the estimates are
$d - d_{\rm s} = 0.44(3)$, $\mu = 0.63(3)$, and $\theta = 0.19(6)$,
whereas for the periodic boundary condition, they are
$d - d_{\rm s} = 0.43(2)$, $\mu = 0.42(3)$, and $\theta = -0.01 (3)$.
The latter set of values are consistent with their preceding
estimates\scite{PalassiniY2000a} whereas the former are not.
%If the droplet picture is the valid one, this result indicates,
%as suggested by the authors of \bcite{Middleton2001} and \bcite{Moore2002},
%that an unknown correction-to-scaling is so large that the exponent $\theta$ 
%cannot be estimated by this method reliablly.
%On the other hand, there was no such inconsistency for the fitting function
%consistent with the RSB picture, while a large sub-leading term has
%to be taken into account in this case for making a reasonable fit to the data.

Another set of numerical results, presented in 
\bcite{MarinariP2000a,MarinariP2000b,MarinariP2001}
seems to contradict to the results of 
\bcite{PalassiniY2000a,PalassiniY1999b} presented above.
As for the calculation based on the small imaginary box 
placed at the center of the system,
it was argued\cite{MarinariP2000a} that
the probability of the state inside the box
being affected by the change in the boundary condition 
should obey the scaling 
\begin{equation}
  P_{\rm change}(L_{\rm box}, L) \sim g(L_{\rm box}/L),
  \label{eq:ProbChangeScaling}
\end{equation}
if the droplet argument is correct.
However, the numerical results in \bcite{MarinariP2000a}
for $L_{\rm box} = 2,3,4$ and $L=12$ did not fit in this scaling.

In \bcite{MarinariP2000a},
the scaling property of the domain wall
induced by the anti-periodic boundary condition 
in the $x$-direction was also examined.
In particular, the probability of the domain wall 
not intersecting a plane perpendicular to the $x$ axis was measured.
It was found that the non-intersecting probability approaches zero
as the system size increases:
% (\Fig{NonIntersectingProbability}):
$$
  P_L(\mbox{"The domain wall does not intersect the plane."}) 
  \propto L^{-\gamma},
$$
with $\gamma = 1.5-2.0$.
Based on this observation they suggested that the domain wall is
space filling, i.e., $d = d_{\rm s}$.
%\deffig{NonIntersectingProbability}{MarinariP2000a_fig2.eps}{0.5}
%{The probability that the domain wall induced by the anti-periodic
%boundary condition in the $x$-direction does not intersect a $yz$ plane.
%The model is the EA model with the Gaussian bond distribution 
%at zero temperature in three dimensions.
%(Adopted from \figcite{MarinariP2000a}.)}

In \bcite{MarinariP2000b,MarinariP2001},
zero-temperature calculations were performed
with the conditions analogous to the ones in 
\bigcite{PalassiniY1999b} and \bigcite{PalassiniY2000a}.
In \bigcite{MarinariP2000b}, the effect of the
anti-periodic boundary conditions, which were imposed in the $x$ direction, 
was compared with the periodic one.
The links perpendicular to the $yz$ plane were treated
separately from those parallel to the $yz$ plane.
The result of the link overlap for the perpendicular links, $q_P$,
and that for the transverse ones, $q_T$, are shown in 
\Fig{MarinariP}(a).
%%%%
%\Fig{MarinariP2000b_fig1}
%\deffig{MarinariP2000b_fig1}{MarinariP2000b_fig1.eps}{0.5}
%\deffig{MarinariP2001_fig1}{MarinariP2001_fig1.eps}{0.5}
\deffig{MarinariP}{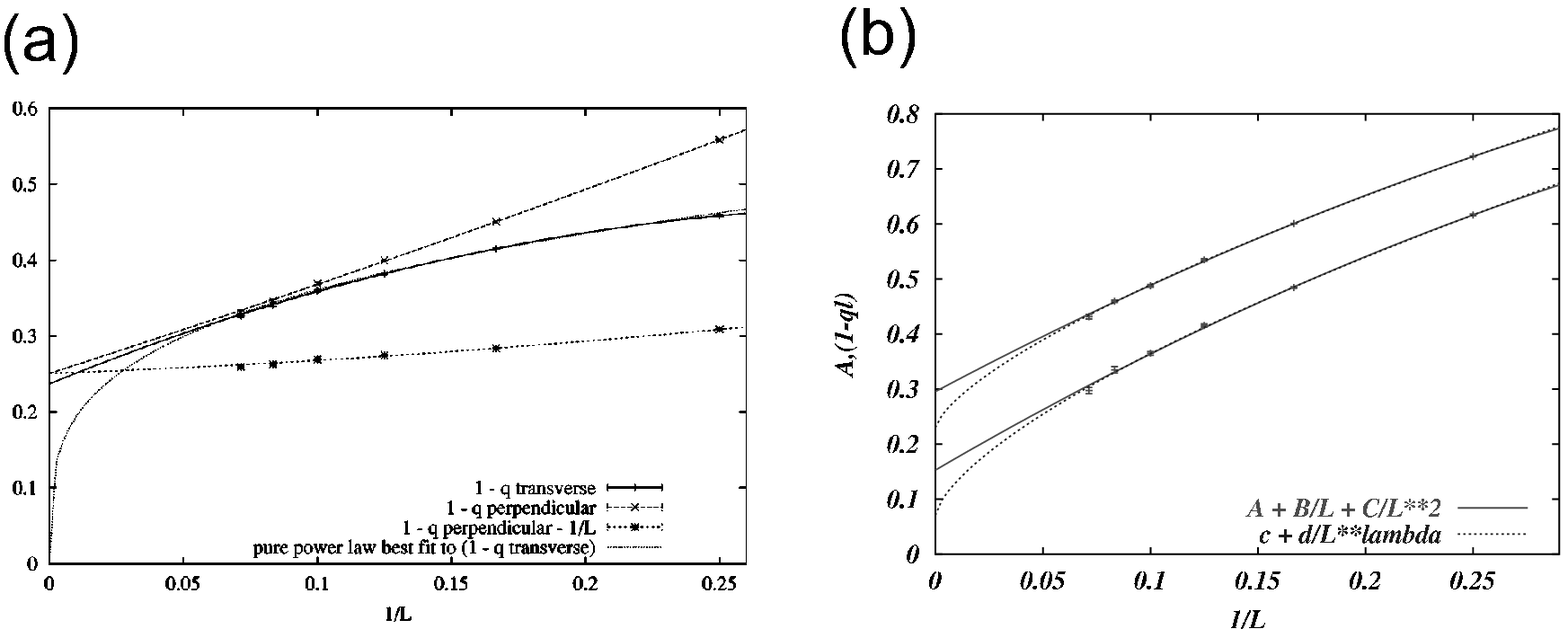}{0.95}
{(a) The link-overlap for the perpendicular links, $q_P$, and
that for the transverse ones, $q_T$, between the two ground states
with and without a twist in the boundary,
and (b) the link-overlap (subtracted from unity)
between the two ground states with and without a bulk perturbation.
(The lower set of data is for the average link-overlap, whereas
the upper one is for the link-overlap restricted to those pairs
of states which have zero mutual overlaps.
For each set of data, the upper fitting curve is a second order 
polynomial in $1/L$ and the lower one is a fractional power in
$1/L$ with an additional constant term.)
For both (a) and (b), the model is the three-dimensional EA model 
with the Gaussian bond distribution at zero temperature.
(From (a) \figcite{MarinariP2000b} and 
(b) \figcite{MarinariP2001}.)}
%%%%
Both kinds of the link overlap can be fitted well by second 
order polynomials in $1/L$ as shown in the figure.
From the zero-th order term, they concluded that
$$
  \lim_{L\to\infty} 1-q_l = 0.245(15),
$$
in contrast to the droplet prediction $1-q_l \to 0$.

In \bigcite{MarinariP2001}, the effect of 
the bulk perturbation at zero temperature was studied.
The perturbation was the same as the one
considered in \bigcite{PalassiniY2000a}.
The link overlap, $q_l(q)$, was computed as a function of 
the bulk overlap, $q$.
In addition to the average link-overlap, $q_l(L)$,
attention was paid to the quantity $A(L) \equiv 1-q_l(q=0)$, 
where $q_l(q=0)$ is the link overlap 
between two states that have zero overlap.
In the droplet picture, both quantities should converge to $0$ as
$L$ goes to infinity.
The numerical results are shown in \Fig{MarinariP}(b).
It was found that 
in both cases of the boundary and the bulk perturbations,
the size dependence of the 
surface-to-volume ratio of the domain wall, i.e., $1-q_l$
can be fitted by a polynomial in $1/L$ with a finite constant term, 
consistent with the RSB picture.

Concerning the results presented in \bcite{MarinariP2000a},
it was pointed out\scite{PalassiniY2000b}
that the probability $P_{\rm change}$ is subject to a
strong correction to scaling and that, taking such a correction term
into account, the data presented in \bigcite{MarinariP2000a} can be fitted
reasonably well with the scaling form \Eq{ProbChangeScaling}.
It was also pointed out that even the two dimensional data, 
which is believed to obey \Eq{ProbChangeScaling},
could not be fitted by this scaling form without a similar correction term.
As for the non-intersecting probability of the domain wall with
the plane, it was simply pointed out that 
the observation in \bcite{MarinariP2000a}
does not necessarily imply $d = d_{\rm s}$ because
the observed behavior can be caused by
domain walls that are rough but not space filling.
%\rem{What was the roughness exponent of the domain wall in 2D?}
In addition, as already mentioned above, their latest 
calculation\scite{PalassiniLJY2002} suggests that we have not yet 
accumulated numerical evidences sufficient to decide between 
the possible scenarios.

Before concluding this subsection, let us mention what is known
about the geometrical nature of droplets.
For this issue, 
a zero-temperature calculation was performed.\scite{LamarcqBMM2002}
Using an efficient heuristic algorithm, the authors
of \bcite{LamarcqBMM2002} obtained ground states
of three dimensional systems of the size $L=6$ and $L=10$.
For each sample, they first obtained the ground state.
Then, a center of the excitation is chosen randomly.
They searched for the droplets of various sizes including 
the central spin.
In their calculation, 
a droplet was defined as
the cluster of spins of the smallest excitation energy 
among those which have a given volume.
In spite of the limited system size, they succeeded in 
obtaining system-size independent numerical results.
They found that 
the linear scale of the droplets is described by
$$
  R(v) \propto v^b\quad (b\sim 0.5),
$$
whereas the volume dependence of the excitation energy obeys
$$
  E(v) \propto v^a\quad (a\sim -0.06).
$$
The first relation means that the droplets are roughly 
two dimensional fractal objects rather than compact ones.
The second relation is rather surprising,
since at first glance it might seem to contradict the existence of
the finite-temperature phase transition.
They claimed that this does not necessarily mean 
the absence of the phase transition because the basic
assumption of the droplet argument may not be valid.
Another possible source of the discrepancy may be the definition 
of the droplet with the very restrictive fixed-volume constraint.

%------------------------------------------------------------------------------
\subsection{Effect of Magnetic Fields}
\ssclabel{EffectOfMagneticFields}
%------------------------------------------------------------------------------
Whereas the RSB picture predicts the persistence of the
spin glass phase against a small but finite magnetic field,
the droplet-picture predicts the contrary ---
the absence of the spin glass phase at any finite magnetic field.
Therefore, one of the possible ways to distinguish between the two 
scenarios may be found in the behavior of short-range 
models in a finite magnetic fields.
However, it is rather hard to perform this task near the 
critical temperature (of the zero-field case).
This difficulty is not only because the dynamics near the critical
point is slow but also because the critical value of the magnetic 
field predicted by the mean-field type argument is small.
To avoid at least the latter difficulty,
a zero-temperature calculation was performed.\scite{KrzakalaHMMP2001a}
Using a heuristic optimization algorithm, 
the ground state of the three dimensional EA model
was computed with the Gaussian bond distribution
up to $L=12$ with various uniform magnetic fields.

In addition to the magnetic field, also some constraints were 
imposed at the same time to probe low-lying excitations.
For example, a spin may be forced to be opposite to 
its natural direction in the original ground state with no constraint.
Then, the lowest-energy excitation among those which
include the chosen spin is excited.
In the RSB picture, where there are infinite number of states
different from each other by a finite fraction of spins within an
${\cal O}(1)$ window of excitation energy, such a excitation should typically
contain a finite fraction of all spins.
(See also \Ssc{Sponge} and \Ssc{TNT} below.)
Therefore, the volume of excited cluster
tends to infinity with increasing system size.
On the other hand, in the droplet scenario, 
the influence of the forced spin cannot reach beyond the
distance determined by the magnetic field $H$ as
$$
  l(H) \propto H^{-2/(d-2\theta)}.
$$
Therefore, when the system size is increased beyond this length, 
the size of the excited clusters should stay constant 
independent of the system size.
Some results of computation along this line of consideration
were presented in \bcite{KrzakalaHMMP2001a}.
They demonstrate the fragility of the ground state in a magnetic field.
For example, the forced spin-flip typically drags as much as
30 percents of all spins at $H=0.2$ for $L=12$.
The size of the excited cluster generally increases as the
system size increases.
However, neither a saturation nor a proportionality 
to $L^d$ could be identified clearly.

Therefore, they tried another constraint
where signs of an array of bonds along a plane 
are switched.
This corresponds to imposing the anti-periodic boundary condition
to the system.
In the RSB picture, the resulting excitation again should
extend to the whole system, whereas in the droplet scenario, 
it should not for systems larger than $l(H)$.
They measured the probability of the excitation wrapping around the
system.
It was found that the wrapping probability 
increases for $H \le 0.6$ as the system becomes larger 
whereas it decreases for $H \ge 0.7$.
Based on this result they suggested that $H_c$ is finite and 
is close to $0.65$ at $T=0$.
While they did not exclude the possibility of $H_c = 0$
considering the finite size corrections that may be present, 
their approach seems promising and it is probably worth 
putting more effort in this direction.

%------------------------------------------------------------------------------
\subsection{Sponge-Like Excitations}
\ssclabel{Sponge}
%------------------------------------------------------------------------------
While the droplet picture presents a clear ``real-space'' image 
for the low-lying excitations, 
we have not discussed how low-lying excitations should look
in the RSB picture.
The authors of \bcite{HoudayerM2000} discussed geometrical properties of 
low-lying excitations that are compatible with the RSB picture.
They proposed that low-lying excitations may have a
``sponge''-like shape.
A sponge-like cluster has a characteristic length scale $l_c$.
They argued that excitations smaller than this scale obey
the droplet scaling whereas beyond this scale the droplet predictions
do not apply.
A sponge-like object is defined by the following properties.
\begin{itemize}
\item[(1)] It is a connected object, and so is its complement.
\item[(2)] It occupies a finite fraction of the whole system's volume.
\item[(3)] It spans the whole system, and so does its complement.
\item[(4)] It has a characteristic length scale different from 
   the system-size and the lattice constant. When coarse-grained beyond 
   this length scale, the object occupies the space uniformly whereas 
   below this scale non-uniformity can be seen.
\item[(5)] Its surface contains a finite fraction of its volume,
   i.e., the surface-to-volume ratio is finite and independent of the
   system size up to a finite size correction.
\end{itemize}
They argued that in finite dimensions, three for instance,
there may be excitations of sponge-like clusters 
with excitation energy of ${\cal O}(1)$, independent of the system size.
At the same time, the energy barrier that must be overcome in 
order to excite such a cluster diverges as the system size increases.
Namely, two pure states in the RSB picture can be transformed into
one another by flipping spins in one or more of the sponge-like clusters.
Since they occupy a finite fraction of the whole system,
as soon as thermal fluctuations are introduced, they give rise to
the non-trivial structure in the overlap distribution function, $P(q)$.
In addition, because of the property (5), 
such excitations yield a non-trivial functional form
for the link overlap distribution $P(q_l)$.

Obviously, the last two properties (4) and (5) above are closely
related to each other, because the surface-to-volume ratio may be 
determined by the characteristic length scale as $S/V \sim l_c^{-1}$.
It is important to estimate the length scale, $l_c$,
in order to check the applicability of the picture to individual cases.
An attempt was made\scite{KrzakalaM2000} to estimate
this ratio numerically by considering the EA spin glass model with
a Gaussian bond distribution.
For each bond realization, the ground state was obtained
using a heuristic optimization technique.
Then, the ground state search was performed once more,
this time under the constraint that a pair of spins (chosen randomly)
have the opposite relative orientation as compared to the one
in the first ground state.

In the droplet picture, the difference of two ground states is 
supposed to be a droplet of a size that is smaller than the distance of 
the two chosen spins, whereas in the sponge picture 
it should be identified with a sponge if one of the chosen spin
is included in the sponge and the other is not.
In the sponge picture, such an event should happen with a 
pfinite and system-size-independent probability
because at the scale larger than $l_c$ the sponge 
is uniform and the fraction of the sponge
is system-size-independent.

Calculations of the system up to $L=11$
were performed.\scite{KrzakalaM2000}
In order to check the property (3),
the authors measured the probability of the event that
the excited cluster spans the whole system.
It was found that the probability does not show a strong
system size dependence, indicating that large excitations 
with ${\cal O}(1)$ excitation energy exists.

%------------------------------------------------------------------------------
\subsection{TNT Picture --- Introduction of a New Scaling Length}
\ssclabel{TNT}
%------------------------------------------------------------------------------
However, their other findings\scite{KrzakalaM2000} indicated that the
sponge picture defined above does not exactly describe the model, either. 
Particularly disturbing was the result of the surface-to-volume ratio of the
system-spanning clusters obtained by the procedure mentioned in
\Ssc{Sponge}; the ratio decreases as a function of the system
size.  A power-law yielded a reasonable fit, i.e.  $S/V \propto
L^{-0.3}$, suggesting $l_c$ (defined as the $V/S$) 
tends to infinity in the limit $L\to\infty$.  
If one compares the ground state with the (sponge) excited state
and focus on a small box located around the center of the system,
the domain boundary $S$ would never pass through this box
in the infinite system size limit.
The link-overlap is therefore always complete (i.e., 1) within the box. 
This is in agreement with the result\scite{PalassiniY1999b,PalassiniY2000a}
mentioned in \Ssc{LowTemperaturePhaseGCM}.
The increase of the characteristic length $l_c$ was also  
confirmed in \bigcite{HoudayerKM2000}.
It follows that the distribution of the link overlap
should have a trivial delta-peak structure
while that of the ordinary overlap may have a
non-trivial structure.
This new scenario with the new system-size-dependent scaling
length $l_c(L)$ is called\scite{KrzakalaM2000} the `TNT' scenario,
an abbreviation of ``Trivial $P(q_l)$ and Non-Trivial $P(q)$''.

A closely related calculation was done
by the authors of \bcite{KatzgraberPY2001},
who performed a Monte Carlo simulation of the EA model 
with the Gaussian bond distribution.
They reached rather low temperatures ($T \sim 0.2J$)
employing the exchange Monte Carlo method.\scite{HukushimaN1996}
The overlap distribution $P(q)$ was measured.
The zero-overlap probability $P(0)$ stays constant,
in contrast to the prediction of the droplet picture
($P(0) \propto L^{-\theta}$ with $\theta \sim 0.2$),
consistent with the RSB picture and the TNT picture.
However, the temperature dependence of $P(0)$ at a fixed
system size agreed with the droplet picture, i.e., 
$P(0) \propto T$.
They also obtained the link-overlap distribution.
Its width turned out to be proportional to
$L^{-\mu_l}$ where $\mu_l$ was estimated as
$$
  \mu_l = 0.76(3)
$$
from an extrapolation to $T=0$.
The TNT picture discussed above assumes the presence
of excitations with size-independent energy,
i.e., $\theta = 0$ whereas the surface-to-volume ratio
of excited clusters goes to zero.
Based on this picture together with the above estimate
of $\mu_l$, they obtained (by setting $\theta = 0$ in
\Eq{TemperatureSizeDependenceOfDeltaQl})
$$
  d - d_{\rm S} = 0.38(2)
$$
which is consistent with the zero-temperature 
calculation\scite{PalassiniY2000a} $ d - d_{\rm S} = 0.42(2) $.

%------------------------------------------------------------------------------
\subsection{Arguments Supporting the Droplet Picture}
\ssclabel{Arguments}
%------------------------------------------------------------------------------
In \bcite{HuseF1987}, the notion of pure states was re-examined 
for the disordered systems.
The relationship between the appearance of $P_J(q)$ and
the number of pure states was discussed.
It was suggested that $P_J(q)$
is an erroneous indicator of the multiplicity of pure states.
In the ferromagnetic two-dimensional Ising model 
below the critical temperature,
one has an example of $P(q)$ being non-trivial while there are
only two pure states.
Namely, when the anti-periodic boundary condition is imposed
in both the directions, because of the arbitrariness of 
the position of domain boundaries,
$P(q)$ includes a continuous part.
On the other hand, in the three-dimensional random field model,
one can see the case where $P_J(q)$ consists of a single delta peak
although there are two pure states.
This is because the free-energy difference between the two pure states,
one with $\langle S_i \rangle = m > 0$ and the other with 
$\langle S_i \rangle = -m$, diverges being proportional to $N^{1/2}$.
Therefore $P_J(q)$ for a given sample consists of only one delta peak.
%It was also argued\scite{FisherH1987} that the number of pure states 
%is likely two in realistic spin glass models.

The argument was elaborated in \bcite{NewmanS1996,NewmanS2003},
in which the authors exploited the translation ergodicity.
They proved that any translationally invariant quantity 
measured for a particular bond realization $J$ is equal to its
bond configuration average, provided that the distribution
of each bond is independent and spatially uniform.
Since $P_J(q)$ is obviously translationally invariant,
it follows that $P_J(q)$ is self-averaging, i.e., does not depend on $J$.
This result is in contrast with what we know for the SK model.
Furthermore, they argued that since $P_J(q)$ is self-averaging it would be
improbable that $P_J(q)$ has a continuous part.
The reason for this is that there are only countably many pure 
states and therefore $P(q)$ consists of countably many delta peaks.
Consequently, we would have to choose many but only countably many 
numbers from the interval $-1\le q \le 1$ to locate these
delta functions.
They argued that existence of such countably many 
``preferred'' locations, yet independent of $J$, are very implausible.
Their argument made it clear that we must be
careful about using the notion of the pure state in the
disordered systems.

However, it was pointed out\scite{MarinariPR1997}
that the very existence of the pure states
that was implicitly assumed in \bigcite{NewmanS1996} is questionable.
This means that the pure states that appear in
\Eq{Symbolic_definition_of_P_of_Q} may have only 
a metaphoric meaning, and should not be taken 
too strictly for disordered systems.
From this point of view, one may say that
the non-existence of the unique thermodynamic limit,
or, more specifically, a chaotic system-size dependence
of various quantities may signify the RSB nature of 
the spin glass systems.

%A counter-argument was presented by Parisi.\scite{Parisi1996}
%It was suggested that Newman and Stein's result
%does not necessarily contradict to the mean-field predictions.
%He argued that it is not the overlap distribution $P(q)$ 
%defined in \bigcite{NewmanS1996} that is non-self-averaging
%in the RSB picture,
%but what is supposed to be
%non-self-averaging is the distribution of the overlap
%between two equilibrium configurations of the same single system.

%\rem{What do you mean by the `new ideas', Heiko? 
%The "Mata-states" things? \quad
%$>$ Mention new ideas of Newman and Stein about the limit $L\to\infty$?}

%==============================================================================
\section{Models in Four or Higher Dimensions}
%==============================================================================
\label{sec:Ising4D}

Numerical simulations are easier in four or higher dimensions than in three.
This is presumably because $d=4$ is well separated 
from the lower critical dimension.
The existence of a finite-temperature phase transition in four dimensions
was established already in an early numerical study
\scite{BhattY1988} where a clear 
crossing in the Binder parameter, defined in \Eq{BinderParameter},
could be observed. The critical point was located at
$T_c = 1.75(5)$ and the critical exponents were estimated to be 
$\nu=0.8\pm 0.15$ and $\eta = -0.3 \pm 0.15$ for the 
Gaussian bond distribution. The scaling relation 
$\gamma/\nu=2-\eta$ yields $\gamma = 1.8(4)$ which agrees well
with the result $\gamma = 2.0(4)$ of a high-temperature series 
expansion\scite{SinghC1986}  for the $\pm J$ bond distribution.
More precise estimates of the critical indices were obtained
in \bcite{age-parisi-lorenzo}:
$$
  T_c = 1.80(1), \quad 
  \nu = 0.9(1), \quad
  \eta = -0.35(5),
  \quad \mbox{and}\quad
  \gamma = 2.1(2).
$$
Based on an off-equilibrium simulation,
they also obtained an estimate of the EA order parameter
below the critical point as a function of the temperature.
A non-vanishing order parameter excludes the possibility that 
the transition is of the BKT-type.
As for the nature of the low-temperature phase,
the observation made in \bcite{RegerBY1990} that $P(0)$ is system-size
independent below the critical temperature indicates
that the droplet picture is inappropriate for
the four-dimensional spin glass.
For a detailed discussion on other results obtained before 1997,
the readers are referred to the review \bcite{MarinariPR1997}.
Here we only mention one of the latest computations for clarifying
the nature of the low-temperature phase in four dimensions.

The replica exchange method was used in 
\bcite{KatzgraberPY2001} to study the 4$d$ spin glass model with 
a Gaussian bond distribution
at rather low temperatures $T \sim 0.2J \sim 0.1 T_{\rm c}$.
The zero-overlap probability $P(0)$ was measured in four dimensions
as well as in three.
In both cases $P(0)$ was found to be independent of the system 
size for a fixed temperature, 
and a roughly linearly dependent on the temperature
for a fixed system size.
The authors of \bcite{KatzgraberPY2001}
summarized their results as being consistent with the
TNT picture mentioned above.
They also estimated the exponent characterizing the size 
dependence of the variance of the link-overlap distribution, namely,
$\theta + 2(d - d_{\rm s})$
where $d_{\rm s}$ is the fractal dimension of droplet boundaries
(see the discussion in \Ssc{LowTemperaturePhaseGCM}).
The extrapolated value to the zero temperature 
was $\theta+2(d-d{\rm s}) = 0.35(6)$ 
while they did not rule out the possibility of this value being zero.

For the ground state properties in four dimensions,
we refer the reader to \bcite{Hartmann1999b}, 
in which the ground states of the model in four dimensions 
up to $L=7$ were computed.
The stiffness exponent was estimated as $\theta_{\rm s} = 0.64(5)$.

Although somewhat misplaced in this section on higher dimensional
models we mention here the one-dimensional spin glass model with
{\it long-range} interactions, i.e.\ $J_{ij}=c_\sigma
\epsilon_{ij}/r_{ij}^\sigma$, where $\epsilon_{ij}$ is a Gaussian random 
number with zero mean and variance one and $r_{ij}$ the Euclidean
distance between spin $i$ and $j$ on a finite ring (of perimeter $L$)
embedded in the two-dimensional space. This system, on the borderline
between finite-dimensional and mean field models, was studied in
\bcite{Katzgraber2003a,Katzgraber2003b} for system sizes up to $L=512$.
Finite temperature simulations as well as the study of low-energy 
excitations yielded results that are inconsistent with the droplet picture
but (partially) consistent with the TNT picture.

\section{Aging}
\label{sec:Aging}

The out-of-equilibrium dynamics of spin glasses has become a very rich
field in the recent years and an excellent review on the intensive
theoretical work that has been performed on it until 1998 can be found
in \bcite{age-review} and an overview over the experimental situation
until that date can be found in \bcite{age-exp1,age-exp2}. Since then a
number of interesting developments have occurred and we will focus on
them with the prerequisites necessary to understand them. 
We start with the theoretical concept of
a length scale that evolves in time and is flexible enough to
account for a number of numerical and experimental results. Then we
focus on two-time quantities that are typically measured
experimentally and their behavior in various
temperature protocols (i.e. aging histories) --- showing effects like
memory and rejuvenation. Finally we discuss the theory for violations
of the fluctuation-dissipation theorem and the numerical and
experimental evidences for it.

\subsection{A Growing Length Scale During Aging?}
\label{ssc:TimeDependentLengthScale}

Spin glasses have an order parameter which is (cum grano salis, i.e.\
disregarding the potential complications arising from a possible RSB)
$q_{EA}=[\langle S_i\rangle^2]$. In comparison to structural 
glasses\scite{angell} this is a very lucky situation, not only as a starting 
point for equilibrium theory, which we have discussed in last
sections, but also for non-equilibrium dynamics. An example is 
the common picture of the dynamical evolution of a system 
out of equilibrium quenched into the phase with non-vanishing $q_{\rm EA}$,
in which larger and larger regions of space become ordered.\scite{bray} 
Thus it appears natural to postulate a time-dependent length-scale $L(t)$ for
these equilibrated domains for spin glasses. We choose
such a picture, motivated by coarsening systems,\scite{bray} 
as our starting point for reviewing the out-of-equilibrium
dynamics of spin glasses.

Within the droplet theory\scite{droplet} the slow out-of-equilibrium
dynamics of a finite-dimensional spin glass is due to thermally
activated growth of locally equilibrated regions. The growth of such
regions is supposed to happen through domain wall
movements as in the context of pinned domain walls in random field
systems\scite{rfim-rev} or in elastic manifolds in a disordered
environment\scite{dir-pol-fisher-huse} via a thermally activated
process overcoming a free energy barrier. It is assumed that the
typical energy barrier $B_{L(t)}$ scales with the typical size $L(t)$ of the
domains reached after a time $t$ as 
\begin{equation}
  B_{L(t)}\sim\Delta(L(t)/L_0)^\psi, \label{eq:PowerLawGrowth}
\end{equation}
where $\psi$ is the barrier exponent characteristic of the particular
system under consideration and $\Delta$ is some constant.
Since the dynamics is activated, the
typical time to overcome a barrier $B$ grows exponentially with
$B/T$, $T$ being the temperature;
$t_{L,\rm activated}\sim\exp(B_L/k_BT)$. 
Hence one would expect that after a time $t$ domains of size
\begin{equation}
L(t)\sim\Bigl(\frac{k_B T}{\Delta}\ln(t/\tau_0)\Bigr)^{1/\psi}
\label{eq:lt}
\end{equation}
are equilibrated ($\tau_0$ being a microscopic time scale), i.e.\ the
typical domain size grows logarithmically slowly with $t$. 
Numerically one can determine the typical domain size $L(t)$ directly
via the spatial two-replica correlation function
\begin{equation}
G(r,t)=\frac{1}{N}\sum_{i=1}^N 
\langle S_i^a(t)S_{i+r}^a(t) S_i^b(t)S_{i+r}^b(t)\rangle\;,
\label{eq:corr-func}
\end{equation}
where $a$ and $b$ denote two replicas of the same system (i.e.\
the same disorder realization). For the first time this spatial correlation
function was studied numerically for the two-dimensional EA spin glass model in
\bcite{rieger-growth-2d}, and for the 3d EA model in 
\bcite{rieger-review,rieger-growth-3d,marinari-growth-3d}, later also in 
\bcite{komori-growth,marinari-num-3d,mod-drop-hukushima,mod-drop-yoshino}. 
The extraction of the typical time-dependent domain-size, $L(t)$, from
the correlation function $G(r,t)$ given in \Eq{corr-func} is itself
a delicate issue. In the first studies in
\bcite{rieger-review,rieger-growth-2d,rieger-growth-3d} the definition
$L_T(t)=\int dr\,G_T(r,t)$ was used where the subscript $T$ 
signifies the temperature dependence.
This form would be justified if
$G_T(r,t)$ is of a pure exponential form
$G_T(r,t)\sim\exp(-r/L_T(t))$. Also a scaling form
$G_T(r,t)\sim\tilde{g}(r/L_T(t))$ has been checked in these early works
using the length scale obtained via the integral method and a good
data collapse was obtained for all parameter values (times and
temperatures) used. A more flexible functional form was assumed in
\bcite{marinari-growth-3d} that was also capable of fitting more
accurate estimates of $G_T(r,t)$ obtained 
later:\scite{marinari-num-3d,age-berthier-bouchaud}
\begin{equation}
G_T(r,t)\sim r^{-\beta(T)}\,\tilde{g}(r/L_T(t)).
\label{eq:spat-corr}
\end{equation}

The best-fit values of temperature exponent $\beta(T)$ were found to
be constant around $0.5$ in three dimensions
and decreasing from $1.6$ to $0.9$ in
the temperature range $1.0$ to $0.5$ in four dimensions
(where $T_c\approx1.8$).
In \bcite{mod-drop-yoshino} only the tail of $G_T(r,t)$ was used
to extract $L_T(t)$ assuming a pure exponential form for it. All in all,
these different forms do not cause large variations in the estimates
of $L_T(t)$, the reason simply being that the typical domain sizes
reached in the times $t$ accessible to Monte-Carlo studies to date are
only a few lattice spacings. It should be noted that these functional
forms imply that $G_T(r,t)$ decays to zero with the distance $r$ even
in the limit $t\to\infty$, whereas in the droplet theory one would
expect a non-vanishing large distance limit:
$\lim_{r\to\infty}\lim_{t\to\infty}G_T(r,t)=$
$q_{EA}^2(T)+{\cal O}(r^{-\theta})$.
Functional forms respecting such an asymptotic behavior can be also
devised to fit the data of $G_T(r,t)$ well.\scite{age-berthier-bouchaud}

Suppose that we have now estimates of $L(t)$ obtained in one way or
the other from numerical simulations or from experiments.
(Experimentally there is no straightforward way, by which one could
possibly measure $G(r,t)$ or its space and/or time Fourier transform
via scattering or similar techniques. 
Still there is an interesting experimental 
development\scite{orbach} on this point which we discuss further below.) 
When fitted to the
logarithmic growth law \Eq{lt} the estimated value of $\psi$ was
approximately $0.7$\scite{rieger-growth-3d} and was only acceptable if
a finite offset length was introduced.  However, a power law form
with a temperature dependent exponent
\begin{equation}
L(t)\propto t^{\alpha(T)}
\label{eq:growth-alg}
\end{equation}
fits the data well, with 
$\alpha(T)\approx0.16\,T/T_c$.\scite{rieger-growth-3d,marinari-growth-3d}
This observation may indicate that we must replace the 
power-law length-dependence of the barriers
\Eq{PowerLawGrowth} by the logarithmic dependence
\begin{equation}
  B_{L(t)} \propto \Lambda \ln L(t)
\end{equation}
as suggested in \bcite{koper,rieger-age},
or it may serve as a motivation to modify 
the simple domain growth picture
discussed so far in a way that has been first discussed in
\bcite{mod-drop-hukushima}, later also in 
\bcite{age-dupuis-vincent,age-jonsson,age-berthier-bouchaud,mod-drop-yoshino}. 
The essential idea in these discussions 
is that the initial coarsening process is still
influenced by critical fluctuations --- which is plausible since (i)
the temperature that are usually studied are not too far from the
critical temperature $T/T_c\ge0.7$ and (ii) in the initial stage the
equilibrated length scales are still very small.

Close to $T_c$ the equilibrium correlation length is given by $\xi(T)\sim
L_0|1-T/T_c|^{-\nu}$, and even at $T<T_c$ the dynamics on length scales
$L<\xi$ might be dominated by critical fluctuations rather than
activated processes. Hence one might expect that as long as $L<\xi$
the typical size of equilibrated domains grows with time $t$ as
$L(t)\sim l_0 (t/t_0)^{1/z}$ ($l_0$ and $t_0$ being microscopic length
and time scales, respectively) and only for $L(t)>\xi$ activated
dynamics obeying \Eq{lt} sets in. Thus a crossover from 
critical to activated dynamics happens at a time of order
$\tau_0(T)/t_0\sim(\xi(T)/l_0)^z$, when the typical domain size
reaches $\xi(T)$:
\begin{equation}
L(t)/\xi(T)\propto g(t/\tau_0(T))\quad
{\rm with}\quad
g(x)\sim\left\{
\ba{lcl}
x^{1/z}      & {\rm for} & x\ll1\\
(\ln x)^{1/\psi} & {\rm for} & x\gg1
\ea
\right.
\label{eq:mod-lt1}
\end{equation}
Thus, if one plots $L(t)\cdot (T_c-T)^\nu$ versus
$t\cdot(T_c-T)^{z\nu}$ one expects a data collapse for the time
dependent typical domain size at different temperatures. Indeed such a
data collapse has been observed in numerical studies of the 4d EA
model.\scite{mod-drop-hukushima}

\begin{figure}[h]
\includegraphics[width=0.48\columnwidth,height=0.37\columnwidth]{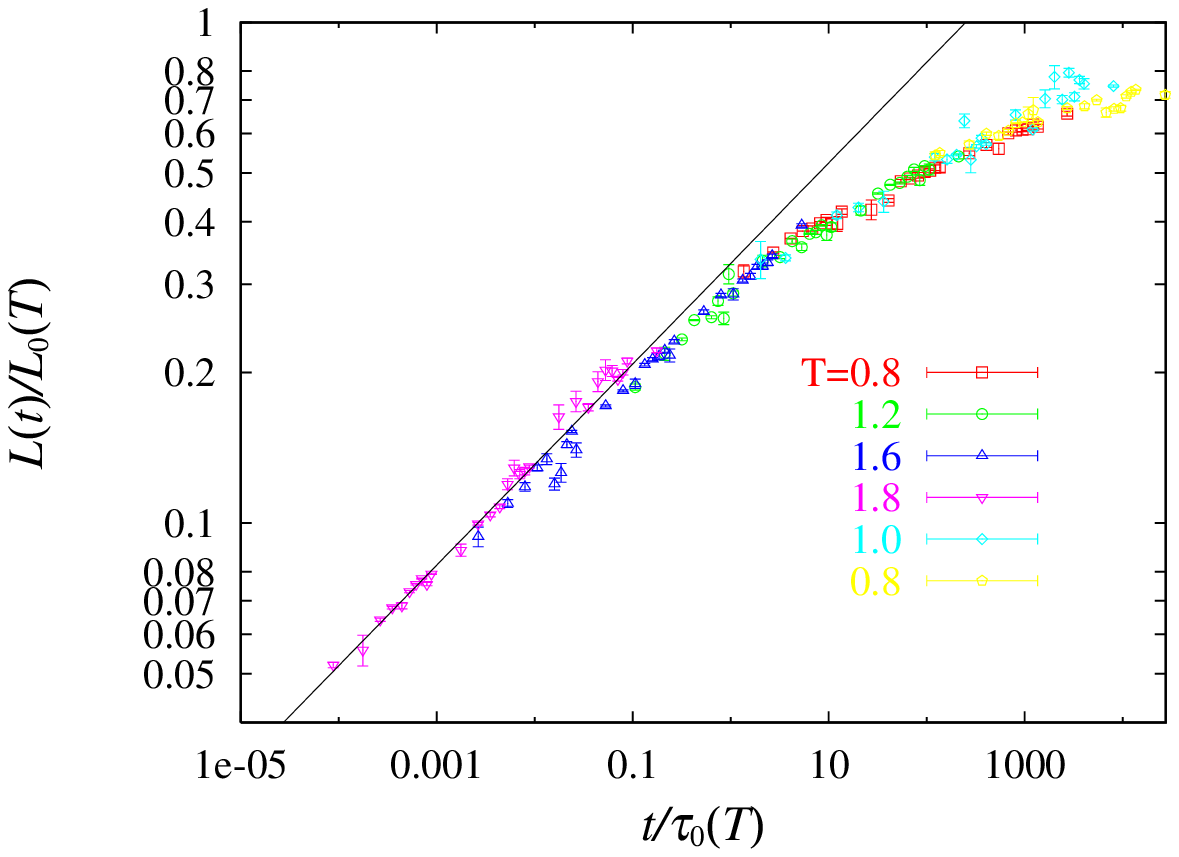}
\includegraphics[width=0.48\columnwidth,height=0.4\columnwidth]{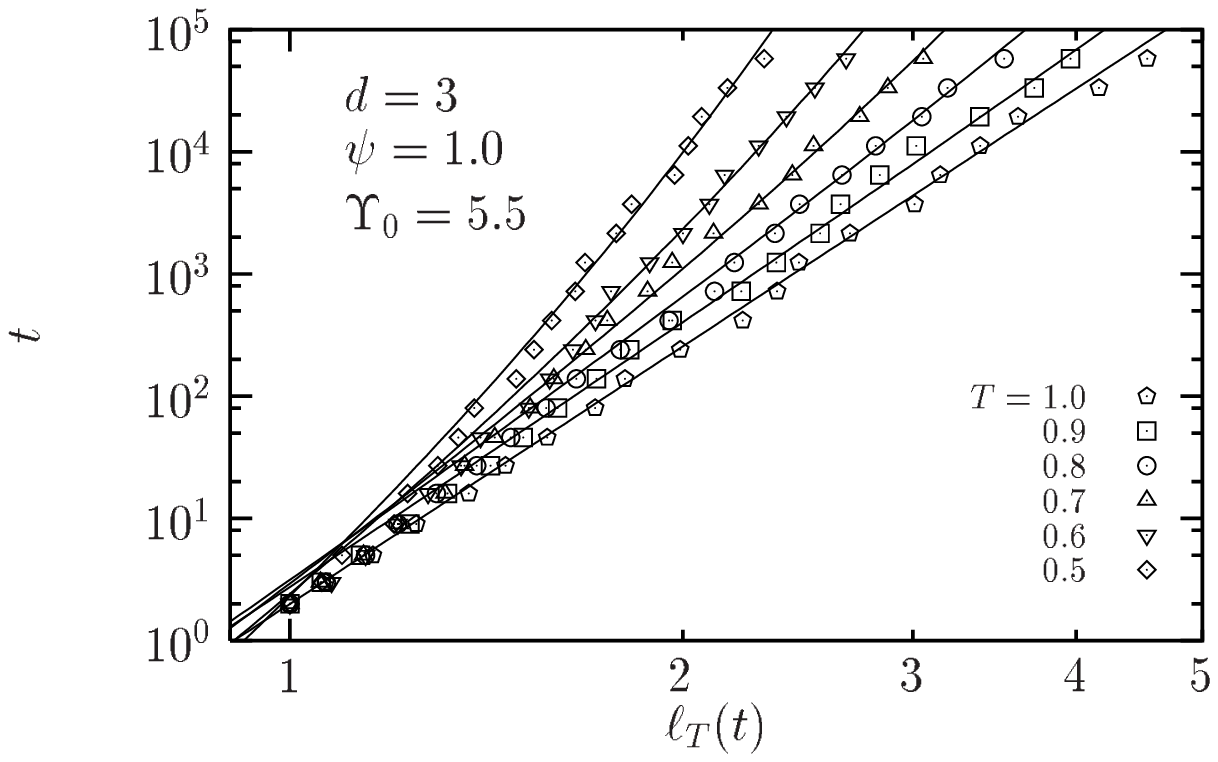}
 \caption{{\bf Left:} Scaling plot of $L(t)$ of the $4d$ EA model
 according to \Eq{mod-lt1}, where the SG transition
 temperature $T_{\rm c}=2.0$ and the critical exponent $\nu=0.93$ are
 fixed, but the dynamical exponent $z$ is obtained to be $4.98(5)$
 from the best scaling. (From \figcite{mod-drop-yoshino}.)
{\bf Right:} Growth laws of the coherence length 
in the $3d$ EA model with fitting curves according to \Eq{mod-lt2}.
(From \figcite{age-berthier-bouchaud}.)
}
\label{lt-scale}
\end{figure}

One can go one step 
further\scite{age-bouchaud-dupuis,age-berthier-bouchaud} and write down 
a more explicit form for the relation between length and time scales,
similar to \Eq{lt} but now taking into account critical fluctuations:
\begin{equation}
t(L)\approx \tau_0 L^z \exp\left(\frac{\Delta(T) L^\psi}{k_{\rm B} T}\right)
\quad\mbox{with}\quad \Delta(T)\approx\Delta_0/\xi(T)^\psi\,
\label{eq:mod-lt2}
\end{equation}
where $\Delta_0$ is an energy scale of order $T_c$. Here, following
\bcite{droplet}, the temperature dependent free energy scale
$\Delta(T)$ is assumed to vanish at $T_c$ since one might expect that,
in analogy to random bond ferromagnets, the surface tension of the
domain walls vanishes at $T_c$. The pre-factor $\tau_0 L^z$
reflects the critical dynamics on short length scales (when the
exponential term is still small) and might be interpreted as a
renormalized microscopic time scale $\tilde{\tau_0}$ due to critical
fluctuations. In this way, we obtain $t(L)/\tilde{\tau_0}\approx f(L/\xi(T))$,
with $f(x)=\exp(c\,x^\psi)$ (and $c=\Delta_0/k_B T$) in reminiscence of
\Eq{mod-lt1}. Note, however, that now the short time critical
dynamics is absorbed into $\tilde{\tau_0}$.

In \bcite{age-berthier-bouchaud} numerical data for the typical domain
size $L(t)$ for the 3d and 4d EA model, obtained via Monte-Carlo
simulations, were successfully fitted to the form \Eq{mod-lt2}. 
For the barrier exponent, $\psi=1.0$ was obtained in three dimensions
using $\nu=1.65$ and $z=7.0$, while in four dimensions
$\psi=2.3$ was obtained using $\nu=0.8$ and $z=5.9$.

As we mentioned above, experimentally it appears to be impossible to
have a direct access to spatial correlations in spin glass models.
However an indirect way to estimate a typical size of correlated
volumes has recently been suggested in \bcite{orbach}.  
When measuring the waiting-time ($t_w$) dependent thermo-remnant magnetization
$M_{\rm TRM}(t,t_w)$, which is a two-time quantity discussed further
below, its logarithmic derivative $S(t)=-d\,M_{\rm TRM}(t,t_w)
/d(\ln t)$ at the time $t$ after the field
has been switched off has a characteristic peak at an {\it effective}
waiting time $t_w^{\rm eff}$. One observes that the peak position
depends on the strength of the field that is applied. Since this time
scale is related, via the thermal activation $\tau_{\rm
relax}\sim\tau_0\exp(\Delta/k_B T)$, to the typical value of the free
energy barriers that can be explored over experimental time scales:
$\Delta(t_w)\sim k_B T\ln(t_w/\tau_0)$, the shift in the peak position
of $S(t)$ contains information on the field dependence of the free
energy barriers: $\Delta(t_w)-E_z\sim k_B T\ln(t_w^{\rm eff}/\tau_0)$,
where in \bcite{orbach} it is assumed that $E_z$ is the the magnetic
(or Zeeman) energy associated with the change in the magnetic
field.The latter is connected to the number $N_s$ of spins that are
involved in the barrier shift induced by the field change via
$E_z\approx N_s\chi_{\rm fc} H^2$, where $\chi_{\rm fc}$ is the field
cooled susceptibility per spin and $H$ the magnetic field strength. If
one assumes that the $N_s$ spins to be effectively locked together one gets
an estimate for a correlated volume $N_s\propto\xi^3$, i.e an estimate
for the coherence length $\xi$ as a function of the waiting time
$t_w$.

By analyzing their data for the peak position of $S(t)$ to obtain
$\xi(t_w)$ in the way just described the authors of \bcite{orbach} find
that it fits to \Eq{growth-alg} with
$\alpha(T)\approx0.169\,T/T_g$, in agreement with the numerical
results reported in \bcite{rieger-growth-3d,marinari-growth-3d}. A fit
to the logarithmic form \Eq{lt} works equally well, but best-fit
value of $\psi$ turns out to be rather large ($\approx5$), which
does not agree with the numerical estimates cited above.  These
experiments probe a time window that is different from the one
accessible to numerical simulations;
each is roughly 6 decade wide but centered around times 
roughly 10 decades apart. 
Thus only the algebraic growth law \Eq{growth-alg}, or the form
\Eq{mod-lt2} that interpolates between critical and activated
dynamics, would consistently match the data of both time windows.

\subsection{Two Time Quantities: Isothermal Aging}
\label{twotimes}

The out-of-equilibrium properties of glassy materials, quenched
rapidly below the glass transition temperature $T_g$ and then aged
{\it isothermally} (i.e. at constant temperature $T<T_g$), manifest
themselves most prominently in two-time quantities, typically response
functions, susceptibilities or correlation functions. The magnetic
response function is defined as
\begin{equation}
R(t+t_w,t_w)=N^{-1}\sum_{i}
\frac{\delta\langle S_i(t+t_w)\rangle}{\delta h_i(t_w)}
\bigg\vert_{h_i=0}
\label{eq:resp}
\end{equation}
In experiments as well as numerical simulations one usually does not
apply field pulses after a waiting time but one switches a constant
magnetic field on or off after a  waiting time $t_w$ --- the
corresponding susceptibility is then related to the response function
via a time integral
\begin{equation}
\chi(t+t_w,t_w)=\int_{t_w}^{t+t_w} dt'\,R(t+t_w,t')
\label{eq:chi}
\end{equation}
If one switches the field on at time $t_w$ after the quench 
and measure the susceptibility, one calls it {\it zero-field-cooled} (ZFC), 
whereas if it is switched off it is called {\it field-cooled} (FC). 
In the former case the magnetization, 
which is within the {\it linear response regime}
simply related to the susceptibility via
$M(t,t_w)=h\cdot\chi(t,t_w)$, increases with the time $t$
spent in the field, whereas in the latter case it decreases with the
time $t$ after the field has been switched off --- it is then also
called the {\it thermo-remnant magnetization} (TRM).

Very useful for the development of theoretical concepts is the
two-time spin autocorrelation function
\begin{equation}
C(t+t_w,t_w)=N^{-1}\sum_{i}\langle S_i(t+t_w)S_i(t_w)\rangle\;.
\label{eq:ctt}
\end{equation}
In {\it equilibrium} it is related to the response function, the
susceptibility and the ZFC/FC magnetization via a {\it
fluctuation-dissipation theorem} (FDT).
(We discuss the violation of the FDT in the subsection \ref{sec-fdt}.)
However, since we are out of equilibrium,
we have first to regard $C$ as being an independent quantity.

The huge amount of experimental data that have been collected are
nicely over-viewed in \bcite{age-exp1,age-exp2}. 
Let us here simply state the main results. 
For the field cooled magnetization one observes 
that the remnant magnetization can be decomposed into a stationary 
part $M_{\rm eq}(t)$ that depends on $t$ only and an aging
part $M_{\rm aging}(t+t_w,t_w)$. Sometimes this decomposition is
assumed to be multiplicative, $M(t+t_w,t_w)=M_{\rm eq}(t)\cdot M_{\rm
aging}(t+t_w,t_w)$, as in the universal short time dynamics for
coarsening at the critical point,\scite{janssen,huse-dyn}
and sometimes
additive, $M(t+t_w,t_w)=M_{\rm eq}(t)+M_{\rm aging}(t+t_w,t_w)$, as in
coarsening dynamics of pure systems\scite{bray} and in the
out-of-equilibrium theory of mean-field spin 
glasses.\scite{age-review}
Since $M_{\rm aging}(t+t_w,t_w)$ is approximately
constant for $t\ll t_w$ both forms usually fit the data well, although
they are fundamentally different. For instance the multiplicative
decomposition is incompatible with the existence of a plateau in
$C(t+t_w,t_w)$ that occurs in mean-field theories of spin glasses at
low temperatures for long waiting times.\scite{age-review}
Analogously
it is also hardly compatible with a non-vanishing long time limit of
$\lim_{t_w\to\infty} C(t+t_w,t_w)$, which is, within the droplet
theory, identical to the order-parameter $q_{EA}$. 

The most remarkable feature of the aging part is that it scales to a
good approximation with $t/t_w$
\begin{equation}
M_{\rm aging}(t+t_w,t_w)\approx\tilde{M}_{\rm aging}(t/t_w)
\end{equation}
In ZFC experiments, instead of applying a constant field after a
waiting time $t_w$, one usually applies a small oscillating field with
frequency $\omega$, which essentially means that one measures the
$t$-Fourier transform $\chi(\omega,t_w)$ of $\chi(t+t_w,t_w)$. 
The frequency $\omega$ plays the role of an inverse observation time
$1/t$, with $t\ll t_w$. 
The waiting time $t_w$ is usually in the range of hours or more,
while $\omega$ is in the range of $0.1$Hz to $100$Hz, i.e.,
$t$ is less than ten seconds. 
Again $\chi(t+t_w,t_w)$ can be decomposed into a stationary
part $\chi_{\rm eq}(\omega)$ and an aging part $\chi_{\rm
  aging}(\omega,t_w)$ that now scales to a good approximation 
with $\omega t_w$:
\begin{equation}
\chi_{\rm aging}(\omega,t_w)\approx\tilde{\chi}_{\rm aging}(\omega t_w)
\end{equation}
corresponding to the aforementioned $t/t_w$ scaling for $M_{\rm FC}$.
Numerically one also has access to the spin autocorrelation function
\Eq{ctt}, which can also be decomposed into two parts.
In \bcite{rieger-growth-3d} it has been shown that
(i) the equilibrium part $C_{\rm eq}(t)$ decays
algebraically with a very small temperature-dependent exponent $x(T)$
and (ii) that $C_{\rm age}(t+t_w,t_w)$ again obeys to a good approximation
the $t/t_w$ scaling:
\begin{equation}
C_{\rm eq}(t)\approx A\, t^{-x(T)}
\quad{\rm and}\quad
C_{\rm age}(t+t_w,t_w)\approx\tilde{c}(t/t_w)\;,
\end{equation}
where the scaling function $\tilde{c}$ behaves as
$\tilde{c}(x)\propto x^{-\lambda(T)}$ with $\lambda(T)$ being
a temperature-dependent exponent much larger than $x(T)$.
This behavior was also found for the 4d EA 
model.\scite{age-parisi-lorenzo}
Later studies\scite{age-berthier-bouchaud}
focused on small but systematic deviations in the aging part from a
simple $t/t_w$ scaling behavior; the values of $C_{\rm age}(t+t_w,t_w)$
for fixed ratio $t/t_w$ show a slight tendency to decrease with
increasing $t_w$, which is called {\it sub-aging} and is interpreted
in terms of an {\it effective relaxation time} $t_{eff}$ that is smaller
than the actual waiting time $t_w$. We shall return to this point
later.

\begin{figure}
\includegraphics[width=0.56\columnwidth]{TRM3a.eps}
\includegraphics[width=0.43\columnwidth]{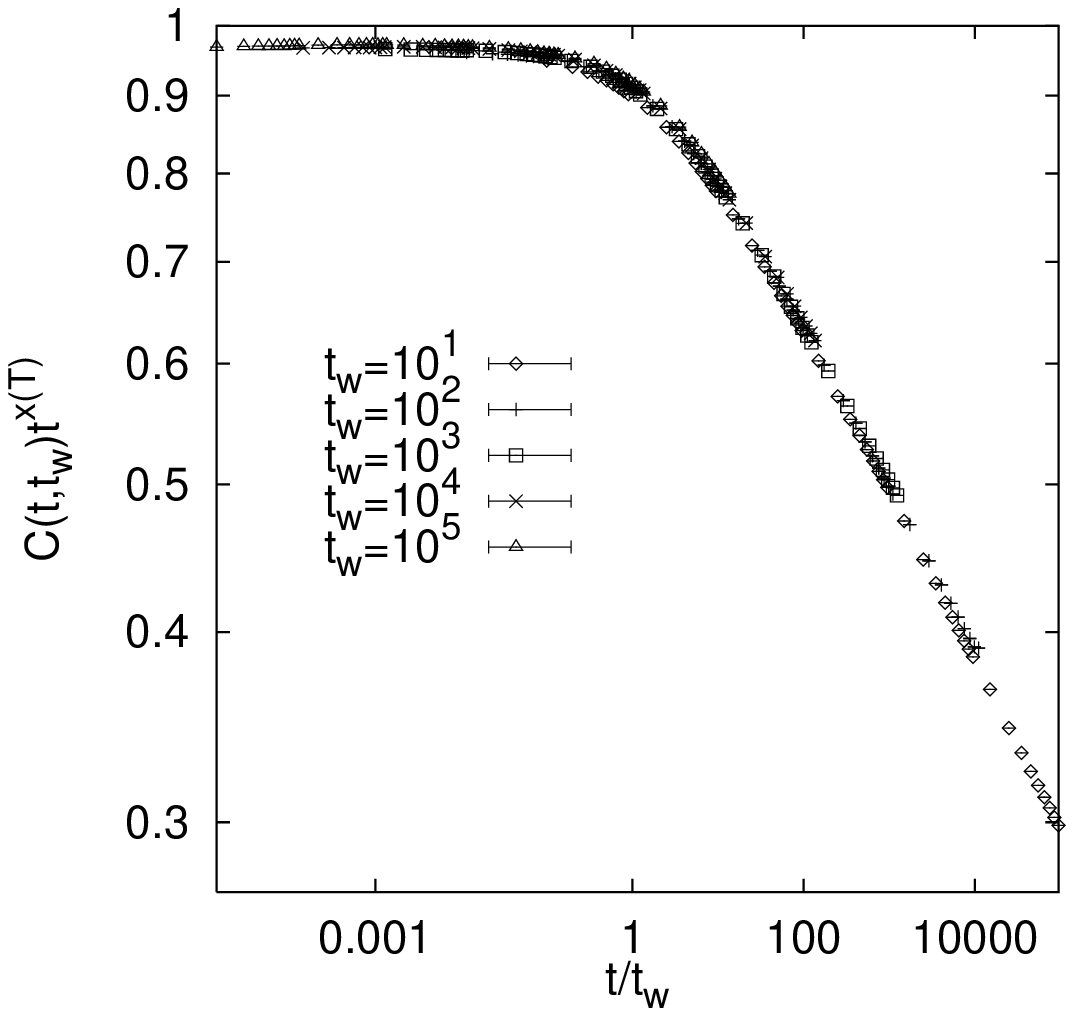}
\caption{{\bf Left:} Aging part of the FC magnetization
$M_{FC}(t+t_w,t_w)$ measured in an AgMn spin glass: the estimated
stationary contribution $M_{\rm eq}(t)$ has been subtracted from the
full measured value. The data is plotted versus $t/t_w$. The
deviations from perfect $t/t_w$ scaling occur for $t\sim t_w$
(sub-aging) and can be accounted for by choosing $h(t+t_w)/h(t_w)$ with
$h(t)$ given in (\protect\ref{eq:ht}) as a scaling variable. (From
\figcite{age-exp1}.)  {\bf Right:} Autocorrelation function
$C(t+t_w,t_w)$, devided by $t^{-x(T)} \propto C_{\rm eq}(t)$, versus
$t/t_w$ for $T=0.4$ in the $3d$ EA model. (From \figcite{rieger-growth-3d}.)}
\end{figure}

Within a domain growth (or droplet) picture of the out-of-equilibrium
dynamics of spin glasses,\scite{droplet}
only ${\it one}$ characteristic length scale $L(t)$ for 
a given time $t$ after the quench is assumed to exist.
One would then expect that two-time quantities
depend only on the ratio of the two length scales that are
equilibrated within the two times $t_w$ and $t+t_w$:
\begin{equation}
O_{\rm aging}(t+t_w,t_w)\sim\tilde{o}(L(t+t_w)/L(t_w))\;,
\end{equation}
where $O(t+t_w,t_w)$ is any two-time observable (such as $R$, $\chi$, $M$ or
$C$) and $\tilde{o}$ the corresponding scaling function. 
The first observation is that an algebraic growth law, 
\Eq{growth-alg} for instance, 
would then be completely compatible with the
$t/t_w$ scaling of the aging part of various observable reported so
far. Note also that as long as $L(t)$ depends algebraically on $t$ the
use of the scaling variable $L(t_w)/L(t)$ is equivalent to using 
$L(t+t_w)/L(t_w)$, both end up to be some function of $t/t_w$.

In the case of logarithmic domain growth \Eq{lt}, on the other hand,
the scaling with $L(t_w)/L(t)$ is different 
from the scaling with $L(t+t_w)/L(t_w)$, and
one should note that droplet scaling,\scite{droplet}
as it is usually
applied to the experimental and numerical data, is assumed to work
with $L(t_w)/L(t)$,\scite{schins,mod-drop-hukushima,age-jonsson}
where
$L(t_w)$ is the typical size of the domains after waiting time $t_w$
and $L(t)$ is the typical size of the droplets being activated in the
time $t$ or being polarized by an ac field of frequency
$\omega=1/t$. The basic physical idea behind this scaling form is that
the presence of a domain wall effectively reduces the excitation gap
of droplets that it touches.

However, assuming the simple logarithmic growth law \Eq{lt},
$L(t)\approx((k_{\rm B} T/\Delta) \,\ln (t/\tau_0))^{1/\psi}$, 
the numerical as well as the experimental data turn out 
not to scale with $L(t_w)/L(t)$ as suggested by the droplet 
theory\scite{droplet,schins} 
(it does not scale with $L(t+t_w)/L(t_w)$, either, 
which is incompatible with the experimental observation of sub-aging).
For some time this observation was
taken as an indication that the droplet theory might be inappropriate
to describe the out-of-equilibrium properties of spin glasses, at
least for the time scales accessible to experiments and to numerical
studies. However, recently it was pointed out\scite{age-jonsson} 
that, taking into account the crossover from critical to activated dynamics
already discussed in the last subsection, the apparent inconsistency
of the data with the droplet theory disappears. There experimental
data for the (imaginary part of the) susceptibility
$\chi(\omega,t_w)$, measured for the 3d Ising spin glass
Fe$_{0.5}$Mn$_{0.5}$TiO$_{3}$ and the 3d Heisenberg spin glass Ag(11\%
at Mn), were presented. 
The aging part $\chi_{\rm aging}''(\omega,t_w)$
(obtained after subtracting a fitted equilibrium part $\chi_{\rm
eq}''$) was shown to scale as $L(t_w)/L(1/\omega)$ with $L(t)$ following
the logarithmic growth law \Eq{lt}, where the microscopic time and
energy scale $\tau_0$ and $\Delta$ replaced by a temperature dependent
characteristic time and energy scale, $\tau_c(T)$ and $\Delta(T)$.
This replacement captures the features of the critical, 
short-time dynamics close to the critical point: 
$\tau_c(T)\sim\tau_m\xi^z$ and
$\Delta(T)\sim\xi^{-\psi}$ ($\xi=|1-T/T_c|^{-\nu}$ being the
correlation length), 
which is in the same spirit as \Eq{mod-lt1} and \Eq{mod-lt2}. 

The authors of \bcite{age-jonsson} found a very good data collapse when
plotting their data for $(\chi''(\omega,t_w)-\chi_{\rm
eq})/\chi''(\omega,t_w)$ versus the scaling variable
$\ln(t/\tau_c(T))/\ln(\omega^{-1}\tau_c(T))$ (which is
$L(t)/L(1/\omega)$, since $\Delta(T)$ and $k_B T$ cancel in this
ratio). Given the critical exponents $\nu$ and $z$ from earlier
studies, the fitting parameters are $\psi$, the whole equilibrium part
$\chi_{\rm eq}$ (for which an explicit functional form predicted by
the droplet theory was used) and a microscopic time scale
$\tau_m$. They found $\psi\approx1.9$ for the Ising spin glass, which
was much larger than what had been reported before in 
experiments\scite{psi-exp} (see also
\bcite{age-dupuis-vincent,age-bouchaud-dupuis}) and in the numerical
simulations discussed in the previous subsection, and $\psi\approx1.3$
for the Heisenberg spin glass. 
The authors of \bcite{age-jonsson} showed that their experimental data
are also compatible to the $\omega t_{\rm w}$ scaling.

So far, we have discussed the compatibility of numerical and 
experimental data with a simple (or simplified) domain growth picture. 
Mean field models, however, show a more 
complicated behavior.\scite{age-review,mft-review} 
The mean field theory for the out-of-equilibrium dynamics predicts 
``ultra-metric behavior'' in the time 
domains\scite{cuku,age-ultra,age-ultra2}:
\begin{equation}
O_{\rm aging}(t+t_w,t_w)=\sum_i{\cal O}_i (h_i(t+t_w)/h_i(t_w))\;,
\label{eq:many-domains}
\end{equation}
where the infinite sum over the index $i$ refers to various large time
sectors\scite{cuku,age-review} defined by the ratio of
$h_i(t+t_w)/h_i(t_w)$ being of order $1$, where the different
(unknown) functions $h_i(t)$ represent different physical mechanisms
at work during aging and need not necessarily to be related to domain
sizes. They are monotonously increasing and grow differently such that
when $0<h_i(t+t_w)/h_i(t_w)<1$ holds for an index $i$ then
$h_j(t+t_w)/h_j(t_w)=1$ for all larger indices $j>i$ --- implying that
if two times $t_1$ and $t_2$ belong to the time sector defined by
$h_i$ and $t_2$ and $t_3$ to the one defined by $h_j$ with $j>i$, then
$t_1$ and $t_3$ also belong to the sector defined by $h_i$. 
One simple example for $h_i(t)$ is
\begin{equation}
h_i(t)=\exp\{(t/t_0)^{1-\mu_i}/(1-\mu_i)\}\;,
\label{eq:ht}
\end{equation}
with $0<\mu_i<1$ ($\mu=1$ yields a $t/t_w$-scaling, $\mu=0$
time-translational invariance.

A consequence of such an Ansatz is for instance a hierarchy in 
these large-time sectors, called {\it dynamic ultrametricity}, which
for the correlation function $C$ means that in the limit of large
times $t_1<t_2<t_3$
\begin{equation}
C(t_3,t_1)={\rm min}\{C(t_2,t_1),C(t_3,t_2)\}
\label{ultra}
\end{equation}
holds.
%Obviously this is fulfilled in simple coarsening models as for
%instance a ferromagnet, which has a long time limit 
%$\lim_{t\to\infty}\lim_{t_w\to\infty}C(t+t_w,t_w)=m^2$, where $m$ is the
%equilibrium magnetization per spin below $T_c$: 
%Suppose that for three very large times we have $C(t_2,t_1)>m^2$
%and $C(t_3,t_2)<m^2$, then for $C(t_3,t_1)>m^2$ the relation
%(\ref{ultra}) holds.
A number of consequences can be drawn from this 
Ansatz.\scite{age-review,mft-review}
Among them is a particular form of the
violation of the fluctuation-dissipation relation, which we discuss
later. Let us here just state that mean-field theory predicts a richer
scenario for glassy out-of-equilibrium dynamics than the domain growth
picture
with only one waiting-time-dependent length-scale. However, as rich as
this scenario is, all experimental data for real spin glasses and all
numerical data for spin glasses with short-range interactions obtained
for isothermal aging up to now can, to our knowledge, be scaled nicely
with one single large-time domain, i.e.\ with only one term appearing
in the infinite sum \Eq{many-domains}. 

For instance, the aforementioned sub-aging property of two-time
quantities that has been observed in experimental 
data\scite{age-exp1,age-picco} (and also in analytically tractable
coarsening, non-spin glass, models\scite{subage-berthier,dyn-monthus})
can easily be accounted for by taking only a single term in the sum
\Eq{many-domains} with $h(t)$ being the function
defined in \Eq{ht} and with the exponent $\mu$ now being a
fitting parameter. This form was actually used in the earliest
experiments on aging in polymer glasses,\scite{struik}
and later also in spin glass experiments.\scite{age-exp1}
The exponent $\mu$ commonly
turns out to be not very different from $1$, accounting for the fact
that the deviations from simple $t/t_w$ scaling are usually very
small. In numerical simulations of $3d$ models the scaling with
$t/t_w$, i.e. $\mu=1$, is almost perfect. In four dimensions
the numerical data
for $C(t+t_w,t_w)$ tend to show 
{\it super-aging},\scite{age-berthier-bouchaud}
i.e.\ the values of $C_{\rm age}(t+t_w,t_w)$
for fixed ratio $t/t_w$ show a slight tendency to increase with
increasing $t_w$. This can be interpreted by an effective relaxation
time that grows faster than $t_w$, and it can be shown that the aging
part $C_{\rm age}(t+t_w,t_w)$ scales nicely with $h(t+t_w)/h(t_w)$,
when $\mu>1$ for $h(t)$ in from \Eq{ht} is chosen (which bears
unfortunately mathematical inconsistencies\scite{kurchan-constr}). 
However, the data can also be scaled with a
domain growth scaling form \Eq{mod-lt2} and a non-vanishing long
time limit for the equilibrium part.\scite{age-berthier-bouchaud}

\subsection{More Complicated Temperature Protocols}

From what we learned from isothermal-aging experiments described in
the last subsection, it follows that the dynamics of a spin glass depends
crucially on the age of the system, i.e.\ the time $t_w$ that it spent
in the glass phase, i.e.\ at $T<T_g$. Moreover, as experiments
have impressively demonstrated,\scite{age-exp1,age-exp2}
it depends
sensitively on changes of temperature (and also field) {\it during}
this waiting time $t_w$. 
A major goal of their systematic study described in this subsection
is to understand the different phenomena that
are measurable by applying different protocols (i.e.\ temperature
variations during aging) and relating them to a theoretical picture,
if not a theory, of the out-of-equilibrium dynamics of spin glasses.

The simplest experiments in this direction are small-temperature-shift
experiments.\scite{hamman-shift}
The protocol is as follows.
First the system is rapidly quenched below $T_c$ to a temperature 
$T_1$ where it is aged for a waiting time $t_w$. 
Then the temperature is shifted to a
new temperature $T_2=T_1\pm\Delta T$ and the measurement is started.
The first systematic numerical study of such a protocol was
performed for the $3d$ EA model in \bcite{komori-shift} and 
a little later in \bcite{age-berthier-bouchaud}. 
It was observed that the
decay of the correlation function is slower for negative shifts
($T_2<T_1$) when compared with the same function aged isothermally 
at $T_1$. For small shifts $\Delta T$ the
functional form of $C(t+t_w,t_w)$ is the same as that for isothermal aging,
and it can be matched with it using an effective waiting time
$t_w^{\rm eff}<t_w$ for $T_2<T_1$. If one assumes that the barriers
that the system can surmount during a time $t_w$ at temperature $T_1$
are of the same size as those that it can surmount during a time
$t_w^{\rm eff}$ at temperature $T_2$, one obtains a good
agreement with the numerical data, only if one replaces the
microscopic time $\tau_0$ by the typical time scale for critical
dynamics as in \Eq{mod-lt2}.\scite{age-berthier-bouchaud}

Obviously such an interpretation of the numerical results implies that
successive aging at two different temperatures will add to each other
in a full accumulative way. The effective age $t_w^{\rm eff}$ is a
monotonically increasing function of $t_w$ and also depends on the
pair $(T_1,T_2)$: $t_w^{\rm eff}=f_{T_1,T_2}(t_w)$, and $t_w$ can be
related to $t_w^{\rm eff}$ by the inverse:
$t_w=f_{T_1,T_2}^{-1}(t_w^{\rm eff})$.  In
\bcite{joensson-shift} twin-experiments --- negative shift from $T_1$
to $T_2$ and positive shift from $T_2$ to $T_1$ --- were considered. A
criterion for purely accumulative aging would be
$f_{T_1,T_2}^{-1}(t)=f_{T_2,T_1}(t)$ and in contrast to the
aforementioned numerical results deviations from it were observed in
these experiments for $\Delta T/T_c\ge0.01$, i.e.\ still very small
shifts. The authors then concluded that positive and negative temperature
shifts both cause a restart of aging and these findings were then
interpreted as a symmetrical {\it temperature-chaos}
effect.\scite{joensson-shift,joensson-shift-comment,joensson-shift-reply}

The temperature chaos in spin glasses is one of the basic ingredients 
of the droplet theory\scite{droplet} and implies that the equilibrium
configurations at two different temperatures $T_1$ and
$T_2=T_1\pm\Delta T$ below $T_g$ are uncorrelated beyond a length
scale called the {\it overlap-length}, $l_{\Delta T}\sim 
(\Delta T)^{-1/\zeta}$, where $\zeta$ is the chaos exponent.  A number of
experimental results, in particular obtained from temperature-shift or
temperature-cycling experiments,\scite{age-exp2}
have been interpreted
within this scenario. Numerical estimates of the value of $\zeta$ are
based on small variations of the {\it couplings} $J_{ij}$ (via
$J_{ij}\to J_{ij}+\delta_{ij}$ with $\delta_{ij}$ small and random) at
$T=0$\scite{chaos-bray,chaos-rieger} rather than variations of the
temperature, simply because ground states of spin glasses, in
particular in $d=2$ can be obtained more easily than equilibrated
configurations at small $T$. The reported estimates are
$\zeta\approx1$ both in two and three dimensions. 
While the same exponent was reported for temperature chaos (via
Monte-Carlo simulations) in the two dimensional case,
recent large scale numerical studies of the
$3d$ EA model did not show any evidence for temperature chaos in spin
glasses on the length and time scales that could be 
probed.\scite{chaos-marinari}
Even if temperature-chaos exists in three dimensional spin
glasses, the overlap length might be much larger than the
length-scales that have been equilibrated during a particular waiting
time, which then makes the chaotic rearrangements due to the
temperature shift invisible. However, there still might be some effect
due to dangerously irrelevant droplets whose free energy gaps are
quite small, as discussed in
\bcite{mod-drop-yoshino,joensson-shift-reply,mod-drop-yoshino2}.

The next temperature protocols that we discuss here are
larger shifts and cycles $T_1\to T_2\to T_1$.\scite{age-exp1}
The first shift experiments was reported in \bcite{cycle}
whereas the cycle experiments can be found in \bcite{shift}.
The first qualitative numerical study
of this kind was reported in \bcite{rieger-cycle}. In the theory
of the asymptotic out-of-equilibrium dynamics of mean-field spin glass
models one can understand these effects on the basis of infinitely
many time scales organized in a hierarchical --- ultra-metric --- way
(which we will discuss in section \ref{sec-fdt}).

A systematic numerical study of large-shift and cycle experiments 
was recently performed for the $3d$ and $4d$ EA model in
\bcite{age-berthier-bouchaud}. The basic message of large temperature
shift experiments is that, independent of the sign of $T_1-T_2$, aging
is ``restarted'' at the new temperature. This {\it rejuvenation
effect} can nicely be observed in experiments as well as in
simulations through the measurement of the susceptibility
$\chi(\omega,t_w)$. Rejuvenation after a negative temperature shift
comes from fast 
modes\scite{cuku-cycling,age-bouchaud-new,age-berthier-bouchaud} which were
equilibrated at $T_1$, but fall out of equilibrium and are slow at
$T_2$. Therefore, one should expect to see this phenomenon if the
spin configurations in the equilibrated regions 
(on length scales $\le L_{T_1}(t_w)$) are
sufficiently different at the two temperatures. This mechanism is
obviously qualitatively different from the interpretation involving
the notion of temperature chaos (see above), which implies that length
scale smaller than the overlap length are essentially unaffected by
the temperature shift, while larger length scales are completely
reshuffled by the shift. In this picture, rejuvenation is thus
attributed to {\it large} length scales and strong rejuvenation
effects therefore require a very small overlap length.

\begin{figure}
\includegraphics[width=0.50\columnwidth,height=0.37\columnwidth]{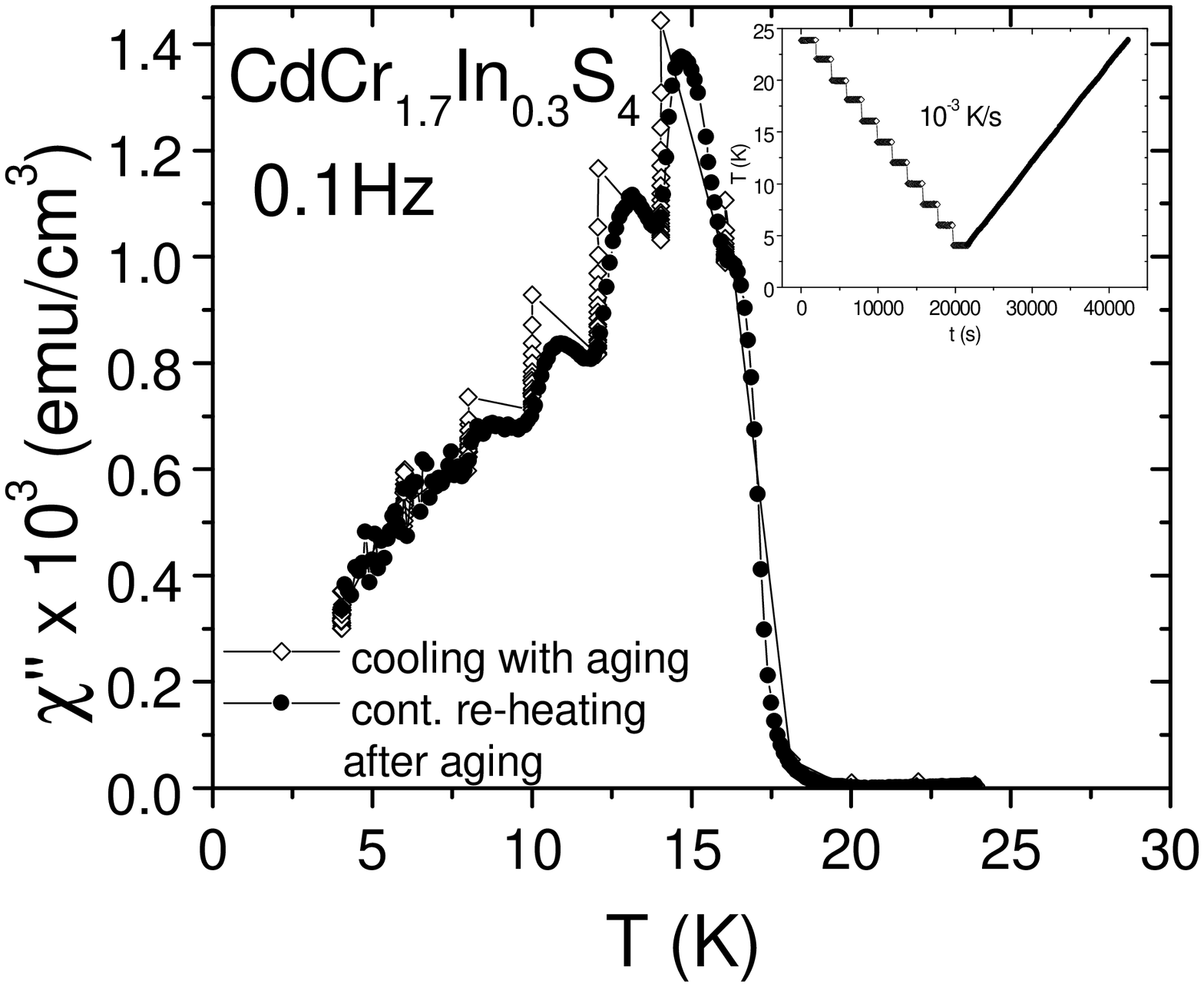}
\includegraphics[width=0.49\columnwidth,height=0.4\columnwidth]{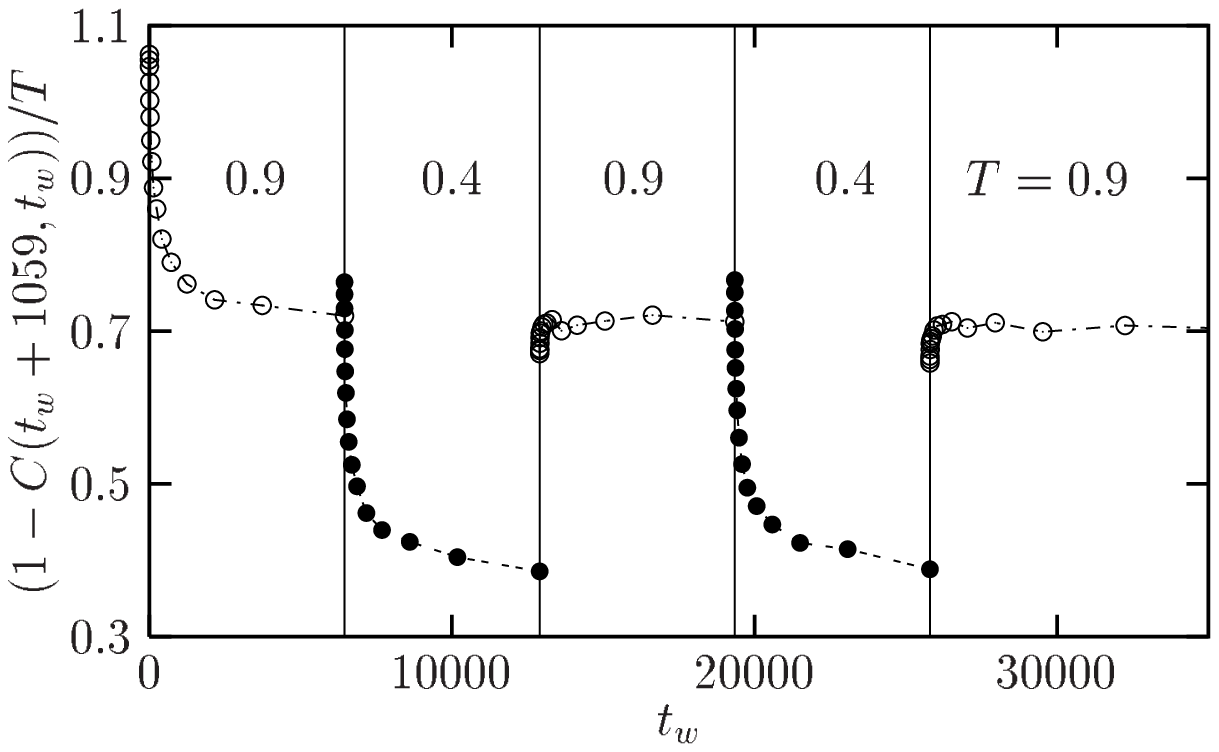}
\caption{{\bf Left:} Series of `dip' imprinted on the
a.c. susceptibility by successive stops at different temperatures
while the system is cooled. Further cooling `rejuvenates' the system
(i.e the susceptibility goes up). However, the dips are one by one
remembered by the system when heated back. For more details, see
\figcite{multiple}. (The figure is adopted from \figcite{age-bouchaud-dupuis}).
{\bf Right:} Evolution of the `a.c.' correlation function
taken from Monte-Carlo simulations of the $4d$ EA model
in the procedure $T=\infty \to T_1=0.9 \to T_2 =0.4 \to T_1 \to T_2$,
showing, as in experiments the coexistence of rejuvenation
and memory effects (from \figcite{age-berthier-bouchaud}).
\label{fig1} }
\end{figure}

No clear rejuvenation effects have ever been observed in simulations
of the $3d$ EA 
model.\scite{rieger-cycle,takayama-chaos,picco-chaos,age-bernardi}
This was
first attributed to the fact that the overlap length was perhaps
numerically large, so that no large scale reorganization could be
observed on the time scale of the simulation. 
However, in the fast-mode mechanism mentioned above,
the crucial ingredient is the small-scale
reorganization due to a temperature shift. When comparing the spatial
correlation function $G_T(r,t_w)$ at two different temperatures and
waiting times such that the length scales
$L_{T_1}(t_{w_1})=L_{T_2}(t_{w_2})$, there is nearly no difference even
for quite large $\Delta T=T_1-T_2$ in three dimensions
but significant differences
in four dimensions,\scite{age-berthier-bouchaud}
which is compatible with the
observation that the exponent $\beta(T)$ in \Eq{spat-corr} remains
roughly constant for a large temperature interval in three dimensions
whereas it varies significantly in four dimensions.
This observation suggests that the $4d$
EA model should be more favorable in studying temperature shift/cycle
protocols. The numerical simulations of the $4d$ EA model show indeed
a clear rejuvenation effect in $\chi(\omega,t_w)$, increasing smoothly
with $\Delta T$, which one would not expect within the temperature
chaos picture. (See Fig.\ref{fig1} (Right).)

The experimental procedure for cycles is $T=\infty\to T_1\to T_2<T_1\to
T_1$. The time spent at $T_1$ is $t_s$ and the time spent at $T_2$ is
$t_s'$. The spectacular {\it memory effect} arises when the
temperature is shifted back to $T_1$. It is observed that although
aging was fully restarted at $T_2$, the system has a strong memory of
the previous aging at $T_1$. The dynamics at $T_1$ proceeds almost as
if no cycle to $T_2$ has been performed.\scite{cycle}
The coexistence
of rejuvenation and memory was made even more impressive in a multiple
step, or {\it dip} experiment,\scite{multiple}
in which the
temperature was decreased to $T_1$, then maintained for some
time, then further decreased to $T_2<T_1$, again maintained, then
further decreased to $T_3<T_2$ etc. At some lowest temperature the
sample was then re-heated with the same rate as it was cooled before,
but now without the interrupts at $T_1,T_2,T_3,\cdots$. 
As shown in Fig.\ref{fig1} (Left), on the way
down to the final temperature the susceptibility $\chi(\omega)$ as a
function of the temperature showed clear dips at $T_1,T_2,T_3,\ldots$,
which is the rejuvenation effect. Most remarkably the same dips
reappear in $\chi(\omega)$ on the way up, which is the memory effect.
The spin glass has memorized the individual temperatures
at which the constant cooling was
interrupted for some time.

According to \bcite{age-bouchaud-dupuis} the memory effect is a simple
consequence of the separation of time and length scales, also
observable in simpler slowly coarsening systems with activated
dynamics.\scite{age-bouchaud-new}
When the system is at $T_2<T_1$,
rejuvenation involves very small length scales as compared to the
length scales involved in the aging at $T_ 1$. Thus when the
temperature is shifted back to $T_1$, the correlation of length scale
$L_{T_2}(t_s')$ nearly instantaneously re-equilibrate at $T_1$. The
memory is just stored in the intermediate length scales, between
$L_{T_2}(t_s')$ and $L_{T_1}(t_s)$. How such a storage mechanism
actually works in a microscopic model for a finite-dimensional spin
glass, like the EA model, is a challenging question. 
Progress has been made for simpler microscopic models, 
such as the $2d$ Mattis model, 
a simple coarsening model,\scite{age-yoshino-lemaitre}
or the directed polymer in a random medium.\scite{age-sales-yoshino}
Also experiments on random {\it
ferromagnetic} systems,\scite{age-ferro-a,age-ferro-b,age-ferro-c,age-ferro-d,age-ferro-e}
on random {\it ferroelectric}
systems\scite{age-ferro-el-a,age-ferro-el-b} and on frustrated systems without a spin
glass transition at finite temperatures\scite{age-frust} show strong
rejuvenation effects and a weak memory effect. This indicates that the
existence of ``many pure states'' as predicted by Parisi's
equilibrium solution of the Sherrington-Kirkpatrick model
and the interpretation of these non-equilibrium effects 
based on it (in terms of a diffusion in a hierarchical
space\scite{age-vincent-review-old,age-exp1}) are not 
necessary prerequisites to observe rejuvenation and memory 
in the out-of-equilibrium dynamics.

Nevertheless the hierarchical picture might be relevant in more
complex situations and a concrete implementation was proposed in
\bcite{age-grem} in terms of a thermally activated random energy model
(for an overview see \bcite{age-review}). Here to each level of a
hierarchical tree a transition temperature is associated, such that
for each level the dynamics is stationary for higher temperatures and
aging for lower temperatures. A small decrease of temperature induces
some rejuvenation by driving out of equilibrium a new level of the
tree, while freezing out the dynamics at the upper levels, thereby
allowing the memory to be conserved. This model was recently studied
further,\scite{age-grem2}
where it was shown numerically that the
rejuvenation/memory effect is indeed already reproduced with two
levels. A real space interpretation of this hierarchical tree was
proposed in \bcite{age-grem,age-balents}, and further developed in
\bcite{age-bouchaud-new} in terms of a multi-scale dynamics. Low levels
of the tree correspond to short wave-length modes, which are only
frozen at low temperature, while large wave-length modes are frozen at
a higher temperature and constitute a {\it backbone}, where memory is
imprinted. In the context of domain walls this picture is particularly
clear\scite{age-bouchaud-new} and should apply directly to disordered
ferromagnets where the slow dynamics comes from the motion of these
pinned domain walls.\scite{age-ferro-a,age-ferro-b,age-ferro-c,age-ferro-d,age-ferro-e}

\subsection{Violation of the Fluctuation-Dissipation Theorem}
\label{sec-fdt}

In a system in equilibrium the response $R$ to an external magnetic
field (Eq.\Eq{resp}) and the autocorrelation function $C$ (Eq.\Eq{ctt})
depend only on the time difference $t$, i.e.,
$R(t+t_w,t_w)=R_{\rm eq}(t)$ and $C(t+t_w,t_w)=C_{\rm eq}(t)$,
and are related to each other through the fluctuation-dissipation
theorem (FDT)
\begin{equation}
R_{\rm eq}(t) = - \frac{1}{T} \frac{\partial C_{\rm eq}(t)}{\partial t}\;.
\label{eq:fdt}
\end{equation}
In a system that is out-of-equilibrium this relation is generally not
valid and the violation of FDT can be parameterized through a violation
factor $X(t+t_w,t_w)$:
\begin{equation}
R(t+t_w,t_w) = - \frac{X(t+t_w,t_w)}{T} 
\frac{\partial C(t+t_w,t_w)}{\partial t_w}\;,
\label{fdt-viol}
\end{equation}
where the differentiation is done with respect to $t_w$, not the time
difference $t$ as in \Eq{fdt}. In analytic studies of mean-field
systems it can be shown\scite{cuku} that for large times
($t_w\to\infty$) $X$ depends on $t$ and $t_w$ only through the value
of the correlation function.
Hence $X(t+t_w, t_w) = X[C(t+t_w,t_w)]$. 
In particular, when
$C>q_{EA}$ it is $X=1$ and the FDT is recovered. In essence, the
asymptotic dependence of $X$ on the values of $C$ alone underlies the
decomposition into asymptotic time domains discussed in subsection
\ref{twotimes} under \Eq{many-domains}. It turns 
out\scite{age-peliti} that the {\it effective temperature} $T_{\rm
eff}(t+t_w,t_w)=T/X(t+t_w,t_w)$ is precisely the temperature which would be
read on a thermometer with response time $t$ (or frequency
$\omega\sim1/t$) when connected to the magnetization at time $t_w$. A
``fast'' thermometer of response time $t\ll t_w$ will then probe the
stationary regime for which $X=1$ and thus measure the heat-bath
temperature.

Different scenarios of the FDT violation can occur.
{\bf (A)} $X(C)=1$ for
all $C$ implies the system is in equilibrium, e.g.\ in the
paramagnetic phase of a spin glass. {\bf (B)} Coarsening systems with
$q_{EA}=m^2$ have $X(C)=1$ for $C>q_{EA}$ and $X(C)=0$ for
$C<q_{EA}$. {\bf (C)} So called {\it discontinuous} spin glasses, 
such as the spherical mean field spin glass model with $p$-spin interactions
($p\ge3$),\scite{cuku} have $X(C)=1$ for $C>q_{EA}$ and
$X(C)=x_1<1$ for $C<q_{EA}$, where $x_1$ is a constant. This
corresponds to the existence of only {\it one} asymptotic time domain
(see \Eq{many-domains}), and the two-time quantities are expected to
scale with only one (possibly unknown) function 
$h(t)$\scite{cuku-pspin} {\bf (D)} So called {\it continuous} spin glasses,
like the SK model, have $X(C)=1$ for $C>q_{EA}$ and $X(C)$
continuously varying, non-constant function of $C$ for $C<q_{EA}$.\scite{cuku}

A convenient way to extract the function $X(C)$ is to measure the
autocorrelation function $C(t+t_w,t_w)$ and the susceptibility
$\chi(t+t_w,t_w)$, \Eq{chi}, for a situation in which a
(infinitesimal) small homogeneous external field $h$ is switched on
after a time $t_w$ (in which case the induced magnetization is
zero-field-cooled and 
$\chi(t+t_w,t_w)=\lim_{h\to0} M_{ZFC}(t+t_w,t_w)/h$). 
In this case we have 
\begin{eqnarray}
\chi(t+t_w,t_w)&=&\int_{t_w}^{t+t_w} dt' R(t+t_w,t')\nonumber\\
&=&\frac{1}{T}
\int_{t_w}^{t+t_w} dt'\,
X[C(t+t_w,t')]\frac{\partial C(t+t_w,t')}{\partial t'}\;.
\label{fdt2}
\end{eqnarray}
As long as the function $C(t+t_w,t')$ is
monotonously increasing with $t'$ for fixed time $t+t_w$, 
the substitution of $X(t+t_w,t')$
by $X[C(t+t_w,t')]$ is legitimate. One should keep in mind that then
$X[C]$ still has a $t_w$-dependence (that we could indicate by
$X_{t_w}[C]$), which however vanishes in the limit
$t_w\to\infty$. Hence one gets (with $C(t+t_w,t+t_w)=1$):
\begin{equation}
\chi(t+t_w,t_w)=\int_{C(t+t_w,t_w)}^1 dC\,X[C]
\qquad{\rm or\;for\;fixed\;}t_w:\quad
\frac{d\chi}{dC}=-\frac{X[C]}{T}
\end{equation}
which implies that for fixed waiting time the slope of a parametric
plot of $\chi(t+t_w,t_w)$ versus $C(t+t_w,t_w)$ yields $X_{t_w}[C]$.
In the limit $t_w\to\infty$ one should obtain the desired
FDT-violation $X[C]=\lim_{t_w\to\infty} X_{t_w}(C)$. 

\begin{figure}[t]
\includegraphics[width=0.48\columnwidth,height=0.42\columnwidth]{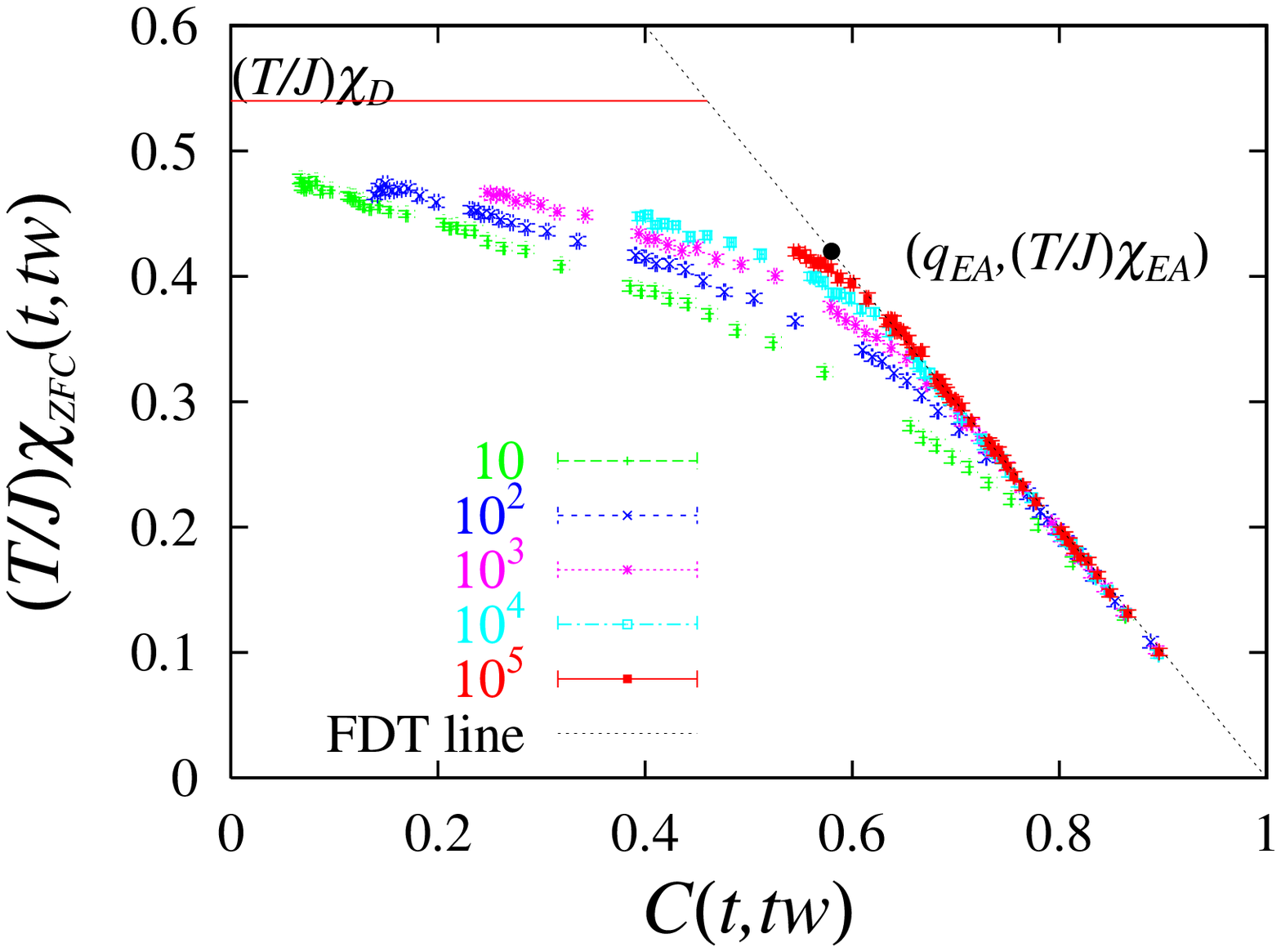}
\includegraphics[width=0.48\columnwidth,height=0.4\columnwidth]{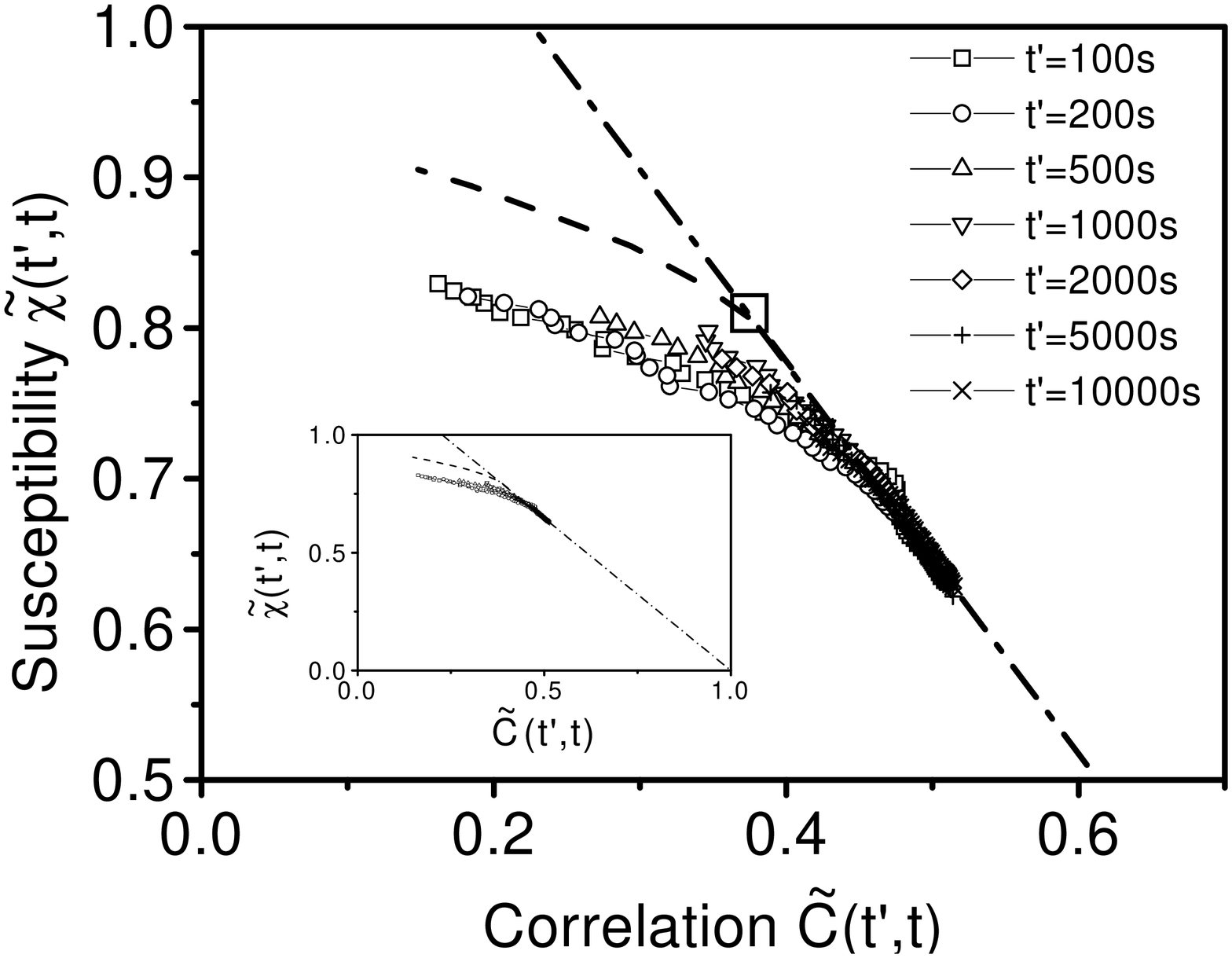}
\caption{{\bf Left:}Parametric plot $(T/J)\chi_{\rm ZFC}(t+t_w,t_w)$ vs
$C(t+t_w,t_w)$ at $T/J=1.2$ for the $4d$ EA model. The
straight tangent lines represents the FDT. The convergence points of
the break points of time translational invariance  
$(q_{\rm EA}, (T/J)\chi_{\rm EA})$ and the
break points of FDT $(q_{\rm D} ,(T/J)\chi_{\rm D})$.  
(from \figcite{mod-drop-yoshino})
{\bf Right:} FDT-plot obtained from the experiments performed in 
\figcite{fdt-herisson-ocio}.
Relaxation measurements are plotted versus correlation
functions for each $t'$. The dot-dashed line (FDT line) is calculated
for $T= 0.8Tg = 13.3 K$, from a calibration obtained with a copper
sample. The dashed line represents the scaling extrapolation for
$t'\rightarrow\infty$. The branching point with the FDT line,
corresponds to {$\widetilde{C}=q_{EA}$} (square symbol, with size
giving the error range). In Inset, the same data in the whole range.
(from \figcite{fdt-herisson-ocio}).
}
\label{fig-fdt}
\end{figure}

The first study of the FDT violation function $X(C)$ in the $3d$ EA
model was performed in \bcite{fdt-rieger} (the first report of an FDT
violation was actually already presented earlier in \bcite{andersson}).
The non-constant part of $X(C)$ still showed a small but systematic
$t_w$-dependence for the small-$C$ branch of the curve $X(C)$. 
Since no tendency to approach a straight line (scenario B or C) could be
observed the data were interpreted as an indication for scenario D,
which implies that the $3d$ EA model has an out-of-equilibrium
dynamics that is similar to the SK model.  Later studies confirmed
this result, in three dimensions\scite{marinari-num-3d} 
as well as in four dimensions.\scite{fdt-4d}
However, in \bcite{mod-drop-yoshino} it was proposed 
that the numerical data for the FDT-violation in the $3d$ EA-model (as
well as those for the spatial correlations and the two-time
quantities) can be interpreted in view of an extended droplet theory
and that for the $3d$ EA model scenario $B$ (essentially slow
coarsening in the spirit of the droplet model) is appropriate. 
(See Fig.\ref{fig-fdt} (Left).)
Since numerical simulations can cover only a few decades of the waiting
time $t_w$ and there is, as mentioned above, still a significant
$t_w$ dependence in $X(C,t_w)$ in the numerical 
data,\scite{berthier-barrat,barrat-berthier}
a definite statement is hard
to make.

For a clear experimental demonstration of the FDT violation one has 
to measure both; response and autocorrelation function (in 
\bcite{cugliandolo-fdt-exp} possible indications of
the FDT violation based only on the response have been discussed).
Such an experimental evidence of the violation of the FDT in spin glasses has
been presented for the first time in \bcite{fdt-herisson-ocio} (for a
structural glass, glycerol, it has been measured in
\bcite{fdt-grigera-israeloff}, for a colloidal glass, laponite, in
\bcite{fdt-bellon-ciliberto}), where autocorrelation function
$C(t+t_w,t_w)$ has been measured via the magnetization fluctuations in
a spin glass sample and the susceptibility $\chi(t+t_w,t_w)$ in the
usual way by applying a magnetic field.
(See Fig.\ref{fig-fdt} (Right).)
The $\chi$ versus $C$ plots
show still a strong waiting time dependence and give some room for
extrapolations to the long time limit --- one that is suggested by the
authors has similarities to the one for the SK-model. On the other
hand, the aging part of their data for $C$ and $\chi$ scale perfectly
with $h(t_w)/h(t)$ where $h(u)=\exp(u^{1-\mu}/(1-\mu))$ with
$\mu=0.85$, which would not correspond to what one expects for the SK
model.

The action for the long-time asymptotic behavior of the generating
functional for the correlation and response functions is invariant under a
time re-parameterization $t\to h(t)$, simply because in the long-time
regime a re-parameterization of time with a monotonous function does
not change the ultra-metric relation in the time ordering. The global
invariance under the re-parameterization group (RpG) for mean-field
models was recently extended\scite{age-chamon,age-castillo} to
its local variant for short range models to treat fluctuations.
In analogy to Heisenberg magnets, where the global spin rotation
invariance lead to the existence of spin waves in finite dimensional
models via the Goldstone modes, the authors proposed that local time
re-parameterizations play a similar role for the asymptotic dynamics
of short-range spin glasses. One consequence of this observation is
that one expects fluctuations in local correlation functions and local
susceptibilities
\begin{equation}
C_r(t+t_w,t_w)=\overline{S}_r(t+t_w)\overline{S}_r(t_w),\quad
\chi_r(t+t_w,t_w)=\overline{S}_r(t+t_w)|_{h_r(t)=h\theta(t)}/h\;,
\end{equation}
where $\overline{S}_r(t)$ means a slightly coarse-grained spin value
at position $r$ and time $t$, i.e.\ it involves a spatial average of
the spin values over a small volume centered around $r$ and a time
average over a small time window centered around $t$. If the volume is
extended to the system size, the usual global quantities discussed
above are recovered. These spatially fluctuating quantities then have 
a joint probability distribution $\rho(C_r,\chi_r)$ that stretches
along the FDT-relation $\chi(C)$ for the global quantities $\chi$ 
and $C$, as has been shown numerically in \bcite{age-castillo}.

\subsection{Hysteresis in Spin Glasses}

Due to the complex energy landscape of spin glasses one expects strong
hysteresis effects, which were studied recently via Monte-Carlo
simulations\scite{hysteresis0,hysteresis1,hysteresis2,hysteresis4}
also using new techniques such as the recently introduced First Order
Reversal Curve (FORC) method.\scite{hysteresis-pike}

In \bcite{hysteresis2},
the zero temperature dynamics of the $2d$ EA spin
glass model was simulated with a varying external field $H$.
The procedure used there was as follows.
The magnetic field is changed in small steps, 
first downward from positive saturation and then upward from a
reversal field $H_R$. 
After each field step, the effective local field
$h_i$ of each spin $S_i$ is calculated: $h_i=\sum_{j} J_{ij}S_j - H$.
A spin is unstable if $h_i S_i < 0$. 
A randomly chosen unstable spin is flipped and then
the local fields at neighboring sites are updated.
This procedure is repeated until all spins become stable. 

\begin{figure}
\centerline{\includegraphics[width=0.6\columnwidth]{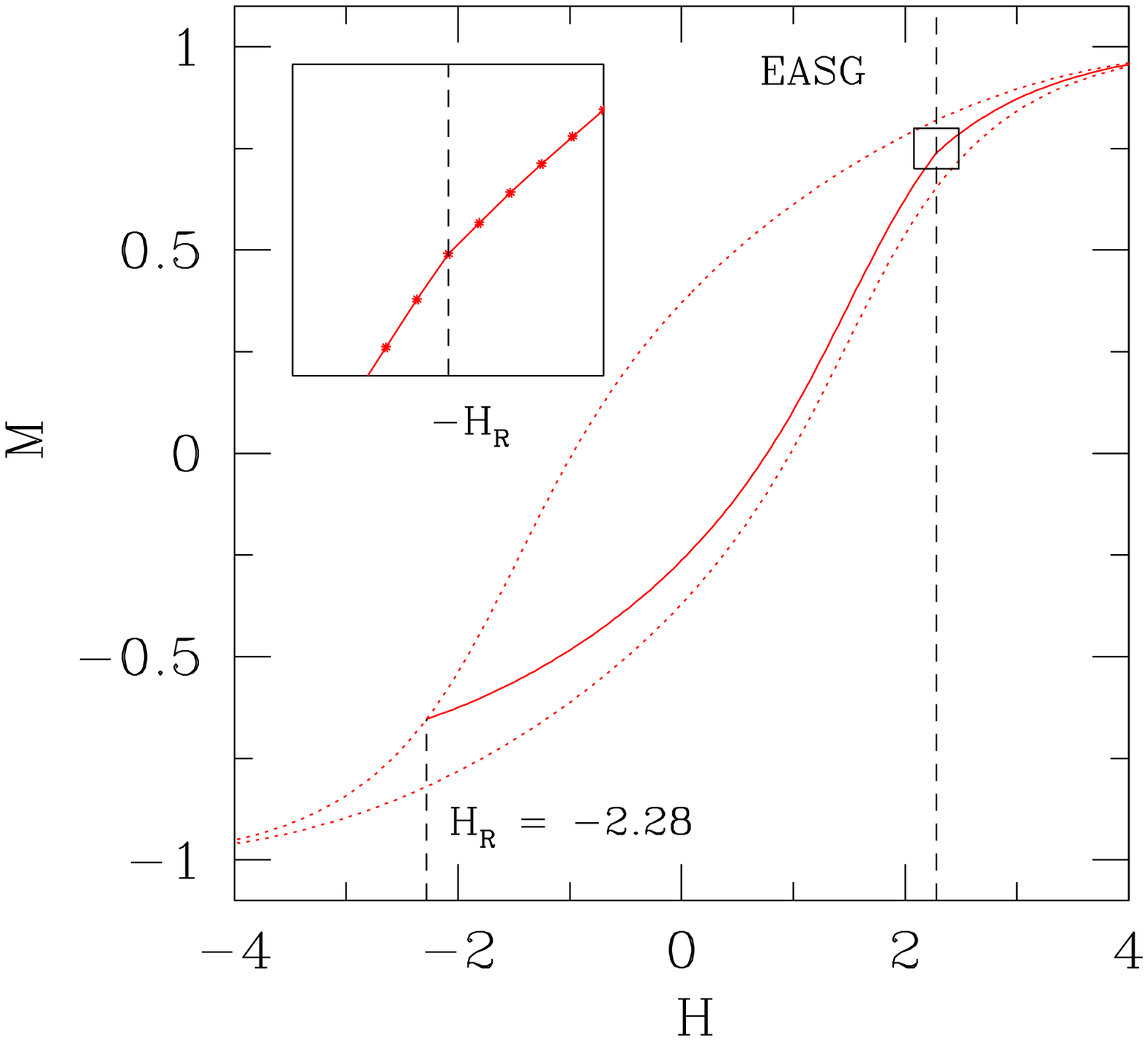}}
\caption{Reversal curve (solid line) and major hysteresis loop (dotted
line) for a two-dimensional (2D) EA spin glass with $10^4$ spins and $H_R =
-2.28$. In the inset a kink is seen around $-H_R$. (From 
\figcite{hysteresis2}).}
\label{hysteresis-kink}
\end{figure}

The first important
observation was of a memory effect in the hysteresis of the $2d$ EA spin
glass model that emerged when the magnetic field was first decreased
from its saturation value and then increased again from some reversal
field $H_R$. It was found that the EA spin glass exhibited a singularity at the
negative of the reversal field, $-H_R$, in the form of a kink in the
magnetization of the reversal curve (See Fig.\ref{hysteresis-kink}).

One can describe this effect within a phenomenological approach
to hysteretic systems, the Preisach model.\scite{hysteresis-preisach}
In the Preisach model a magnetic system is described as a collection
of independent two-state ($\pm 1$) switching units, or
``hysterons''. Unlike Ising spins, which always align with their local
field, the hysteron's state changes from $-1$ to $+1$ at a field
$H_b+H_c$, different from the field $H_b-H_c$, required to switch the
hysteron from $+1$ to $-1$.  Different systems are distinguished by
their different distributions $\rho(H_b, H_c)$ of hysterons of a given
bias $H_b$ and coercivity $H_c$.  Here $\rho(H_b, H_c)$ is the
so-called ``Preisach function''.

This function was extracted by a tool 
developed for analyzing experimental data of hysteretic
systems.\scite{hysteresis-pike}
A family of First Order Reversal Curves (FORCs)
with different $H_R$ was generated, 
with $M(H, H_R)$ denoting the resulting
magnetization as a function of the applied and reversal fields.  
Computing the mixed second order derivative $\rho(H, H_R)= -(1/2)
[{\partial}^2 M/{\partial} H {\partial} H_R]$ and changing variables to
$H_c=(H-H_R)/2$ and $H_b=(H+H_R)/2$, the local coercivity and bias,
respectively, yielded the ``FORC distribution'' $\rho(H_b, H_c)$. For
phenomenological Preisach models, the FORC distribution is equal to the
Preisach function. 
However, the definition of the FORC distributions is more general, because
it is extracted from numerical or experimental data, 
and thus is model-independent.

\begin{figure}
\includegraphics[width=\columnwidth]{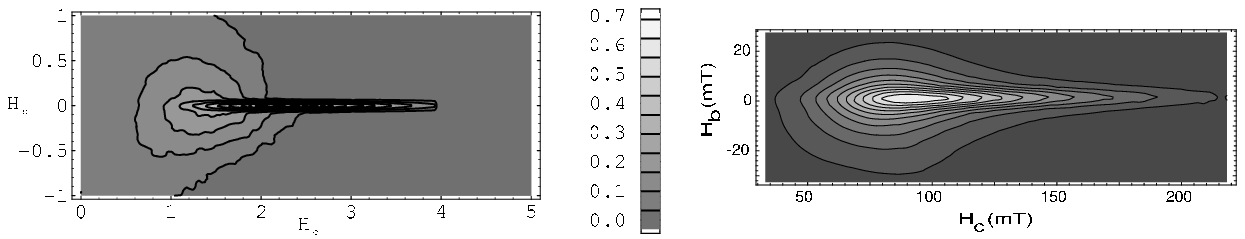}
\caption{{\bf Left:} FORC Diagram of the EASG. Note the ridge along
  the $H_c$ axis. {\bf Right}: Experimental FORC diagram of a Kodak
  sample. Note the similarity to the FORC diagram of the EASG shown in
  the left diagram. (From \figcite{hysteresis2}).}
\label{hysteresis-forc}
\end{figure}

Fig. \ref{hysteresis-forc} shows the FORC diagram of the EA spin glass.
The ridge along the $H_c$ axis in the range $1.5 < H_{c} < 4.0$
corresponds to the kinks of Fig.~\ref{hysteresis-kink}. Thus FORC
diagrams capture the reversal-field memory effect in the form of a
ridge along the $H_c$ axis. In Fig. \ref{hysteresis-forc} also the
experimentally determined FORC diagram of thin films of well-dispersed
single-domain magnetic Co-$\gamma$-Fe$_2$O$_3$ particles provided by
Kodak Inc is shown. It clearly exhibits the horizontal ridge
associated with the reversal-field memory effect. This striking
similarity between the experimentally determined FORC diagram of the
Co-$\gamma$-Fe$_2$O$_3$ films and the numerically determined FORC
diagram of the EASG indicates not only that Co-$\gamma$-Fe$_2$O$_3$
films exhibit reversal-field memory but also that frustration may be a
component of the physics of the Co-$\gamma$-Fe$_2$O$_3$ films.
Note that for instance the $2d$ random field Ising
model does NOT exhibit a reversal-field memory and that 
its FORC diagram has a vertical ridge rather than a horizontal 
one.\scite{hysteresis2}

%%%%%%%%%%%%%%%%%%%%%%%%%%%%%%%%%%%%%%%%%%%%%%%%%%%%%%%%%%%%%%%%%%%%%%%%%%%%%%%
\section{Equilibrium Properties of Classical $XY$ and Heisenberg Spin Glasses}
%%%%%%%%%%%%%%%%%%%%%%%%%%%%%%%%%%%%%%%%%%%%%%%%%%%%%%%%%%%%%%%%%%%%%%%%%%%%%%%
\label{sec:ContinuousSpinModels}

%==============================================================================
\subsection{Continuous Spin Models in Three Dimensions}
%==============================================================================

Compared to Ising spin glass models, less effort has been devoted to
the models with continuous spins. One of the reasons might be that 
the minimal model of spin glasses, the Ising spin glass model, 
already shows a highly non-trivial behavior and 
poses serious difficulties to reach a final conclusion. 
Nonetheless, the continuous spin models are worth
studying for several reasons. An obvious reason is that, in many
real spin glass materials such as AuFe and CuMn, the magnetic
anisotropy is much smaller than the (isotropic) exchange interaction.
For these materials, therefore, the Heisenberg spin glass models would
be more realistic than the Ising spin glass models.

However, the continuous spin models in three dimensions had long been
thought of as models without a spin glass transition, which is another
reason why the continuous spin models have not been studied so
extensively until recently.  The scenario that seemed to explain this
apparent disagreement between the models and the materials was that
the anisotropy is a relevant perturbation and, no matter how small it
may be, it changes the system to an Ising-like spin glass, giving rise
to a finite temperature phase transition.  This is quite analogous to
what happens when a small amount of Ising anisotropy was introduced
to the isotropic Heisenberg model (with no disorder) in
two dimensions.
If this scenario is valid, the critical behavior of any spin glass
material should be the same and fall into the Ising-spin-glass
universality class. However, there are some qualitative differences
between experimental results for materials with a small anisotropy
and numerical results for the Ising spin glasses. 
In particular in a series of recent torque experiments 
\bcite{campbell1,campbell2}
of Heisenberg spin glass samples with varying degree of anisotropies
indications for a freezing transition even within an external field
were found (implying the applicability of a mean field or RSB scenario
rather than a droplet scenario to these Heisenberg spin glass
systems). More surprisingly, the transition in spin glass materials with
lower anistropy (more Heisenberg-like) appeared to be more robust
against the application of an external field than those with higher
anisotropy (more Ising like), which is in contrast with the
traditional view that higher anisotropy implies a stronger tendency to
order. These observations call for further theoretical investigations,
and here we report on recent theoretical efforts to scrutinize the
existence of a transition in pure Heisenberg spin glasses without
anisotropy.
Below we present two sets of evidences that seem to contradict
each other.
(The conflict has not been solved yet.)
While the first set of evidences supports 
a novel phase transition via the chirality scenario (see below), 
the other suggests an apparently more familiar spin glass transition.

The first one was proposed by Kawamura and coworkers.
(For a review, see \bcite{Kawamura2001}.)
It was suggested that a phase transition may exist in
three dimensional Heisenberg model even if there is no anisotropy,
and belong to a new universality class different from that 
of the Ising spin glass model.  
In a series of numerical works (e.g., \bcite{Kawamura1998}), the
origin of the disagreement between the experimental and numerical
results was scrutinized by reconsidering the role of the magnetic
anisotropy in the mechanism of the spin glass transition.  
The resulting hypothesis was that there is a phase transition
already in the isotropic system but the spins do not show
long-range correlations even below this critical point. 
It is therefore difficult, if not impossible, to detect 
an anomalous behavior in spin-spin correlations
which had been focused on in most preceding numerical simulations.
Only the anisotropy gives rise to the long range
correlations in spins, not the transition itself, by coupling them to
other degrees of freedom that have been already ordered.  
It was proposed that these degrees of freedom relevant for
the transition are the chiralities defined as
$$
  \chi_i = \vect{S}_i\cdot(\vect{S}_{i+\delta}\times\vect{S}_{i+\delta'})
$$
where $\delta$ and $\delta'$ are two distinct unit lattice-vectors.

The dynamical behavior of the 
three-dimensional Heisenberg model
with and without a uni-axial spin anisotropy
was studied\scite{Kawamura1998}
and a clear aging phenomena, for both the isotropic and the
anisotropic cases, was found in the autocorrelation function 
of the chirality at sufficiently low temperatures. 
For the isotropic Heisenberg spin glass model in three dimensions with 
the Gaussian bond distribution,
the critical temperature and the exponent were estimated as
$$
  T_{\rm ch} = 0.157(1),\quad \beta_{\rm ch} = 1.1(1).
$$
Such a clear aging phenomena, however, was observed
in the bare spin degrees of freedom only when the anisotropy exists. 
It was therefore suggested\scite{HukushimaK2000a} 
that this phase transition is not accompanied by a freezing of spins
in the isotropic case.
As can be seen in \Fig{HukushimaK2000a_fig1}, 
the estimate of the Binder parameter
for the bare spins decreases as the system size increases at
any temperature and does not show any crossing.
In contrast, the Binder parameter of the chirality crosses.
\deffig{HukushimaK2000a_fig1}{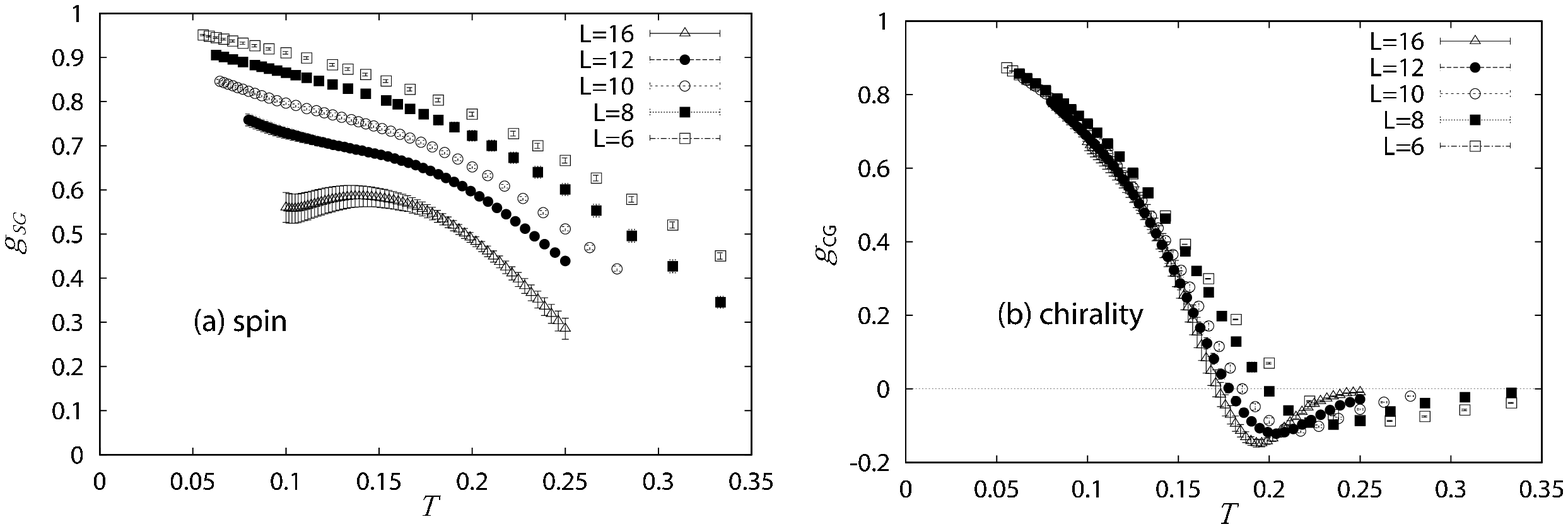}{0.9}
{The binder parameter for the $\pm J$ Heisenberg spin glass
model in three dimensions. The left panel is for the
spin overlap distribution and the right panel is for
the chirality overlap distribution.
(From \figcite{HukushimaK2000a}.)}

An evidence of a finite temperature phase transition was
also found for the plane-rotator model (i.e., the $XY$ model with
two-component spins) in three dimensions.
In \bcite{KawamuraL2001} an equilibrium 
Monte Carlo simulation was performed and a crossing in the Binder
parameter defined in terms of the chirality was observed
while, again, no such crossing was observed for the Binder parameter defined
with the bare spins.
For the $3d$ plane-rotator model with $\pm J$ bond-distribution
their estimates are
$$
  T_{\rm ch} = 0.39(3),\quad 
  \nu = 1.2(2), \quad
  \eta_{\rm ch} = 0.15(20), \quad
  z_{\rm ch} = 7.4(10).
$$

It was further argued that, if a small but finite 
anisotropy exists, the spins and the chiralities are coupled, 
and therefore the spin sector, 
being dragged by anomalies in the chirality sector, 
also shows anomalies at the transition point.
It follows that, as a function of the anisotropy, 
the transition temperature continuously approaches its isotropic limit.
Moreover, the singular part of dragged quantities, i.e., those which diverge
when the system is anisotropic and do not diverge when it is isotropic,
are proportional to the quantity that drags them, namely, the chirality.
The spin glass susceptibility, for instance, is given by
$$
   \chi_{\rm SG}(T,D) \sim D^4 (T-T_c)^{-\gamma_{\rm SG}^{\rm chiral}}
$$
The exponent $\gamma_{\rm SG}$ characterizing the 
spin glass susceptibility is the same as the one characterizing 
the chiral spin glass susceptibility.
Their estimates of critical indices were compared with those of
real materials, and many experimental estimates are closer
to those of chirality transition than the Ising spin glass
transition (See, \Tab{ExponentsInThreeDimensions}).

The second set of evidences contradicts to these findings.
In a few recent reports on the $XY$ and the Heisenberg
spin glasses in three dimensions, it was suggested that 
spins are ordered at the same finite temperature as
the chiral degrees of freedom even in the isotropic case.
The first such evidence was obtained\scite{MaucourtG1998}
through the computation of 
the domain-wall energy of the $XY$ model in two and three-dimensions.
The authors of \bcite{MaucourtG1998}
computed the spin and the chiral domain-wall energies and
estimated the corresponding stiffness exponents.
%While they estimated for the two-dimensional case as
%$$
%  \nu_{\rm spin} = 1.29(2)\quad\mbox{and}\quad \nu_{\rm chiral} = 2.57(3),
%$$
In the three-dimensional case, they obtained a positive value
for the spin stiffness exponent,
$$
  \theta_S^{{\rm spin}} = 0.056(11),
$$
while they obtained a much larger
stiffness exponent for the chiral domain-wall energy.
The positivity of the stiffness exponent was reconfirmed by
another computation based on the Coulomb gas representation.\scite{AkinoK2002}
However, since the estimated value is close to zero and the
actual increase observed in the domain-wall energy 
was only about 10 percent for the whole range of system size explored,
these results on the stiffness exponent alone could not be taken
as a conclusive evidence.

Additional evidences came from Monte Carlo simulations at finite
temperature,\scite{MatsubaraES2000,EndohMS2001,MatsubaraSE2001} in
which the Binder parameters for the magnetization and the chirality
of the three-dimensional $\pm J$ Heisenberg model were computed. A
common crossing of the curves for different system sizes was found in
both Binder parameters. The two crossing temperatures, one for the
spin and the other for the chirality, were close to each other.  Based
on this result, the presence of a finite temperature
phase transition at which both the spin and the chiral degrees of
freedom are ordered was suggested.

This result was confirmed in \bcite{NakamuraE2002} where
again the $\pm J$ Heisenberg spin glass in three dimensions
was studied. There the relaxation of several quantities at temperatures
close to the critical one were measured,
starting from a completely random initial configuration.
In the time regime where the time-dependent length
scale discussed in \Ssc{TimeDependentLengthScale}
is smaller than the system size, the size dependence was absent
in the auto-correlation function
and the observed time-dependence could be regarded as
the one in the thermodynamic limit.
Right at the critical temperature, this time dependence
was well described by an algebraic function for diverging quantities.
For example, for the spin glass susceptibility, we have
\begin{equation}
  \chi_{\rm SG} \propto t^{\lambda}
  \label{eq:AsymptoticRelaxation}
\end{equation}
with $\lambda = \gamma_{\rm SG}/z\nu$.\scite{SadicB1984}
Using this asymptotic form, the critical temperature as well as the critical
exponents (divided by $z\nu$) can be determined through
the observation of the time dependence of some quantity.
The exponent $z\nu$ can be obtained from
the temperature dependence of the characteristic time
$$
  \tau(\Delta T) \propto (\Delta T)^{-z\nu},
$$
where $\Delta T \equiv T - T_{\rm sg}$.
In addition, $z$ was estimated from the results of relatively
small system sizes for which the correlation length was larger than
the system size.
In such a setting, the size dependence of the correlation time 
near the critical point should be described by
$$
  \tau(\Delta T = 0) \propto L^z.
$$
The estimated critical temperature and exponents
for the Heisenberg spin glass model with $\pm J$ 
bond-distribution in three dimensions were
$$
  T_{\rm sg}/J = 
  0.21 \begin{array}{l}{\tiny +0.01} \\ {\tiny -0.03}\end{array},
  \quad
  \gamma_{\rm sg} = 1.9(4),
  \quad
  \nu = 1.1(2),
  \quad
  \beta = 0.72(6).
$$
It should be noted that this estimate of $\nu$ agrees with some of
the experiments on spin glass materials that appear not to be
in the Ising universality class.

A clearer demonstration of the existence of the spin ordering
was presented in \bcite{LeeY2003}, in which an 
equilibrium Monte Carlo simulation was performed 
covering not only the critical point
but also a fairly large region in the low temperature phase.
Instead of the Binder parameter, 
which is an indirect probe of the presence of the long range order,
they used the correlation length itself 
as in the estimation of the transition 
temperature of the Ising spin glass model\scite{BallesterosETAL2000} 
(See \Fig{BallesterosETAL2000_fig3}).
For both the XY and the Heisenberg spin glass with the Gaussian bond
distribution in three dimensions,
a clear crossing in the correlation length divided by
the system size as a function of the temperature
was observed (\Fig{LeeY2003_fig2_and_fig4}).
The critical indices were estimated as
$$
  T_{\rm sg} = 0.34(2),\quad \nu = 1.2(2) \quad\mbox{(XY)}
$$
and
$$
  T_{\rm sg} = 0.16(2),\quad \nu = 1.1(2) \quad\mbox{(Heisenberg)}.
$$
\deffig{LeeY2003_fig2_and_fig4}{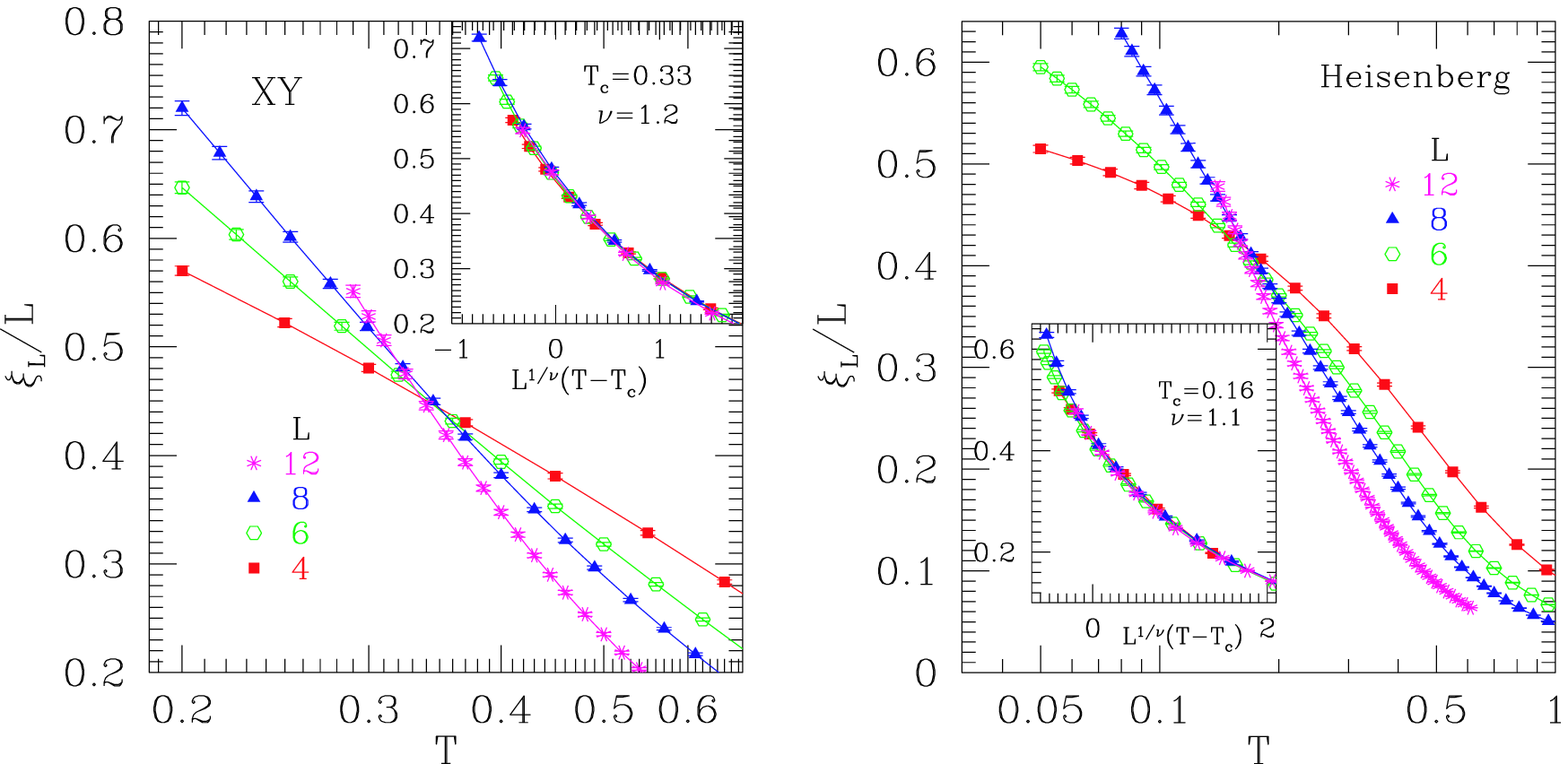}{0.9}
{Correlation lengths of bare spins of the $XY$ model (left) 
and of the Heisenberg model (right)
divided by the system size. (From \figcite{LeeY2003}.)}

It is rather puzzling to have two conflicting sets of numerical evidences,
represented, e.g., by \Fig{HukushimaK2000a_fig1} and \Fig{LeeY2003_fig2_and_fig4},
each would pass as a convincing evidence if the other did not exist.
This puzzling situation has not been resolved yet.
It is at least clear that there is an unexpectedly large correction 
to scaling and that we need to answer why one of the 
seemingly legitimate methods for detecting the long-range 
order failed in the present case.

%==============================================================================
\subsection{Continuous Spin Models in Higher Dimensions}
%==============================================================================

The Heisenberg spin glass model with Gaussian couplings was studied in
four and in five dimensions via equilibrium Monte Carlo simulations in
\bcite{ImagawaK2003}. As in the study of the three-dimensional 
case\scite{HukushimaK2000a} the Binder parameter
for spins and chiralities was calculated. In the four dimensional
case, it was found that the spin Binder parameters for various system
sizes do not intersect while the chirality Binder parameters do.  In
contrast, in the five dimensional case, both kinds of the Binder
parameters intersect.  Based on these results it was suggested that the
spin chirality separation occurs only in four or lower dimensions
while in five or higher dimensions an ordinary phase transition, at
which both degrees of freedom freeze, takes place.
In five dimensions, the transition temperature was estimated as 
$T_c = 0.60(2)$ and the critical indices as
$$
  \alpha = -1.0(3), \ 
  \beta_{\rm SG} = 0.7(3), \ 
  \gamma_{\rm SG} = 1.7(2), \ 
  \nu_{\rm SG} = 0.6(2).
$$
Indications were found that the chiral order phase is characterized by
one-step replica symmetry breaking unlike the mean-field model.
The findings for the four dimensional model were not as clear as those
for the five dimensional model.  In four dimensions, the chiral
transition temperature was estimated to be $T_c = 0.38(2)$ but no
reliable estimates for the critical indices for
the four dimensional model could be obtained.
This may be because four dimensions is
marginal or nearly marginal and the critical region is rather 
narrow.

For these numerical evidences based on the Binder parameters, however,
one has to be aware that the same ambiguities might be present in four and
five dimensions as in the three-dimensional case discussed in the 
previous subsection. 
Therefore, while the existence of a phase transition has been established 
in four and five dimensions as well as in three, we still cannot make
a conclusive statement concerning the nature of the phase transition
in continuous spin models for the same reason as 
in the three-dimensional case.

%==============================================================================
\subsection{Potts Spin Glasses}
%==============================================================================
Potts spin variables are discrete variables that can take on $q$
different states $S_i\in\{1,2,\ldots,q\}$ and that interact with one
another in such a way that only two cases are discriminated: either
both interacting spins are in the same state or they are in different
states. The Hamiltonian of the $q$-state Potts glass is given by 
\begin{equation}
H=-\sum_{(i,j)} J_{ij}\, ( q\delta_{S_i,S_j}-1 )\;,
\end{equation}
where the coupling constants $J_{ij}$ are quenched random variables
and the sum runs over all interacting spin pairs (i.e., all possible
spin pairs in a mean-field model). 

The $q$-state infinite range Potts glass with 
$q>4$\scite{potts-kirk1,potts-kirk2} displays a dynamical phase transition
at a temperature $T_d$, where the dynamics freezes and, for instance,
the spin auto-correlations do not decay any more (i.e. they reach
asymptotically a non-vanishing plateau value).  The second, static,
transition takes place at a lower temperature $T_c$, below which the
EA spin glass order parameter has a non-vanishing value and the
replica symmetry is broken. On the mean-field level there is a close
connection between Potts glasses for $q\ge4$, and Ising spin glass
models with $p$-spin interactions for $p\ge3$, which again display
similar self-consistency equations as the mode-coupling equations
describing structural glasses above the glass transition 
temperature.\scite{goetze1,goetze2}
Therefore Potts spin glasses can also be regarded as simple 
prototypical models for structural glasses. 

The question unanswered to date is, whether this analytically
well-established mean-field scenario with the two separate
transitions is also valid in finite-dimensional models with short
range interactions. Recently, the $q$-state Potts glass was thoroughly
investigated via Monte Carlo simulations in three 
dimensions,\scite{potts-brangian1,potts-brangian2} on the simple cubic 
lattice, with $q=10$ for a discrete, bimodal bond-distribution as well as
a Gaussian distribution. In both cases, the first two moments of the
distribution were chosen such that no ferromagnetic ordering of the
Potts spins could occur. It was found that for all temperatures
investigated the spin glass susceptibility remained finite, 
the spin-glass order-parameter remained zero, 
and that the specific heat had
only a smooth Schottky-like peak. These results could be understood
quantitatively by considering small but independent clusters of
spins. These observations imply that there is no static phase
transition at any nonzero temperature. Consistent with these findings,
only very minor size effects were observed, which implied that all
correlation lengths of the models remained very short.

Moreover, the auto-correlation function $C(t)$ of the Potts spins was
computed. While in the Gaussian model $C(t)$ shows a smooth uniform
decay, the correlation for the $\pm J$ model has several distinct
steps. These steps correspond to the breaking of bonds in small
clusters of ferromagnetically coupled spins (dimers, trimers,
etc.). The relaxation times follow simple Arrhenius laws, with
activation energies that are readily interpreted within the cluster
picture, giving an evidence that the system does not have a dynamic
transition at a finite temperature. Hence one can conclude that
all the transitions known for the mean-field version of the
model are completely wiped out in three dimensions.

It should also be mentioned that the mean-field model of the Potts
spin glass was studied numerically via Monte-Carlo 
simulations.\scite{potts-mf-brangian1,potts-mf-brangian2,potts-mf-brangian3}  
It turned out the it is extremely difficult to control the finite size
effects in the mean-field model and that the fourth order cumulant
$g_4(N,T)$ and the Guerra parameter $G(N,T)$ could not be used to
locate the static transition temperature for the system sizes
investigated. Also the spin-autocorrelation function $C(t)$ showed strong
finite size effects and did not display a plateau even for
temperatures around the dynamical critical temperature $T_d$.

%%%%%%%%%%%%%%%%%%%%%%%%%%%%%%%%%%%%%%%%%%%%%%%%%%%%%%%%%%%%%%%%%%%%%%%%%%%%%%%
\section{Weak Disorder}
%%%%%%%%%%%%%%%%%%%%%%%%%%%%%%%%%%%%%%%%%%%%%%%%%%%%%%%%%%%%%%%%%%%%%%%%%%%%%%%
\label{sec:WeakDisorder}
While there are only very few exactly known facts 
about spin glass models, they serve as helpful guides
to numerical studies.
This is particularly the case in the study of random spin system 
with weak disorder.
An example is the findings based on the gauge invariance
(a brief review can be found in \bcite{Nishimori2002c}).
A line in the phase space, commonly referred to as
the Nishimori line (simply N-line, hereafter), 
was found on which the free energy can be computed exactly.
The line is associated with the crossover from 
the high-temperature ferromagnetic region to 
the low-temperature disorder-dominated region.
The results derived from the gauge invariance
seem to have a close connection
to a number of renormalization group studies.
In the following, we present a gross survey of 
these findings and related works.

%==============================================================================
\subsection{Phase Diagram of the Discrete Spin Models}
%==============================================================================
\label{ssc:PhaseDiagramDescreteSpins}
It was pointed out\scite{Nishimori1981} that the gauge invariance
of the partition function under the transformation $S_i\to\sigma_i S_i$
(with an arbitrary choice of $\sigma_i\in\{-1,+1\}$ for every site)
can be used to derive a number of exact relations among various
quantities in spin glass models. 
This gauge transformation relates thermal fluctuations to
the geometrical properties of the bond configurations.
Consequently, a {\it geometric temperature}, $T_p$, can be defined
that depends only on the bond distribution.
This temperature appears in various exact relationships derived from this
gauge invariance.
An example is the identity for
correlation functions in $\pm J$ models:
\begin{equation}
  [\langle S_i S_j \rangle_T]
  = [\langle S_i S_j \rangle_T \langle S_i S_j \rangle_{T_p}]
  \label{eq:NishimoriCorrelation}
\end{equation}
where $\langle \cdots \rangle_T$ is the thermal average at temperature $T$.
In particular, $\langle \cdots \rangle_{T_p}$ means the thermal average
with the real temperature and the geometrical temperature being equal
to each other.
The symbol $[\cdots]$ denotes, as usual, the bond-configuration average.
The geometric temperature $T_p$ is defined as
$$
  p = \left(1 + {\rm tanh} (J/T_p) \right)/ 2,
  \quad \mbox{or}\quad 
  T_p = \frac{2J}{{\rm log} \frac{p}{1-p}}
$$
where $p$ is the concentration of the ferromagnetic bonds.
The higher the randomness ($p\to 1/2$), 
the higher the geometric temperature ($T_p\to \infty$).
The line in the $p-T$ phase diagram defined by $T_p = T$ is the N-line.
Along this line, \Eq{NishimoriCorrelation} infers $m=q$, i.e.,
the spin-glass order is always accompanied by the ferromagnetic order.
Therefore, no part of the line lies inside the pure spin-glass phase.
(Here, we define the spin-glass phase by
the conditions $m=0$ and $q>0$.)
Another consequence of \Eq{NishimoriCorrelation} is that
the ferromagnetic long-range order is absent at any temperature 
if the concentration $p$ is such that
the point $(p,T_p)$ falls in the paramagnetic region in the $p-T$ plane.
This can be understood by an inequality directly derived from 
\Eq{NishimoriCorrelation}:
\begin{equation}
  \left|\left[\langle S_i S_j \rangle_T\right]\right| 
  \le 
  \left[\left|\langle S_i S_j \rangle_{T_p}\right|\right].
  \label{eq:CorrelationInequality}
\end{equation}
\rem{No need for the absolute value symbol on the right hand side?}
If the point $(p,T_p)$ belongs to the paramagnetic region,
the right hand side of the inequality is zero in the limit of
$R_{ij}\to\infty$, which makes the left hand side also zero, 
meaning the absence of ferromagnetic order at $(p,T)$.
Assuming that the topology of the $T-p$ phase diagram is as depicted in 
\Fig{PhaseDiagram2D}(a),
this fact implies that any phase in the region $p<p_N$ (or $T_p>T_N$)
cannot have a ferromagnetic long-range order where $p_N$
is the ferromagnetic bond concentration at the N-point.
Here the N-point $(p_N,T_N)$
is defined as the point at which the paramagnetic
phase boundary intersect with the N-line.
It follows that the phase boundary separating the ferromagnetic
phase from the other low-temperature phase (paramagnetic or spin-glass)
must be strictly vertical or bend toward the ferromagnetic side.
(See \Fig{PhaseDiagram2D}.)

The inequality \Eq{CorrelationInequality} also means
that the ferromagnetic order is maximal on the N-line 
when $p$ is fixed. 
When $T_p < T_N$, as we decrease the temperature departing 
from the N-line, the magnetization decreases and the spin
glass order parameter becomes larger than the square of the 
magnetization squared. 
This hints a cross-over from the purely ferromagnetic region to
the randomness-dominating region at the temperature $T=T_p$.

It was further argued\scite{Nishimori1986} that the phase boundary
below the N-line is strictly vertical.
The argument is based on the fact that
the free energy on the N-line is
the same as the geometric entropy, i.e., the entropy of the 
frustration distribution at $p$:
$$
  S(p) \equiv -\sum_F P(F)\, {\rm log}\, P(F)
$$
where $F$ is a configuration of frustrated plaquettes and
$P(F)$ is the probability of $F$ being realized.
Since the free energy (or its derivative with respect to some
external field) must have some singularity at the N-point
when one moves along the N-line,
so does the geometric entropy.
The latter singularity is, however, solely due to the geometric 
properties of the bond configuration, in which the temperature does not
play any role.
Therefore, it is reasonable to suspect that
the same singularity with the geometric origin may affect the spin
system defined on it at $p = p_{\rm N}$ regardless of the
temperature.
This leads to a temperature-independent transition point, i.e.,
the strictly vertical phase boundary (See \Fig{PhaseDiagram2D}).
Note that the argument applies also to models with continuous degrees of
freedom, such as the $XY$ model and the Heisenberg model, if the
ferromagnetic phase exists in the pure model.
It was claimed\scite{Nishimori1986} that the vertical phase boundary
between the ferromagnetic phase and the other is universal.
\deffig{PhaseDiagram2D}{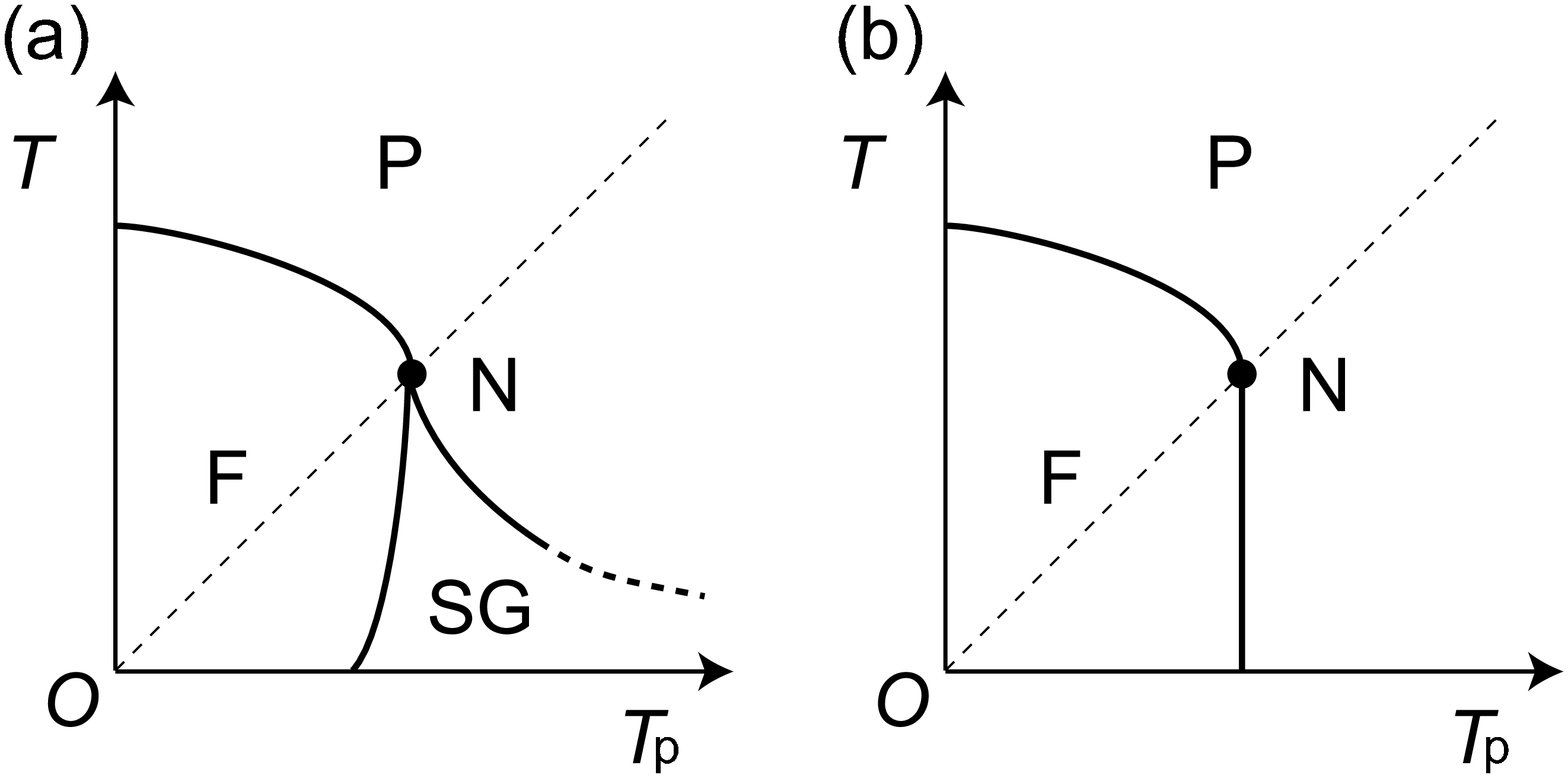}{0.8}
{The schematic $p-T$ phase diagram the $\pm J$ Ising model
for (a) the generic case and (b) the two-dimensional case.
The thin dashed lines represent the N-line.}

In order to substantiate this argument, 
the authors of \bcite{NishimoriFO2002} 
considered the random $Z_q$ model with gauge symmetry,
$\Ham = -\sum_{(ij)} V(S_i-S_j+J_{ij})$ with $S_i, S_j, J_{ij}$ takes
on a value $0,1,2,\cdots$ or $q-1$, and $V(\cdots)$ is a periodic 
function of the period $q$.
This model includes the $\pm J$ Ising spin glass model as a special case.
The sample-to-sample fluctuations of the energy along 
the N-line was computed via the exact relationship
between the sample-to-sample fluctuations and the thermal fluctuations:
$$
  [(\Delta E)^2] \equiv [E^2]-[E]^2
  = N_{\rm B} (\{ V^2 \} - \{ V \}^2) 
    - [\langle H^2 \rangle - \langle H\rangle^2].
$$
Here, $E \equiv \langle \Ham \rangle$, $N_{\rm B}$ the number of 
nearest-neighbor pairs, and
$\{ \cdots \}$ denotes a single-bond average defined by
$
  \{ Q(l) \} \equiv \sum_{l=0}^{q-1} \exp(-\beta V(l)) Q(l) 
$ 
$
  / \sum_{l=0}^{q-1} \exp(-\beta V(l)).
$
The sample-to-sample fluctuations are proportional to the 
`geometric specific heat', i.e., the energy differentiated by 
$K_p\equiv 1/T_p$.
The results showed that $[(\Delta E)^2]$ takes its maximum at the N-point.
The authors of \bcite{NishimoriFO2002} 
argued that this indicates the existence of a singularity in the
geometric nature of the system
at the N-point, though the nature of the singularity
is not known possiblly because of the size limitation in their calculation.

It is interesting to note that the correlation length exponent $\nu$ at 
the zero-temperature phase transition from the paramagnetic phase to
the ferromagnetic phase in two dimension was found\scite{SinghA1996,KawashimaR1997} 
to be very close to (or perhaps exactly equal to)
the one for two-dimensional percolation.
This fact, together with the geometric mechanism of the transition,
makes it very attractive to speculate that the transition is due
to a percolation of something that is defined geometrically.
However, this ``something'' has not yet been identified.

%==============================================================================
\subsection{Dynamical Properties}
\label{ssc:DynamicalProperties}
%==============================================================================
Based on the gauge invariance of the model,
several interesting exact relations were derived\scite{Ozeki1997}
for the dynamics.
Particularly important is the one that relates 
the auto-correlation function with the totally aligned initial state
and that with the initialization at $T=T_p$.
Specifically,
\begin{equation}
  [\langle S_i(t+t_w) S_i(t_w) \rangle^{\rm (F)}_T]
  =
  [\langle S_i(t+t_w) S_i(T_w) \rangle^{T_p}_T].
\end{equation}
The superscripts specify the initial spin configurations:
``(F)'' indicates that the initial state is the 
totally aligned ferromagnetic state,
and ``$T_p$'' an equilibrium spin configuration at $T_p$.
The left hand side is the autocorrelation function of a
spin $S_i$ at the $t_w$-th Monte Carlo step (MCS) 
after the initialization (at $t=0$) and itself at the 
$(t+t_w)$-th step.
At $t>0$ the system evolves with the dynamics of a finite
temperature $T$.
The right hand side is the autocorrelation
function after a quenching from the temperature $T_p$ to $T$
(or a sudden heating up, depending upon whether $T>T_p$ or $T<T_p$).
Of particular interest is the relation obtained by setting $t_w=0$
and $T=T_p$ in the above
$$
  [\langle S_i(t) \rangle^{\rm (F)}_{T_p}]
  =
  [\langle S_i(t) S_i(0) \rangle^{T_p}_{T_p}].
$$
This indicates the equivalence of 
the equilibrium auto-correlation function on the N-line
to the non-equilibrium relaxation of the single-spin expectation 
value starting from the all-aligned condition.
By using this relation, we can simply measure the value of a
spin without equilibrating at all to obtain the equilibrium 
autocorrelation function.
This is a considerable advantage from the computational point of view.

%==============================================================================
\subsection{The Renormalization Group Approach for the Discrete Models}
%==============================================================================

The Harris criterion\scite{Harris1974} is well-known as 
the criterion by which one can decide whether introduction of 
a weak disorder to a pure system is relevant or irrelevant:
If the specific heat diverges at the
critical point of the pure system, i.e., the specific heat exponent
$\alpha$ is positive, the universality class of the transition in the 
disordered system will be different from the one of the pure system.
In particular, it was argued that 
if the disorder is relevant the specific heat divergence 
is smeared out by the effect of the quenched disorder.

The argument was elaborated further with the help of the
renormalization group theory.\scite{GrinsteinL1976}
The analysis of
the randomly diluted $m$-vector model showed that the quenched
disorder is relevant around the ``pure'' fixed point (denoted as ``P''
hereafter), if the specific exponent $\alpha$ is positive at P.  As a
result, the existence of another fixed point, called the random fixed
point (denoted as ``R''), was suggested, and the renormalization group
flow leads the system to R when starting in the vicinity of P.
However, because of the small cross-over exponent, i.e., $\alpha$ in
this case, the critical region where the true critical behavior of R
can be observed may be very narrow in many cases, such as the random
Ising model in three dimensions ($\alpha \sim 0.11$).  For instance
results of Monte Carlo simulations of the
site-disordered Ising system\scite{Heuer1993} 
were interpreted in terms of a non-universal
behavior, i.e., continuously varying critical exponent depending
on the strength of the disorder. It was argued that this apparent
non-universal behavior can be interpreted as a cross-over from P to R.

As for the critical indices at R, 
the following values for three dimensional systems
were reported\scite{Mayer1989}
based on the four-loop order field-theoretic
renormalization group calculation:
\begin{equation}
  \beta = 0.348,\ 
  \gamma = 1.321,\ 
  \nu = 0.671,\ 
  \eta = 0.032.
\end{equation}

In \bcite{LeDoussal1988} the role of the gauge invariance in the
renormalization group method was scrutinized. It was found that the
defining condition of the N-line, $T_p = T$ is invariant under a
certain renormalization group transformation.  
More specifically, in an extended phase space 
an invariant manifold exists that intersects with 
the $p-T$ plane with the intersection being the N-line.
A fixed point is located on this manifold and
can be identified with the one corresponding 
to the multi-critical point that is shared by three phases: 
paramagnetic, ferromagnetic and
spin-glass.  One of the directions of the two scaling axes is also
found to be parallel to the temperature axis while the other is
tangent to the N-line.  The fixed point is unstable in both
directions.

The authors of \bcite{BallesterosETAL1998} studied the site-diluted
Ising model with Monte Carlo simulation.
Their estimates for the critical indices were
$$
%  \alpha = -0.051(7),\ 
  \beta = 0.3546(18),\ 
  \gamma = 1.342(5),\ 
  \nu = 0.6837(24),\ 
%  (y = 1.463(5),)\ 
  \eta = 0.0374(36).
$$
These are very close to the
corresponding estimates of the indices for the pure model
such as:\scite{LeGuillowZ1980}
$$
  \beta = 0.3250(15),\ 
  \gamma = 1.241(2),\ 
  \nu = 0.6300(15).
$$
Therefore it is technically rather difficult to discriminate a
critical point corresponding to the random fixed point from the
pure fixed point.
\deffig{FlowDiagram3D}{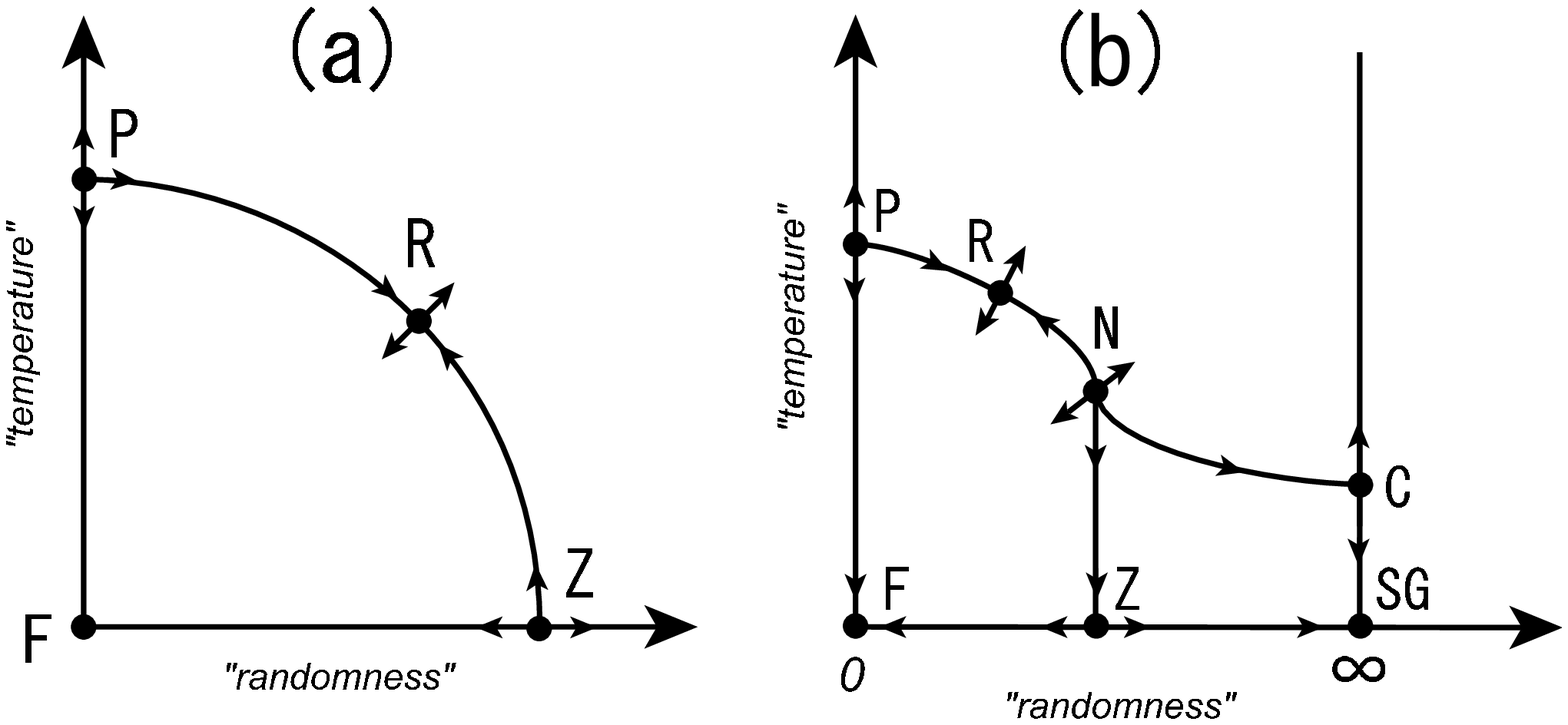}{0.9}
{A schematic RG flow diagram for (a) non-frustrated models 
(e.g., the randomly diluted ferromagnet) and (b) frustrated
models (e.g., the $\pm J$ model) in three dimensions.}

The numerical renormalization group approach based on Monte Carlo
simulation\scite{Hukushima2000} suggested the simplest
flow diagram consistent with all of these predictions.
The variance and the mean of the domain wall free energy
were computed and analyzed with the method of the domain-wall 
renormalization group.\scite{McMillan1984a}
For the randomly diluted spin systems in three dimensions
the flow diagram turned out 
to be the simplest one consistent with preceding theoretical
predictions (see \Fig{FlowDiagram3D}).
The RG flow diagram for the bond-diluted model 
consists of two finite temperature fixed points
(the pure fixed point (P) and the random fixed point (R))
together with two zero-temperature fixed points
(the ferromagnetic fixed point (F) and the zero-temperature 
percolation fixed point (Z)).
%The random fixed point was located at
%\begin{equation}
%  \tilde T = 0.63(2) \ \mbox{and}\ \tilde D = 1.77(5)
%  \label{eq:RandomFixedPoint}
%\end{equation}
%for the bond-diluted system
%where $\tilde T$ is the renormalized temperature and
%$\tilde D$ is the renormalized bond-randomness.
The two eigen values of the linearized RG transformation
were evaluated at R.
The positive one turned out to be
$$
  y_1 = 1.47(4)
$$
which is consistent with the previous estimates of $\nu=1/y \sim 0.68$
mentioned above. The other negative one was
$$
  y_2 = -1.3(4).
$$
A similar calculation for the site-diluted system yielded
the result consistent with these.

When in addition the system is randomly frustrated, 
the simplest flow diagram
must contain at least three more fixed points:
the multi-critical or Nishimori fixed point (N),
the spin-glass critical fixed point (C), and the spin-glass
zero-temperature fixed point (SG).
The findings in \bcite{Hukushima2000}
were indeed consistent with this simplest
flow diagram (\Fig{FlowDiagram3D}).
In this case the exponent at R was estimated as
$$
  y_1 = 1.52(2),\quad \mbox{and}\quad y_2 = -0.42(13).
$$
While the positive one agrees with the estimate for the
non-frustrated system, the other does not agree.
The reason for this has not been clarified yet.
%As for the random fixed point, its location in terms of
%the renormalized temperature and randomness was found to be
%$$
%  \tilde T = 0.66(2) \ \mbox{and}\ \tilde D = 1.74(1)
%$$
%which agrees with that of the non-frustrated system 
%\Eq{RandomFixedPoint}.

%==============================================================================
\subsection{The Location of the Multi-Critical Point}
%==============================================================================
For the location of the multi-critical point, or the N-point,
of the $\pm J$ model, there are a number of numerical estimates.
In particular, the non-equilibrium simulation was employed effectively
to deduce the equilibrium properties.
The basic idea of the method can be traced back to the dynamical scaling 
hypothesis.\scite{Suzuki1977,Suzuki1978}
Various scaling relations, such as \Eq{AsymptoticRelaxation}, 
can be derived from it.
It is assumed that a generating function, which is an extension of
the equilibrium free energy, exists and that the time-dependence of
various quantities can be obtained as its derivatives.
The hypothetical generating function has the form
\begin{equation}
  f(\Delta T,h,L,t) = L^{-d} \tilde f(\Delta T L^y, h L^{y_h}, t L^{-z})
\label{eq:DynamicalScalingGeneratingFunction}
\end{equation}
analogous to the ordinary form of the finite size scaling.
While this form is presumably valid in the near-equilibrium time regime,
it was suggested\scite{janssen} that there exists an initial
time regime where the dynamics is governed by another critical exponent
independent of those which appear in 
\Eq{DynamicalScalingGeneratingFunction}.
\rem{
Is this O.K., Heiko?
I don't know why Nakamura and Endoh didn't observe
what Janssen et al predicted.
Has anyone observe that yet?
In any event, I think \Eq{AsymptoticRelaxation} is the correct form
at least in the last stage of the relaxation.
$>$ Explain. Cite Janssen, Schaub and Schmittmann,
refer to the necessary scaling relations between the 
equilibrium exponents
}

Using the scaling relations derived from 
\Eq{DynamicalScalingGeneratingFunction},
the non-equilibrium relaxation of the system
was studied.
The authors of \bcite{OzekiI1998}
measured the magnetization 
$m(t) \equiv [\langle S_i(t) \rangle^{(F)}_T]$ on the N-line, 
which is proved to be the same as the equilibrium autocorrelation
function $q(t)$ as mentioned above (see \Ssc{DynamicalProperties}). 
Then they used the following scaling form
$$
  m(t) = L^{-\beta y} \tilde q((T-T_{\rm N}) L^y, t L^{-z})
$$
that can be derived from \Eq{DynamicalScalingGeneratingFunction}.
In particular, the long-time asymptotic form at $T = T_{\rm N}$
$$
  m(t) \sim t^{-\beta y/z}
$$
was used to obtain the effective exponent at a finite time,
$$
  \lambda(t) \equiv - \frac{d\,{\rm log}\, m(t)}{d\,{\rm log}\, t},
$$
which should converge to $\lambda \equiv \beta y / z$ as $t\to\infty$
at the critical point.
A good convergence to a finite value was observed around
the N-point for the two and three dimensional models,
although the convergence in the latter case seemed more
unstable. They quoted the values
\begin{equation}
  p_{\rm N} = 0.8872(8), \quad \lambda = 0.021(1)
\end{equation}
for the two dimensional model and
\begin{equation}
  p_{\rm N} = 0.7673(3), \quad 
  \lambda = 0.090(3)
\end{equation}
for the three dimensional model.

In two dimensions,
the exact location of the N-point was predicted by
\bcite{NishimoriN2003}.
The authors studied the duality transformation formulated 
in \bcite{WuW1976},
applying it to a random model with $Z_q$ symmetry.
The model is defined in terms of variables, $\xi_i$, 
each taking one of $q$ values ($\xi = 0,1,2,\cdots,q-1$).
The partition function is defined as 
$$
  Z = \sum_{\{ \xi_i \}} e^{\sum_{(ij)} V(\xi_i-\xi_j+J_{ij})}
$$
where $V(l)$ is a periodic function defined on integers
with the period $q$ and $J_{ij}$ is a quenched random variable that takes
on $0,1,2,\cdots,q-1$.
They obtained an equation that yields a (possiblly exact) value of
the critical concentration $p_c$.
The resulting values of $p_c$ agreed with numerical estimates
for various models such as the $\pm J$ Ising model
and the three state Potts gauge glass.
In addition, a similar equation derived for the random Ising model
with Gaussian bond distribution yielded a value very close to the
numerically obtained critical value of $J/\Delta J$ where $J$
and $\Delta J$ are the mean and the standard deviation of the
bond distribution, respectively.

The replica method was employed in the derivation of the key equation;
the duality relation of the $n$ replica system was considered.
As reported in \bcite{Nishimori1979},
a quantity, which we denote as $\alpha$, defined with 
the physical parameters, such as $p$ and $T$ for the Ising model, 
has a very simple transformation rule under the duality transformation:
$$
  \alpha \rightarrow q^n/\alpha,
$$
Therefore, $\alpha = q^{n/2}$ defines a manifold that is
invariant under the duality transformation.
The virtue of this equation is that we can explicitly take
its $n\to 0$ limit, resulting in
\begin{equation}
  \sum_{l=0}^{q-1} p_l\, {\rm log}\left( \sum_{\eta=0}^{q-1} 
  e^{V(\eta + l) - V(l)} \right) = \frac12 {\rm log}\, q,
  \label{eq:SelfDualLine}
\end{equation}
where $p_l$ is the probability of $J_{ij}$ being $l$.
If a well-defined duality equation in the $n\to 0$ limit exists 
and if a self-dual point lies on our real $p-T$ plane, 
the condition \Eq{SelfDualLine} must be satisfied on it.
The problem is that we do not know what the
$n\to 0$ self-dual equation is nor
whether a self-dual point lies on our $p-T$ phase diagram.

However, with the assumption that a self-dual point
coincides with the N-point, it becomes possible to obtain
a number, which may be an exact value of $p_c$.
Specifically, the intersection of 
the N-line and the line defined by \Eq{SelfDualLine}
for the $\pm J$ Ising model is located at
$$
  p_c = 0.889972\cdots
$$
which agrees with preceding numerical estimates of the N-point
such as
$$
  p_c = 0.8905(5)
$$
in \bcite{AaraoReisQS1999}.
This remarkable agreement, together with similar
agreements for a few other models, seems a little too much
to regard as a mere coincidence;
the exact transition points of these models
may have been derived from the duality.
%in some random systems.
%It is not known, however, whether \Eq{SelfDualLine} is the
%sufficient and necessary condition for the self-duality or it is
%only a necessary condition.

%==============================================================================
\subsection{Phase Diagram of the Random $XY$ Model in Two Dimensions}
%==============================================================================

The effect of weak randomness in the
$XY$ model in two dimensions is of special relevance
due to its relationship to various other models,
such as the Coulomb gas and 
Josephson junction arrays.
The model discussed most frequently is the random phase $XY$ model
$$
  \Ham = - \sum_{(ij)} J_{ij} \cos( \theta_i - \theta_j - \phi_{ij} )
$$
with $J_{ij}$ being a uniform constant $J > 0$ and $\phi_{ij}$
a quenched random variable.
However, other models, such as the one with
random $J_{ij}$ and uniform $\phi_{ij}$,
most likely have essentially the same property.

In an early study\scite{RubinsteinSN1983}
the random phase model was studied and it was suggested that 
a sufficiently strong
disorder destroys the quasi-long-range order of the pure $XY$ model.
More strikingly, 
the amount of the disorder sufficient to destroy the
quasi-order is vanishing as we approach the zero temperature.
Consequently, the system undergoes a re-entrant phase transition
as the temperature is lowered with the magnitude of the disorder being
fixed at a sufficiently small value.

The latter prediction concerning the re-entrance phase transition 
was corrected later by a number of 
groups.\scite{NattermannSKL1995,ChaF1995,Tang1996,Scheidl1997,CarpentierD2000}
Here we follow the heuristic argument given by the authors of \bcite{ChaF1995}.
They argued that there is a disorder induced phase transition 
at zero temperature in the two-dimensional XY model.
Generalizing the Kosterlitz argument for the KT transition, 
they considered the balance between the energy-cost for creating a 
topological excitation (vortex) and the energy-gain due to the interaction
of the vertex with the disordered potential.
They estimated the probability $p_{\rm defect}$
of creating a vortex on a given site 
being energetically favored.
It depends algebraically on the area of the system $A$
with an exponent $\alpha$:
$$
  p_{\rm defect} \sim A^{-\alpha}
$$
where $\alpha$ depends on the magnitude of the disorder and the
temperature.
The ordered phase is stable against the introduction of disorder when 
the total number of sites on which the creation of the
vortex is favored is zero in the thermodynamic limit.
That is, $Ap_{\rm defect} \to 0$.
Therefore, $\alpha = 1$ defines the phase boundary.
This result contradicts preceding studies such as the 
one mentioned above\scite{RubinsteinSN1983,Nelson1983}
that suggest that the ground state of the pure 
system is unstable at any finite disorder.

They argued that the preceding renormalization group theories
failed to capture the correct physics because they neglected the
fluctuation in the local energy gain due to disorder potential,
which can be very large with a small but finite probability.
In particular, we cannot neglect such a fluctuation
when it exceeds the thermal fluctuation.
This means that the previous argument may fail
near and below the N-line because the N-line can be
regarded as the cross-over line below which the geometrical
fluctuation dominates (see \Ssc{PhaseDiagramDescreteSpins}).

They also discussed a finite temperature phase diagram introducing the
thermal fluctuations in their argument.
This yields another interesting feature of their results; the straight
phase boundary between the ordered phase and the disordered phase
in the $T-T_p$ phase diagram, in agreement with the Nishimori's claim
(see \Ssc{PhaseDiagramDescreteSpins}).
Generalizing the simple argument based on the energy-balance, 
they computed the exponent $\alpha$ for finite temperature:
\begin{equation}
  \alpha = \left\{
  \begin{array}{ll}
    \frac{T^{\ast}}{T_p} & (T_p > \lambda T) \\
    \frac{T^{\ast}}{T_p}\left( \frac{T_p}{\lambda T}
    \left( 2 - \frac{T_p}{\lambda T} \right)
    \right) & (T_p < \lambda T) 
  \end{array}
  \right.
\end{equation}
where $T^{\ast}$ and $\lambda$ are model-dependent parameters.
The phase boundary is again given by $\alpha = 1$.
Therefore, for the region where the disorder fluctuation dominates,
($T_p > \lambda T$), the phase boundary is $T_p = T^{\ast}$,
independent of the temperature,
whereas it depends on the temperature in the
region where the thermal fluctuation dominates ($T_p < \lambda T$).
If one identifies, quite naturally,
the line $T_p = \lambda T$ with the N-line,
the result is perfectly in parallel with Nishimori's picture based
on the gauge invariance.

In fact, the gauge invariance can be used for deriving exact results for
continuous spin models, such as the $XY$ models and the Heisenberg models.
In \bcite{Nishimori2002c}, the author derived
a number of exact equations for the Villain model on the N-line.
For example, the exact solution for the energy along the 
N-line was obtained.
As we have seen in \Ssc{PhaseDiagramDescreteSpins}
for the discrete models, one can argue, based on the
exact results, that the N-line ($T=T_p$)
is the cross-over line separating the
purely ferromagnetic region and the disorder dominant region.
Interestingly, the renormalization group theory on the $XY$ 
model\scite{NattermannSKL1995} indicates that there is a freezing transition
or a cross-over at $T = 2T_p$ with the difference of a factor $2$
from the N-line.
It is plausible that this difference is only due to the
approximation involved in the renormalization group theory,
and they both reflect the same physics.

\section{Quantum Spin Glasses}
\label{sec:QuantumSpinModels}

\begin{figure}
\centerline{\includegraphics[width=0.6\columnwidth]{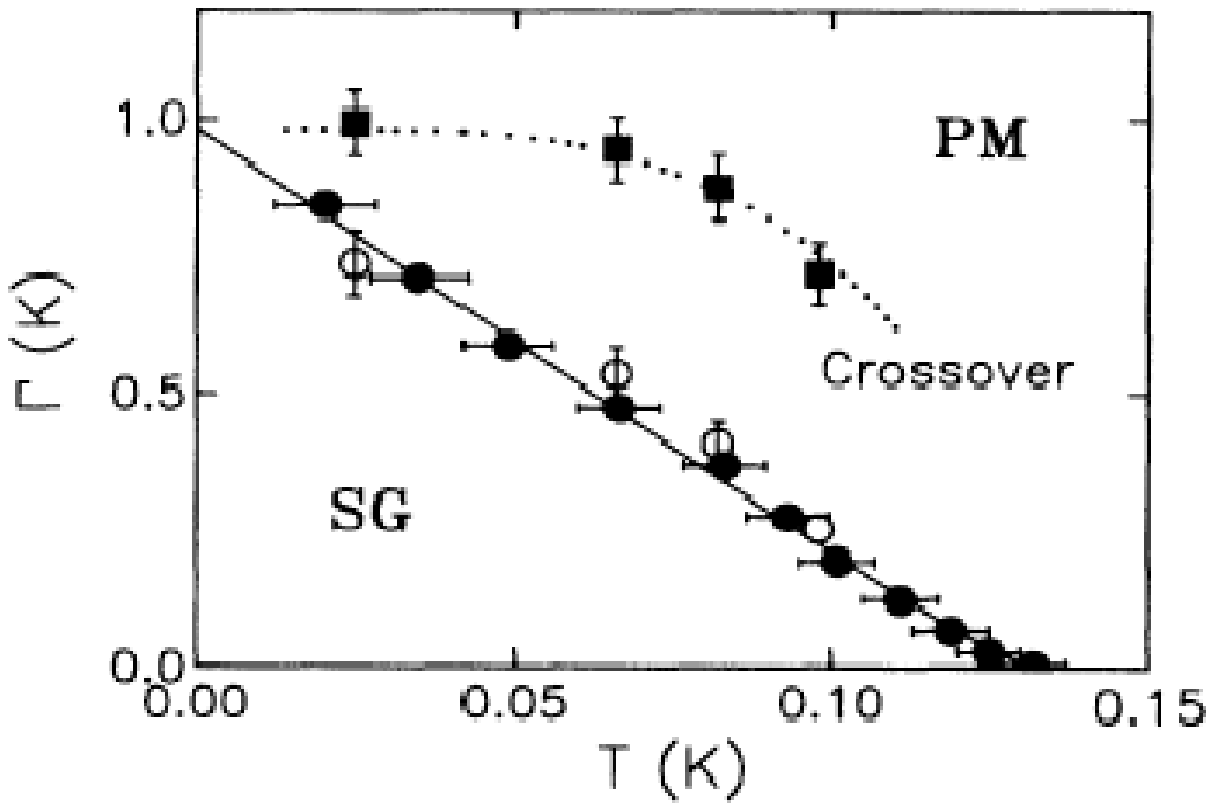}}
\caption{Phase diagram of LiHo$_{0.167}$Y$_{0.833}$F$_4$ 
according to the measurement of the nonlinear susceptibility.
From \figcite{qsg-exp1}.}
\label{qsg1}
\end{figure}

A quantum spin glass is a magnetic system that can be described by a
quantum mechanical Hamiltonian with spin-glass like features
(randomness and frustration).
In such a system, a spin glass phase may exist while
at the same time quantum fluctuations play an important role,
possibly a dominant role, in particular, in the absence of thermal 
fluctuations at zero temperature. 
Such a Hamiltonian is, for instance, the spin-1/2 Heisenberg spin glass
\begin{equation}
H=\sum_{(ij)} J_{ij} 
(\sigma_i^x\sigma_j^x+\sigma_i^y\sigma_j^y+\sigma_i^z\sigma_j^z)\;,
\label{q-heisen}
\end{equation}
where $\sigma^{x,y,z}$ are Pauli spin-1/2 operators, $J_{ij}$
random exchange interactions (e.g., Gaussian), 
and the sum runs over all nearest neighbors on some $d$-dimensional lattice. 
Another example is the Ising spin glass in a transverse field
\begin{equation} H=-\sum_{(ij)} J_{ij} \sigma_i^z\sigma_j^z
+\Gamma\sum\sigma_i^x\;,
\label{q-ising}
\end{equation} 
where $\Gamma$ denotes the transverse field strength. This
Hamiltonian becomes diagonal if $\Gamma$ is zero, in which case it
reduces simply to the classical Ising spin glass that we have discussed in
the previous sections. Thus the role of the parameter $\Gamma$ is to
tune the strength of quantum fluctuations, they do not play a role in
the equilibrium statistical physics of a diagonal Hamiltonian. An
important experimental realization of this model Hamiltonian is the
system LiHo$_x$Y$_{1-x}$F$_4$,\scite{qsg-exp1}
an insulating magnetic
material in which the magnetic ions (Ho) are in a doublet state due to
crystal field splitting. The interactions between Ho ions can be
described by an Ising model with dipolar couplings. For $x=1$
the system is a ferromagnet with a critical temperature of $T_c=1.53$ K
at $\Gamma=0$ and as $x$ is reduced the critical temperature
decreases. For concentrations below 25\% Ho and above 10\% Ho a thermal
phase transition to a spin glass phase occurs indicated by a diverging
nonlinear susceptibility (for instance at $x=0.167$ the spin glass
transition temperature is $T_g=0.13$K at $\Gamma=0$).  If a
transverse field is applied ($\Gamma>0$) the spin glass transition
temperature decreases monotonically to zero (see
Fig. \ref{qsg1}). This particular point, at zero temperature and at a
critical field strength is what we denote as a 
quantum-phase-transition point.\scite{qpt-sachdev}
%Remarkably with increasing transverse field the
%divergence of the nonlinear susceptibility was drastically suppressed
%indicating even the absence of a divergence at zero temperature.

Earlier reviews on quantum spin glasses and in particular the Ising
spin glass in a transverse field can be found in
\bcite{qsg-rev1,qsg-rev2,qsg-rev3}. Here we try to focus on a number of
new developments that have been made since then.

\subsection{Random Transverse Ising Models}
\label{rtim}

\begin{figure}
\includegraphics[width=0.44\columnwidth]{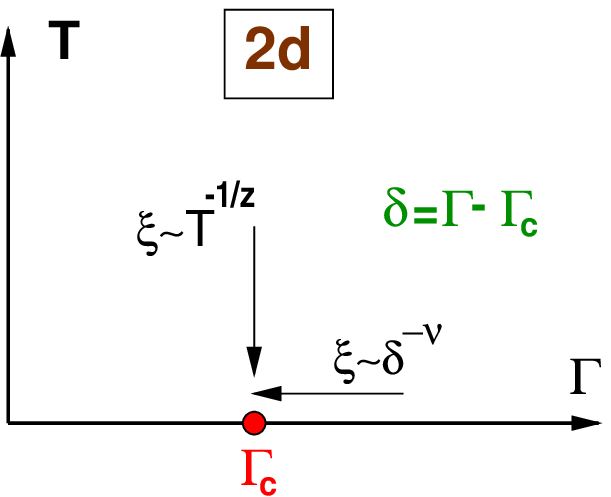}
\includegraphics[width=0.49\columnwidth]{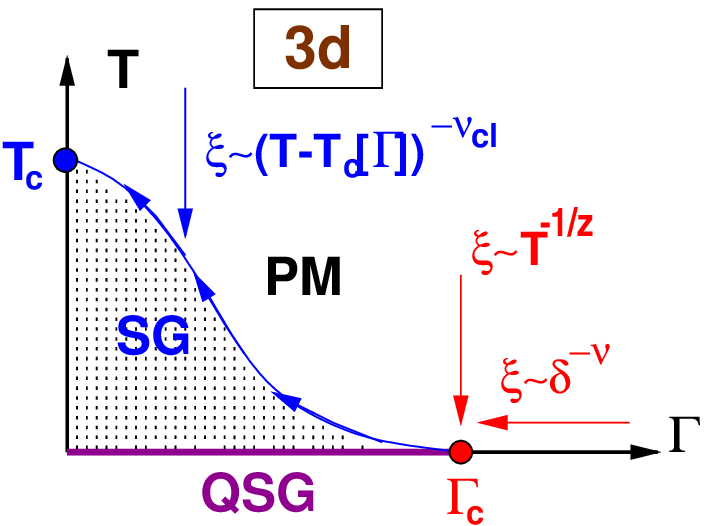}
\caption{{\bf Left:} Generic phase diagram for the two-dimensional
Ising spin glass in a transverse field $\Gamma$. Since no spin glass
phase is present in $d=2$ for $T>0$, only a quantum spin glass phase
and a quantum phase transition at $T=0$ exists. Approaching the quantum
critical point at $\Gamma_c$ by decreasing the temperature $T$, the
correlation length diverges like $T^{-1/z}$, where $z$ is the dynamical
critical exponent (if $z$ is formally infinite, it increases
logarithmically). {\bf Right:} Generic phase diagram of a
three-dimensional Ising spin glass in a transverse filed. The
classical transition temperature (at $\Gamma=0$) is $T_{\rm c}$ and the
corresponding classical correlation length exponent is $\nu_{cl}$.}
\label{qsg2}
\end{figure}

The generic phase diagram for the EA Ising spin glass model in a
transverse field $\Gamma$ is shown in Fig.\ \ref{qsg2} for two dimensions
% (in which case there is no spin glass phase at finite temperature $T>0$)
and for three dimensions. 
In the three-dimensional case, starting from the classical spin glass transition
temperature $T_c$ for $\Gamma=0$ the critical temperature decreases
monotonically with increasing transverse field strength $\Gamma$ until
it reaches $T=0$. One expects that the universality class of the
transition at any non-vanishing temperature is the same as the one
of the classical Ising spin glass transition at $T_c$. The zero-temperature
quantum phase transition, however, establishes a new universality
class. This transition exists in any dimension, 
including one and two dimensions.
A critical value $\Gamma_c$ for the transverse field strength
separates a disordered or paramagnetic phase for $\Gamma>\Gamma_c$
from an ordered phase for $\Gamma<\Gamma_c$. This transition is
characterized by a diverging length scale
$\xi\sim|\Gamma-\Gamma_c|^{-\nu}$ and a vanishing characteristic
frequency $\omega\sim\Delta E\sim\xi^{-z}$. The latter is the quantum
analog of ``critical slowing-down'' in the critical dynamics of
classical, thermally driven transitions. The new and most important
property occurring at zero temperature in the random transverse Ising
model is the {\it infinite randomness fixed point} (IRFP) that
governs the quantum critical behavior at the critical value $\Gamma_c$
of the transverse field.\scite{fisher-irfp}
One feature of the IRFP is
that the dynamical exponent $z$ is formally infinite, the relation
between length and energy scales is not algebraic but
exponential: $\Delta E\sim\exp(-A\xi^\psi)$.

To describe this scenario we generalize the discussion of the
transverse Ising spin glass by including also random {\it
ferromagnetic} interactions $J_{ij}>0$, because many more analytical
and numerical results are available for the ferromagnetic rather than
the spin glass case and the same main features are expected to hold in
both cases.

Let us start with a review of the one-dimensional case, in which the sign 
(if it can be negative) of the nearest neighbor couplings can be gauged away 
so that all interactions are ferromagnetic and the resulting
model is the random Ising chain in a transverse field or a random 
transverse-field Ising model (RTIM) in one dimension:
\begin{equation}
\Ham = -\sum_i J_i \sigma_i^z \sigma_{i+1}^z+\sum_i h_i \sigma_i^x\;.
\qquad (J_i > 0)
\label{hamilton}
\end{equation}
A uniform transverse field is represented by $h_i=\Gamma$ for all
sites. Since this case and the case of random transverse fields turn
out to belong to the same universality class, we also consider random
transverse field here.  The couplings $J_i$ and the transverse
fields $h_i$ are random variables with distributions $\pi(J)$ and
$\rho(h)$, respectively. The Hamiltonian in (\ref{hamilton}) is
closely related to the transfer matrix of a classical two-dimensional
layered Ising model, which was first introduced and studied by McCoy
and Wu.\scite{mccoywu}
Extensive researches on this model were initiated
by D.~Fisher\scite{fisher} with an application of the Ma-Dasgupta-Hu
renormalization group scheme,\scite{dasgupta-ma}
followed by numerical and analytical 
work.\scite{r1da,r1db,r1dc,r1dd,r1de,r1df,r1dg,r1dh}
We briefly summarize the results. The quantum control-parameter of the
model is given by
\begin{equation}
\delta={[\ln h]_{av}-[\ln J]_{av} \over \rm{var}[\ln h]+\rm{var}[\ln J]}\;.
\label{delta}
\end{equation}
For $\delta<0$ the system is in the ordered phase with a non-vanishing
average magnetization, whereas the region $\delta>0$ corresponds to
the disordered phase. There is a phase transition in the system at
$\delta=0$ with rather special properties, which differs in several
respects from the usual second-order phase transitions of pure
systems.  One of the most striking phenomena is that some physical
quantities are not self-averaging, which is due to very broad,
logarithmic probability distributions. 
As a consequence the {\it typical value} 
(which is the value in an frequent event)
and the {\it average value} of such quantities can be drastically different. 
Thus the critical behavior of the system is primarily determined by rare events
that give dominating contributions to the averaged 
values of various observables.

The average bulk
magnetization is characterized by an exponent $\beta$, 
which is $\beta=2-\tau$
where $\tau=(1+\sqrt{5})/2$ is the golden-mean. The average spin-spin
correlation function $C(r)=[\langle\sigma_i^z
\sigma_{i+r}^z\rangle]_{av}$ involves the average correlation length
$\xi$, which diverges at the critical point as $\xi \sim
|\delta|^{-\nu_{\rm av}}$, and $\nu_{\rm av}=2$.
On the other hand, the typical correlations have a faster decay, since
$\xi_{\rm typ}\sim |\delta| ^{-\nu_{\rm typ}}$ with $\nu_{\rm typ}=1$.

Close to the critical point the relaxation time $t_r$ is
related to the correlation length as $t_r \sim \xi^z$, where $z$ is
the dynamical exponent. The random transverse-field Ising spin chain
is strongly anisotropic at the critical point, since according to
the RG-picture\scite{fisher} and to numerical results\scite{youngrieger}
\begin{equation}
\ln t_r \sim \xi^{1/2}\;,
\label{scales}
\end{equation}
which corresponds to $z=\infty$. On the other hand the relaxation time
is related to the inverse of the energy-level spacing at the bottom of
the spectrum $t_r \sim (\Delta E)^{-1}$. Then, as a consequence of
(\ref{scales}), some quantities (such as specific heat, bulk
and surface susceptibilities, etc.) have an essential singularity at
the critical point, and the correlation function of the critical
energy-density has a stretched exponential decay, in contrast to the
usual power law behavior.

Away from the critical point in the disordered phase the rare events
with strong correlations still play an important role, up to the point
, $\delta = \delta_G$.
Above this point, all transverse-fields are
bigger than the interactions. In the region $0<\delta < \delta_G$,
which is called the Griffiths-McCoy phase, the magnetization is a
singular function of the uniform longitudinal field $H_z$ as $m_{\rm
sing} \sim |H_z|^{1/z}$, where the dynamical exponent $z$ varies with
$\delta$. At the two borders of the Griffiths-McCoy phase it behaves
as $z\approx 1/2\delta\times(1+{\cal O}(\delta))$\scite{fisher} 
as $\delta \searrow 0$ and $z=1$ as $\delta \nearrow \delta_G$,
respectively.

All these results could be obtained and understood by the application
of a Ma-Dasgupta-Hu renormalization group scheme,\scite{fisher}
in which strong bonds or fields are successively decimated either by
elimination of spins (in case of large transverse fields) or formation
of strongly coupled clusters (in case of large ferromagnetic bonds).
With decreasing energy scale $\Delta$ of the bonds and fields to be
decimated the typical size $L$ of these strongly coupled clusters
increases as
\begin{equation}
L \sim |\ln\Delta|^{1/\psi}
\end{equation}
defining an exponent $\psi$ that is $1/2$ in the random transverse-field
Ising chain. Such a cluster typically contains 
\begin{equation}
\mu\sim L^{\phi\psi} (=|\ln\Delta|^\phi)
\end{equation}
spins that essentially behave collectively (for instance in response
to the application of a longitudinal magnetic field $H$ --- and thus
generating a huge contribution to the spin susceptibility). This
defines another exponent $\phi$, which is $(1+\sqrt{5})/2$ in the
RTIM. Finally there is the correlation length exponent $\nu$ that
defines the characteristic length scale of spin-spin correlations away
from the critical point. 

The RG runs into a fixed point that is fully determined by the
geometrical features of the clusters that are generated asymptotically
--- very much in reminiscence of the percolation fixed point in
conventional percolation. This picture is expected to hold also for
higher-dimensional RTIMs, and even for the spin
glass case. Therefore we summarize its essence here. 
The distribution of the random bonds and fields not yet decimated during
the RG procedure becomes broader and broader.
Hence the name, {\it infinite randomness fixed point} (IRFP). 
It is characterized by the three exponents $\psi$, $\phi$ and
$\nu$ and the critical behavior of the physical observables is
determined by them. For instance the correlation function (at
criticality) for two spins at site $i$ and $j$ with a distance $r$
from each other is simply given by their probability to belong to the same
cluster of size $r$: $[C_{ij}]_{\rm av} \sim |{\bf r}_i-{\bf
r}_j|^{-2(d-\phi\psi)}$. Other relations follow straightforwardly from
this scheme\scite{fisher-irfp}:

\bc
\begin{tabular}{lccl}
lowest energy scale: & 
  $-\ln\Delta$ & $\sim $ & $L^\psi$ \\
magnetic moment:     & 
  $\mu$ & $ \sim $ & $(-\ln\Delta)^\phi$\\
average correlations: &
  $[C_{ij}]_{\rm av}$ &
  $ \sim $ & $|{\bf r}_i-{\bf r}_j|^{-2(d-\phi\psi)}$\\
typical correlations: &
  $-[\ln C_{ij}]_{\rm av} $ & $ \sim $ &
  $\kappa_{ij}|{\bf r}_i-{\bf r}_j|^\psi$\\
finite $T$-susceptibility: &
  $\chi$ & $\sim$ & 
  $T^{-1}(-\ln T)^{2\phi-d/\psi}$\\
finite $H$-magnetization: &
  $M$ & $\sim$ & 
  $(-\ln H)^{-\phi+d/\psi}$
\end{tabular}
\ec
Away from the critical point ($\delta\ne0$) the correlation length is
finite and its average and typical value scale differently:

\bc
\begin{tabular}{lccl}
average correlation length: & $\xi_{\rm av}$ & $\sim$ & 
$\delta^{-\nu}$\\
typical correlation length: & $\xi_{\rm typ}$ & $\sim$ &
$\xi_{\rm av}^{1-\psi}$\\
spontaneous magnetization: & $M_0$ & $\sim$ &
$(-\delta)^{\nu(d-\phi\psi)}$
\end{tabular}
\ec

In spite of the finiteness of the average correlation length away from
the critical point still arbitrarily large strongly-coupled clusters
exist --- though with an exponentially small probability --- leading
to algebraically decaying correlations in imaginary time.
Phenomenologically, one can see that as follows.\scite{thill-huse,rieger-qsg2}
Let $L$ be the size of a region of {\it strongly coupled} spins.
In a random system in the paramagnetic phase
they occur with an exponentially small probability $P(L)\propto
\exp(-\lambda L^d)$.
For instance in the diluted ferromagnet strongly
coupled regions are connected clusters and their probability is $p^V$,
where $V$ is the region's volume and 
$p$ is the site occupation probability ($0<p<1$).
Then, $\lambda$ is given by $\lambda=|\ln p|>0$. 
The special feature of transverse-field Ising systems is that in
first order perturbation theory the gap of a finite system 
containing $L^d$ spins is exponentially small: $\Delta_0\sim\exp(-s
L^d)$. An exponentially small gap means an exponentially large tunneling
time, and combining the two observations on cluster probability and
relaxation time one obtains an algebraical decay for the spin-spin
correlation function:
$C(\tau)=[\langle\sigma_i(\tau)\sigma_i(0)\rangle]_{\rm av}
\sim\tau^{-\lambda/s} =\tau^{-d/z(\delta)}$. The parameter
$z(\delta)=s/d\lambda$ is called the {\it dynamical exponent} in the
Griffiths phase and it varies continuously with the distance from the
critical point. The consequences, e.g., for the susceptibility are
dramatic: $\chi(\omega=0)=\int_0^{1/T} d\tau\,C(\tau)\propto
T^{-1+d/z(\delta)}$ which implies that for $z>d$ the susceptibility
diverges for $T\to0$ even away from the critical point. Since in
random transverse-field Ising system $z(\delta)$ grows without bounds for
$\delta\to0$ (and thus merging with the critical dynamical exponent at
$\delta=0$, which is infinite), there is always a region around the
critical point, where the susceptibility diverges.

In general the dynamical exponent $z(\delta)$ introduced above
is expected to determine all singularities occurring in the
Griffiths-McCoy phase close to an IRFP\scite{fisher-irfp}:

\bc
\begin{tabular}{lccl}
dynamical exponent:  & $z(\delta)$ & $\propto$ &
$\delta^{-\psi\nu}$\\
lowest energy scale: & $\Delta$ & $\sim $ & 
$L^{-z(\delta)}$\\
finite $H$-magnetization: & $M$ & $\sim$ & 
$H^{1/z(\delta)}$\\
susceptibility: & $\chi(\omega=0)$ & $\sim$ & 
$T^{-1+d/z(\delta)}$\\
nonlinear susceptibility: & $\chi_{nl}(\omega=0)$ & $\sim$ & 
$T^{-1+d/3z(\delta)}$\\
specific heat: & $c$ & $\sim$ & 
$T^{d/z(\delta)}$\\
\end{tabular}
\ec

\begin{figure}
\centerline{\includegraphics[width=0.6\columnwidth]{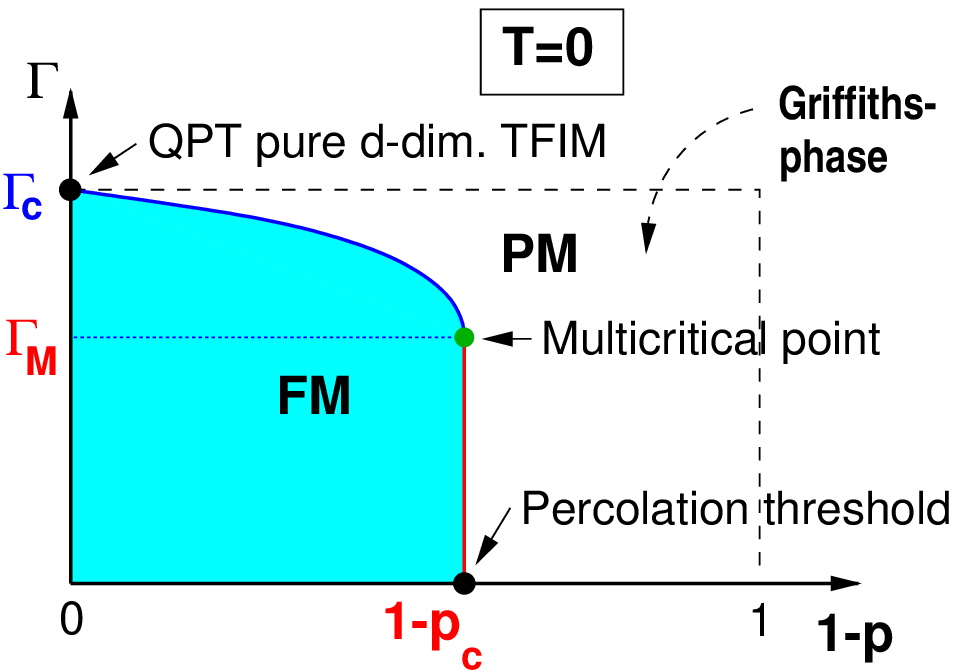}}
 \caption{Phase diagram of the diluted Ising ferromagnet in an
 transverse field $\Gamma$ at zero temperature $T=0$.}
\label{dilfm}
\end{figure}

The last three tables summarize the scaling predictions at and close
to a IRFP and in $1d$ they have been confirmed many times,
analytically and 
numerically.\scite{r1da,r1db,r1dc,r1dd,r1de,r1df,r1dg,r1dh}
In higher dimensions
$d\ge2$ the randomly diluted Ising-ferromagnet in a transverse field is 
a show-case for a quantum phase transition governed by an IRFP. The
site diluted model is defined by the Hamiltonian
\begin{equation}
H=-J\sum_{(ij)}\varepsilon_i\varepsilon_j \sigma_i^z\sigma_j^z
-\Gamma\sum_i\varepsilon_i\sigma_i^x
\end{equation}
and the bond diluted model by
\begin{equation}
H=-J\sum_{(ij)}\varepsilon_{ij}\sigma_i^z\sigma_j^z
-\Gamma\sum_i\sigma_i^x
\end{equation}
where $\varepsilon_i$ and $\varepsilon_{ij}$ are random variables that
take on the values $1$ with probability $p$ and $0$ with probability
$1-p$. Its phase diagram is depicted in Fig. \ref{dilfm}

Along the vertical line starting from the point $(p,\Gamma) = (p_c,0)$
up to the multi-critical point the transition from the paramagnetic to the
ferromagnetic phase is a {\it classical percolation transition}.\scite{senthil,miyashita}
Denoting the distance from the critical
point with $\delta=p_c-p$ the connectivity correlation length diverges
upon approaching the percolation point $1-p_c$ as
$\xi\sim|\delta|^{-\nu_{\rm perc}}$. The number of spins $M$ in the
percolating cluster at $p=p_c$ scales with the linear system-size $L$
as $M\sim L^{D_{perc}}$, where $D_{perc}$ is the fractal dimension
of the percolating clusters. For small values of the transverse field
$\Gamma$ one expects the percolating cluster to be fully magnetized,
which implies that the gap scales as $\Delta\sim \exp(-L^{D_{perc}})$.
This means that $\psi=D_{perc}$ in the IRFP scenario described above.
Moreover, the connectivity correlation
function at the percolation threshold $p_c$ decays as
$C(r)\sim r^{-(d-2+\eta_{\rm perc})}$, 
which means that the exponent $\phi$ is given
by the relation $2(d-\psi\phi)=(d-2+\eta_{\rm perc})$. 
To summarize the exponents characterizing the IRFP in the randomly 
diluted ferromagnet in a transverse field are related to the classical 
percolation exponents (which are exactly known in dimensions $d=2$ and in $d>6$)
via:

\begin{equation}
\nu=\nu_{\rm perc}\,,\quad
\psi=D_{\rm perc}\,,\quad
\phi=(d+2-\eta_{perc}))/D_{\rm perc}\,.
\end{equation}

%\begin{tabular}{lcl}
%$\nu$  &=& $\nu_{\rm perc}$\\
%$\psi$ &=& $D_{\rm perc}$\\
%$\phi$ &=& $(d+2-\eta_{perc}))/D_{\rm perc}$
%\end{tabular}

For the random bond ferromagnet in a transverse field in dimensions
$d\ge2$ one has to rely only on numerical calculations: In the $2d$
ferromagnetic case quantum Monte-Carlo 
simulations\scite{rieger-tim2d,rieger-tim2db} provided 
an evidence for an infinite
randomness fixed point with $\psi\approx0.42$ and $\phi\approx2.1$.
Later a numerical implementation of the Ma-Dasgupta-Hu RG scheme indeed
provided another evidence for an infinite randomness fixed 
point\scite{motrunich} with $\psi=0.42\pm0.06$, $\phi=2.5\pm0.4$, which
agrees with the QMC estimate within the error bars, and
$\nu=1.07\pm0.15$. For random Ising ferromagnets in a transverse field
the existence of the IRFP dominating the quantum critical behavior
thus appears to be confirmed for finite dimensions. Strictly speaking
detailed numerical studies have only be performed for $d=1$ and $d=2$
up to now, but there seems to be no strong argument against the
existence of the IRFP also in higher, finite, dimensions although
one expects the numerical and experimental visibility of the IRFP to
diminish for increasing dimension $d$. In the mean field limit
($d\to\infty$) the quantum phase transition is {\it not} described by
an IRFP and obeys conventional scaling laws.
In particular $z$ is
finite and Griffiths-McCoy singularities are absent.

What about the spin glass case? Quantum Monte Carlo simulations 
on the Ising spin glass with a transverse-field
have been performed for the cases $d=2$\scite{rieger-qsg,rieger-qsg2} and
$d=3$,\scite{bhatt-qsg,bhatt-qsg2} 
they are reviewed in \bcite{qsg-rev2,qsg-rev3}. 
The main result is that the numerical data
appeared to be compatible with a finite value for the dynamical
exponent in $d=2$ and $3$ and that the critical behavior can be
described by conventional scaling laws. However, the existence of a
Griffiths-McCoy phase away from the critical point has been uncovered,
with a continuously varying dynamical exponent describing the
singularities of the susceptibility and non-linear susceptibility.  In
contrast to the quantum Monte-Carlo simulations of the random bond
ferromagnets no cluster-algorithm could be used in the quantum spin
glass case, which restricted the system sizes and in particular the
temperatures to rather small values (note that anisotropic finite size
scaling demands that the temperature has to decrease exponentially
with the system size at a quantum critical point described by an
IRFP). In addition a homogeneous rather than a random transverse field
has been used, which causes strong cross-over effects and the true
asymptotic scaling behavior might be more difficult to extract.
Therefore it might very well be that the indications found for the
absence of a IRFP in the $2d$ and $3d$ quantum spin glass are still
pre-asymptotic and that studies using larger system sizes and more
sophisticated simulation methods could detect evidence for the IRFP
also here.

Finally a word about the consequences of the aforementioned
theoretical developments for the experiments. There it was observed
that upon approaching the quantum critical point the divergence of the
non-linear susceptibility was drastically suppressed indicating even
the absence of a divergence at zero temperature. The numerical
results, on the other hand, hint at a strong divergence of the
non-linear susceptibility at the quantum critical point --- even more than
the IRFP scenario. Up to now no clear reason for the discrepancy has
been pinned down. The possibility of a second-order transition turning
a first-order one at low temperatures has been 
raised,\scite{qsg-cugliandolo-grempel-santos}
but this possibility can
definitely be ruled out for a system that can be described by the
Hamiltonian (\ref{hamilton}) that we discussed here. We do not think
that dipolar interactions of a magnetically diluted system cause
substantial modifications of the picture that emerged for short range
interactions. At this point one cannot rule out the possibility that
the transverse field Ising Hamiltonian with quenched disorder is
simply not a sufficiently detailed description of
LiHo$_{0.167}$Y$_{0.833}$F$_4$.

\subsection{Mean-Field Theory}

As a mean-field model of quantum Ising spin glass, we consider
the Sherrington-Kirkpatrick model in a transverse field
\begin{equation}
H=-\sum_{(i,j)} J_{ij}\sigma_i^z\sigma_j^z-\Gamma\sum_i\sigma_i^x.
\label{qsg-sk}
\end{equation}
The first sum is over all pairs of spins and the couplings $J_{ij}$
are quenched random variables that obey the Gaussian distribution with
zero mean and variance $J^2/N$, where $N$ is the number of spins.
$\Gamma$ is the strength of the transverse field.  Although no exact
solution has been found for finite $\Gamma$, the phase diagram
of this model has been well delineated.
At zero transverse field the
transition is the well-known classical transition of the SK model at
$T_c(\Gamma=0)=J$. For sufficiently high temperature and/or
sufficiently large $\Gamma$, thermal and/or quantum fluctuations
destroy the spin glass order, yielding a paramagnet.\scite{qsg-fedorov}
For low $T$ and small $\Gamma$ one finds a SG
ordered phase, apparently with broken replica symmetry.\scite{qsg-goldschmidt}
Monte Carlo calculation, numerical spin
summation\scite{qsg-usadel} and perturbation 
expansion\scite{qsg-ishii} in $1/\Gamma$ have determined the phase boundary to
some precision. As in the classical model, the infinite range
interactions apparently wipe out the Griffiths singularities discussed
in the last subsection. The critical behavior along the line
$T_c(\Gamma)$ is expected to be the same as the classical critical
behavior, i.e., the non-linear susceptibility diverges as
$\chi_{nl}\sim (T-T_c(\Gamma))^{-\gamma}$ with $\gamma=1$,
the specific heat exponent is $\alpha=0$, etc.

The zero temperature quantum critical point $\Gamma_c(T=0)$ is in a
different universality class and has been studied in
\bcite{qsg-miller-huse,qsg-sachdev-read1,qsg-sachdev-read2}. The static
approximation --- the approximation usually applied to small field values
in which the imaginary time correlation function
$C(\tau)=\langle\sigma_i(\tau)\sigma_i(0)\rangle$ is assumed to be
time independent --- is not valid at $T=0$ (large fields) and the full
functional form of $C(\tau)$ has to be explored. 
The dynamical self-consistency equations obtained via 
standard manipulations\scite{qsg-fedorov,qsg-bray-moore} was
analyzed at $T=0$ at the quantum critical point in
\bcite{qsg-miller-huse,qsg-sachdev-read1,qsg-sachdev-read2},
and it turned out that the quantum critical point is
located at $\Gamma_c\approx0.7J$.
%, the precise estimate depending on the approximation scheme applied. 
At $\Gamma>\Gamma_c$ (and zero
temperature) $C(\tau)$ decays exponentially with $\tau$ as
$\tau\to\infty$, indicating a gap $\Delta$ in the corresponding
spectral density; at $\Gamma=\Gamma_c$, $C(\tau)$ decays as
$1/\tau^2$, and in the ordered phase, $C(\tau)\to q_{\rm EA}$. The Fourier
transform of $C(\tau)$ has the form $C(\omega)\sim{\rm
const.}-\sqrt{\omega^2-\Delta^2}$ for $\Gamma\ge\Gamma_c$, which is
responsible for the $1/\tau^2$ behavior at $\Gamma_c$ and it turned out
that the correlation time diverges as
$\xi_\tau\sim1/\Delta\sim[(\Gamma-\Gamma_c)^{-1}\ln(\Gamma-\Gamma_c)]^{1/2}$.
Thus we can define an exponent $z\nu$, anticipating anisotropic
scaling in space and time in the short range model, which takes the
value $z\nu=1/2$ in the infinite-range model. Since
$C(\tau\to\infty)=q_{\rm EA}$ is the Edwards-Anderson order parameter,
we may also define $q_{\rm EA}=(\Gamma_c-\Gamma)^\beta$ and it was
found that $\beta=1$. At $\Gamma=\Gamma_c$ one expects
$C(\tau)\sim\tau^{-\beta/z\nu}$, which is satisfied with the values
obtained. The non-linear susceptibility diverges as $1/\Delta$, which
implies with $\chi_{nl}\sim(\Gamma-\Gamma_c)^{-\gamma}$ that
$\gamma=1/2$. Studying Gaussian fluctuations around the saddle-point
solution valid for infinite range one finds\scite{qsg-sachdev-read2}
for the correlation length exponent above the upper critical
dimension (i.e.\ $d\ge8$) that $\nu=1/4$ and therefore $z=2$.
Moreover $\eta=0$ in mean field theory. The complete collection of
critical exponents obtained so far in comparison with the classical
model ($T>0$, where we assume to cross the phase boundary under a
non-vanishing angle) are as follows:
\begin{equation}
\begin{array}{l||c|c|c|c}
                        & \quad\beta\quad  & \quad\gamma\quad &
  \quad\nu\quad  & \quad z\quad \\
\hline
{\rm quantum}\;   (T=0) &   1/2  &    1/2   &  1/4 & 2 \\
\hline
{\rm classical}\; (T=0) &   1    &    1   &  1/2 & -
\end{array}
\end{equation}
Note that as a consequence of the absence of Griffiths-singularities
in mean-field models the dynamical exponent $z$ is finite in contrast
to the IRFP scenario that is supposedly valid for the finite-dimensional models.
In a longitudinal field one obtains, in analogy to the classical case,
an AT manifold in the $T,\Gamma,h$ phase diagram below which replica
symmetry is broken and the system is in the SG phase.

The dynamics of the model (\ref{qsg-sk}) in the paramagnetic phase has
been studied in \bcite{qsg-rozenberg-grempel}, where the dynamical
single-site self-consistency equations have been iteratively solved
using a quantum Monte Carlo scheme developed in
\bcite{qsg-grempel-rozenberg}. They mapped the spin-glass transition
line in the $\Gamma$-$T$ plane using the stability criterion
$1=J\chi_{\rm loc}$, where $\chi_{\rm loc}=\int_0^\beta d\tau\,C(\tau)$ is the
local susceptibility. They found a second-order transition line ending
at a quantum critical point at $T=0$ in agreement with the argument
presented above. 
Going down in temperature to $T\sim0.01J$ and extrapolating the
results to $T=0$ they determined a precise value for the critical
field $\Gamma_c=0.76\pm0.01$, which lies between previous 
estimates.\scite{qsg-goldschmidt,qsg-miller-huse}
The asymptotic form of
$C(\tau)\sim\tau^{-2}$ found in \bcite{qsg-miller-huse} was also
confirmed. A comparison of the results for the low-frequency
susceptibility with the experimental curves obtained for 
LiHo$_{0.167}$Y$_{0.833}$F$_4$ in \bcite{qsg-exp1} yields 
a good agreement.

A different class of mean-field spin-glass models has been studied in
\bcite{qsg-cugliandolo-grempel-santos} --- simplified in so far as
spherical spins rather than Ising spins were considered and
more general in so far as $p$-spin interactions were considered. 
The quantum fluctuations are introduced via a kinetic energy rather 
than the transverse field.
The corresponding quantum spherical $p$-spin-glass
Hamiltonian is defined by
\begin{equation}
H=\frac{1}{2M}\sum_{i=1}^N \hat{p}_i^2 -
\sum_{i_1,\ldots,i_p} J_{i_1,\ldots,i_p}s_{i_1}\cdots s_{i_p}
\label{eq:pspin}
\end{equation}
where $s_i$ are ``soft-spins'' fulfilling the spherical constraint
$\sum_{i=1}^N s_i(t)^2=N$ for all times $t$. 
Quantum mechanics is introduced into the classical $p$-spin
glass via the canonical momenta $\hat{p}_i$ that fulfill the commutation
relation $[\hat{p}_i,s_j]=-i\hbar\delta_{ij}$. The multi-spin 
coupling constants are taken from a Gaussian distribution with zero mean and
variance $\tilde{J}p!/(2N^{p-1})$ with $\tilde{J}$ being a constant
of ${\cal O}(1)$.

Before we discuss this model we want to clarify the connection to the
SK model in a transverse field discussed above. The replacement of
Ising spins $S_i=\pm1$ by continuous spins $s_i\in[-\infty,+\infty]$
is often performed in the theory of critical phenomena --- the
discrete symmetry is then enforced by a quartic term $\sum_i s_i^4$ in
the Hamiltonian (this is just the usual $\Phi^4$ Ginzburg-Landau
theory for critical phenomena with a discrete symmetry), which also
restricts automatically the spin length. Analytically the
quartic term causes extra complications in all computations, saddle
point evaluations, RG calculations, dynamical formalism etc. --- for
which reason one often skips it and replaces it by a spherical
constraint (either strictly or via a Lagrangian parameter having the
same effect as a chemical potential). Unfortunately the classical 
spherical mean-field spin-glass model with the usual 2-spin
interactions does not have a non-trivial spin glass phase.
Therefore, generalizations to $p$-spin interactions are 
sometimes considered.\scite{cuku}
At this point a clear connection to the original magnetic
system of interest is already lost.
Nevertheless, one might expect
that one can learn something about possible scenarios. 

Finally spherical spins cannot be quantized in the same way as Ising
spins via the application of a transverse field.
Therefore they are usually quantized via the introduction of a kinetic 
energy term as in \Eq{pspin}. 
In addition, various analytical techniques available for
interacting soft spins with kinetic energy, such as the Schwinger-Keldysh
formalism,\scite{qsg-cugliandolo2} 
are not available for spin operators. 
The microscopic details of the quantum dynamics described
by either a transverse field or a kinetic energy term might be very
different, on large timescales, however, one expects a similar
behavior for the following reason. 
To see this, let us consider a model that consists of two terms;
an arbitrary classical Hamiltonian, $H_{\rm cl}$, 
that is diagonal in the $z$-representation of the spins, 
and the transverse-field term.
Performing a Trotter decomposition of the partition function of 
this model, one obtains
\begin{eqnarray}
&& {\rm Tr} e^{-\beta(\Gamma\sigma^x+H_{\rm cl}(\sigma^z))}
=\lim_{\Delta\tau\to0}\prod_{\tau=1}^{L_\tau}
\left\langle S_\tau \left| e^{-\Delta\tau[\Gamma\sigma^x+H_{\rm cl}(\sigma^z)]}
\right|S_{\tau+1}\right\rangle\nonumber\\
&&\propto\lim_{\Delta\tau\to0}\sum_{S_1,\ldots,S_{L_\tau}}
\exp\left(-\Delta\tau\Bigl[\sum_{\tau=1}^{L_\tau} 
K(S_\tau-S_{\tau+1})^2+H_{\rm cl}(S_\tau)\Bigr]\right)
\label{kinetic1}
\end{eqnarray}
where $L_\tau$ is the number of Trotter slices in the imaginary time
direction, $\Delta\tau=\beta/L_\tau$ and $K$ given by
$e^{-2K}=\tanh(\Delta\Gamma)$.
For $\Delta\tau \ll1$ it is
$K=|\ln(\Delta\tau\Gamma)|/2$. In the last step we neglected a constant
factor $\cosh(\Delta\tau\Gamma)^{L_\tau}$.  
If we choose $\Delta\tau$ as a small time cut-off 
(representing the typical spin flip time) 
we can approximate the last Trotter sum as the
imaginary time path integral
\begin{equation} 
Z\approx \int{\cal D}S(\tau)\,\exp\left(
\int_0^\beta d\tau\,
\left[
\frac{M}{2}\left(\frac{\partial S}{\partial\tau}\right)^2 +H_{\rm cl}(S(\tau))
\right]
\right)
\label{kinetic2}
\end{equation} 
where $M=2K\Delta\tau=\Delta\tau|\ln(\Gamma\Delta\tau)|$. 
The first term in the integral of the action is identical to 
what one would obtain for the kinetic energy if one writes down the
imaginary time path integral for the partition sum of the Hamiltonian
\Eq{pspin}. 
In this way, the transverse-field term and the kinetic-energy term 
are related.

\begin{figure}
\includegraphics[width=0.5\columnwidth]{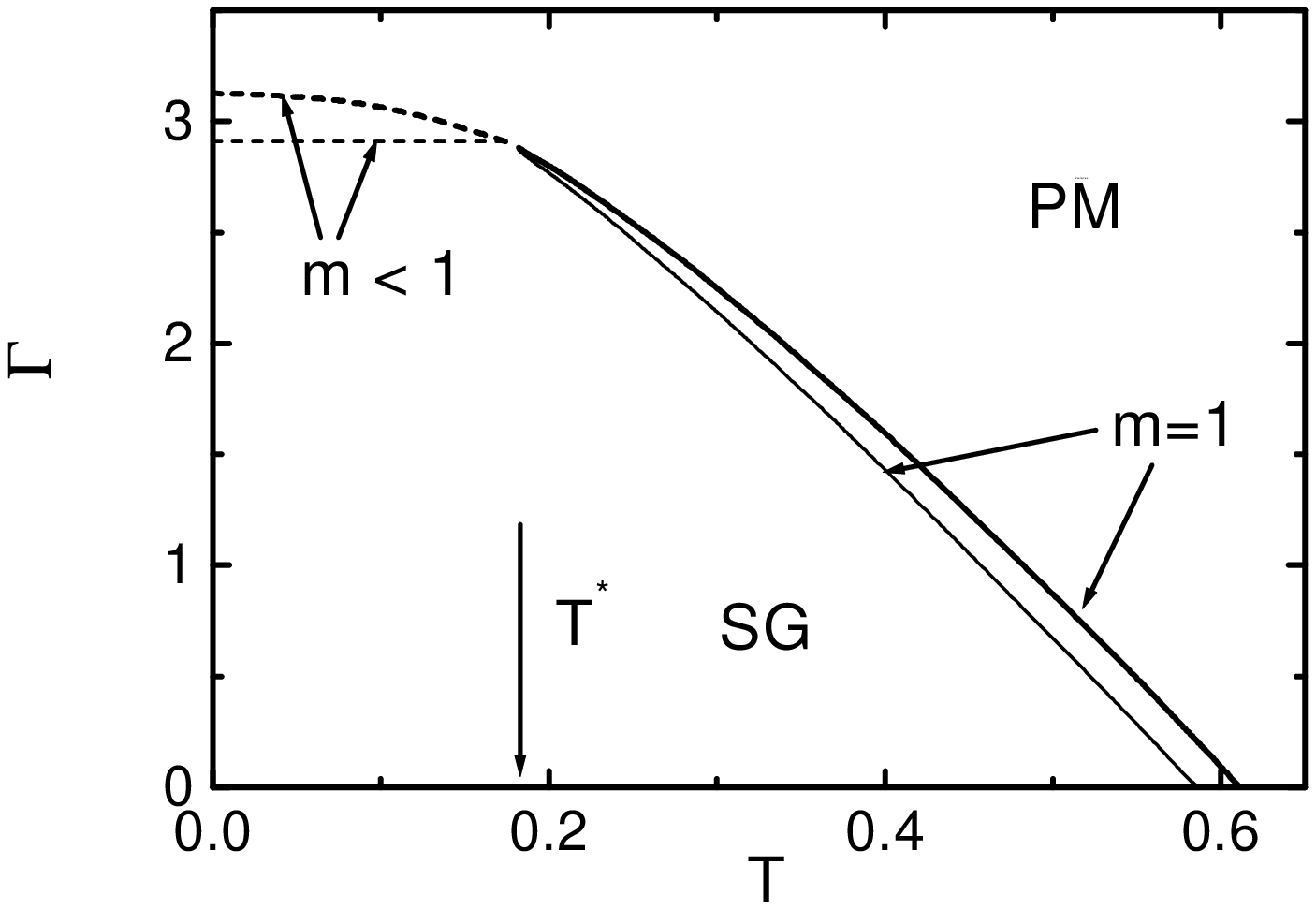}
\hspace{0.05\columnwidth}
\includegraphics[width=0.43\columnwidth]{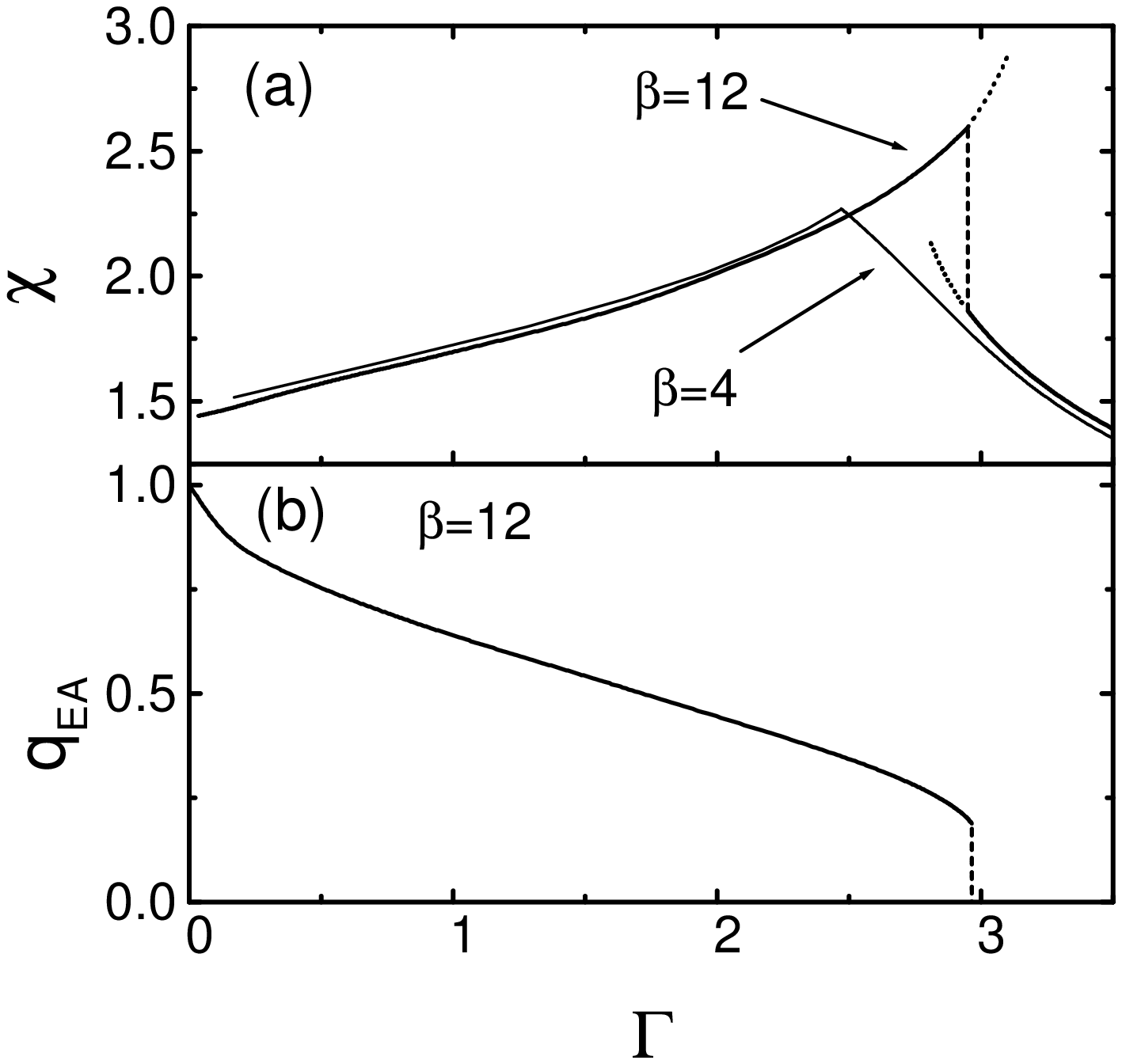}
 \caption{{\bf Left:} Static (thin lines) and dynamic (thick lines)
phase diagrams of the $p$-spin model for $p=3$. Solid and dashed lines
represent second and first order transitions, respectively.
{\bf Right:} Magnetic susceptibility (a) and Edwards-Anderson order
parameter (b) of the $p$=3 model. (From
\figcite{qsg-cugliandolo-grempel-santos}).
\label{fig-qsg-pspin}}
\end{figure}

In \bcite{qsg-cugliandolo-grempel-santos} the equilibrium properties of
the model were obtained using a replicated imaginary-time path
integral formalism\scite{qsg-bray-moore} and analyzing the dynamical
self-consistency equations for the spin auto-correlation function
$C(\tau)$ arising in the limit $N\to\infty$ from a saddle point
integration. The result for the phase diagram, EA order-parameter and
linear susceptibility in the case $p=3$ are depicted in Fig.\
\ref{fig-qsg-pspin}, where the parameter $\Gamma=\hbar^2/(JM)$ has
been used --- resembling the transverse field strength (since for
$\Gamma\to0$ one recovers the classical case).
Above a temperature $T^*$ one has a continuous
transition at a critical point $\Gamma=\Gamma_c(T)$ from a paramagnetic
phase with vanishing EA order parameter to a spin glass phase with
$q_{EA}\ne0$ and one-step replica-symmetry-breaking (1RSB). Although
the EA order-parameter jumps discontinuously the transition is second
order: there is no latent heat (as in the classical case $\Gamma=0$)
and the susceptibility has only a cusp. This is due to the fact that
the parameter $m$ characterizing the Parisi order parameter function
$q(x)$ (which is a step function with a single step at $x=m$) is unity
at the transition. However, for temperatures below $T^*$ this
parameter jumps at the transition, too, and the transition becomes
discontinuous; for $T<T^*$ the transition is of the first order with latent
heat and a discontinuous susceptibility (see Fig.\ \ref{fig-qsg-pspin}).

\subsection{Mean-Field Theory --- Dissipative Effects}

An important question that arises for interacting quantum spins at low
temperatures are the effects of a dissipative 
environment.\scite{tls-leggett,tls-weiss}
This is usually described in terms of
its collective excitations, lattice vibrations, spin or charge
fluctuations, etc., which may be thought of as an ensemble of
independent quantum harmonic 
oscillators.\scite{tls-caldeira,tls-bray-moore,tls-chakravarty,tls-feynman,tls-prokofev}
A concrete example of a single quantum degree of freedom, a spin-1/2
or a so-called two-level-system (TLS), coupled to a bath of bosons is
the well-known spin-boson-model:\scite{tls-leggett,tls-weiss}
\begin{equation}
H=H_S+H_B+H_{SB}
\end{equation}
where $H_S$, $H_B$ and $H_{SB}$ denote the Hamiltonian of the system,
the bath and their coupling, respectively. These are given by
\begin{equation}
\begin{array}{lcl}
H_S & = & -\Gamma\sigma^x\\
H_B & = & {\displaystyle \frac{1}{2}\sum_n (p_n^2/m_n + m_n\omega_n^2x_n^2)} \\
H_{SB} & = & {\displaystyle -\sum_n c_n x_n \sigma^z}
\end{array}
\label{spin-boson}
\end{equation}
where $\Gamma$ is the transverse field (or tunneling matrix element in the
context of TLSs), $n$ the index enumerating an infinite number of
harmonic oscillators with coordinates and momenta, $x_n$ and $p_n$, and
mass and frequency, $m_n$ and $\omega_n$, respectively. 
The constant $c_n$ is the
coupling between oscillator $n$ and the spin. 
The spectral density of the environment, 
$I(\omega)=\pi\sum_n (|c_n|^2/(m_n\omega_n)\delta(\omega-\omega_n))$,
is commonly assumed to take the standard form\scite{tls-weiss}
\begin{equation}
I(\omega)=2\alpha\hbar(\omega/\omega_{\rm ph})^{s-1} \omega e^{-\omega/\omega_c}\;,
\end{equation}
where $\alpha$ is a dimensionless coupling constant, $\omega_c$ a
high frequency cut-off (which can be set to $\omega_c=\infty$ if
$0<s<2$), and $\omega_{\rm ph}$ a phonon frequency necessary in the
non-ohmic ($s\ne1$) case to keeps $\alpha$ dimensionless.

With standard techniques\scite{tls-grabert,tls-leggett} one can
integrate out the oscillator degrees of freedom to express the
partition function of the system solely in terms of the spin variables
\begin{equation}
Z={\rm Tr}\,e^{-\beta H} = \int {\cal D}\sigma(\tau)\,{\cal T}\exp(-S/\hbar)\;,
\end{equation}
where $\int {\cal D}\sigma(\tau)$ denotes a path integral over all
spin configurations (in time), 
${\cal T}$ is the imaginary time ordering operator and the action is
\begin{equation}
S=-\int_0^{\beta\hbar}d\tau\,\Gamma\sigma^x(\tau)
-\frac{1}{2}\int_0^{\beta\hbar}d\tau\,\int_0^{\beta\hbar}d\tau'\,
K(\tau-\tau')\sigma^z(\tau)\sigma^z(\tau')\;.
\label{bathaction}
\end{equation}
The kernel $K(\tau)$ is related to the spectral density $I(\omega)$
and is for the ohmic case ($s=1$) essentially an inverse square
$K(\tau-\tau')\propto\alpha/(\tau-\tau')^2$. The effect of the
dissipative environment is therefore a long range interaction of the
quantum spin in imaginary time. In analogy to the Ising model with
inverse square interactions\scite{yuval} depending on the strength of
the coupling constant $\alpha$, the system is
ferromagnetically ordered or paramagnetic in the imaginary time
direction; for large $\alpha$ the spin is frozen and for small
$\alpha$ the spin will tunnel.

Indeed, for the ohmic case, at zero temperature, there is a phase
transition at $\alpha=1$.\scite{tls-bray-moore,tls-chakravarty}
For $\alpha<1$ there is tunneling and two distinct regimes develop. If
$\alpha<1/2$, the system relaxes with damped coherent oscillations; in
the intermediate region $1/2<\alpha<1$ the system relaxes
incoherently. For $\alpha>1$ quantum tunneling is suppressed and
$\langle\sigma^z\rangle\ne0$, signaling that the system remains
localized in the state in which it was prepared. These results also
hold for sub-Ohmic baths while weakly damped oscillations persist for
super-Ohmic baths.\scite{tls-leggett}
At finite temperatures (but low
enough such that thermal activation can be neglected) there is no
localization but the probability of finding the system in the state it
was prepared decreases slowly with time for $\alpha>\alpha_{\rm c}$.

These conclusions hold for a single spin interacting with a bath. The
question then arises as to which are the effects of the interplay
between the spin-spin interactions and the spin-noise coupling
in the physics of the interacting system. 
In \bcite{qsg-cugliandolo2} the effect of a dissipative bath on a
mean-field spin glass model with $p$-spin interactions has been
investigated. They studied the dissipative spin-boson system
(\ref{spin-boson}) for $N$ interacting spins
$H=H_S+H_B+H_{SB}$, where the bath Hamiltonian is the same,
the coupling Hamiltonian gets an additional sum over the spin index
$i$ and $H_S$ is now the $p$-spin Hamiltonian with transverse field
\begin{equation}
H_S=-\Gamma\sum_{i=1}^N \sigma_i^x
-\sum_{i_1,\ldots,i_p} J_{i_1,\ldots,i_p}\sigma_{i_1}^z\cdots\sigma_{i_p}^z.
\end{equation}
The second term, namely, the multi-spin interaction term is 
the same as the one in \Eq{pspin}. For the reason
explained in the last section it is analytically easier to study
spherical spins instead of quantum spin-1/2 degrees of freedom and
the quantization of the spherical spins is done via the introduction of a
kinetic energy term. 
The partition function then reads
\begin{equation}
Z=\int {\cal D}\sigma(\tau)\,\exp(-S/\hbar)\;,
\end{equation}
with the action 
\begin{eqnarray}
S=&&
\int_0^{\hbar\beta}d\tau\,
\biggl[\frac{M}{2}\sum_i
\biggl(\frac{\partial s_i(\tau)}{\partial\tau}\biggr)^2
-\sum_{i_1<\ldots<i_p} 
J_{i_1,\ldots,i_p}s_{i_1}(\tau)\cdots s_{i_p}(\tau)\nonumber\\
&&\;+\;z\sum_i[s_i^2(\tau)-1]\biggr]
-\int_0^{\hbar\beta}d\tau\,\int_0^{\hbar\beta}d\tau'\,
K(\tau-\tau')s_i(\tau)s_i(\tau')\;,
\label{diss-action}
\end{eqnarray}
where the first term is the kinetic-energy term already motivated in
(\ref{kinetic1}-\ref{kinetic2}) replacing the transverse-field term,
the second is the $p$-spin interaction term, the third a term with the
Lagrangian multiplier $z$ enforcing the spherical constraint and the
last term is the long range interaction imaginary time
(\ref{bathaction}) that is generated by the integration over the bath
variables.

Starting from (\ref{diss-action}) the saddle point equations for the
self-consistent single-spin dynamics were 
derived\scite{qsg-cugliandolo2} and the phase diagram computed. 
Analogous to
the non-dissipative case discussed in the previous subsection a critical
line with a second-order section (close to the classical critical
point $(T_d, \Gamma=0)$) and a first-order section (close to the
quantum critical point $(T=0, \Gamma_d)$) was obtained in the presence
of a dissipative environment. The second order critical line is
determined by the condition $m=1$, the first order critical line is
defined as the locus of the points where a marginally stable solution
first appears with decreasing $\Gamma$ for $T$ fixed.  For each
$\Gamma$ and $\alpha$ this defines a {\it dynamic} transition
temperature $T_d(\Gamma,\alpha)$.  The qualitative features of the
phase diagram, similar to those found for the isolated system, see the
discussion in the previous section.  Notice that the line
$T_d(\Gamma,\alpha)$ lies always {\it above} $T_s(\Gamma,\alpha)$, the
static critical line that we shall discuss below.

\begin{figure}
\includegraphics[width=0.49\columnwidth]{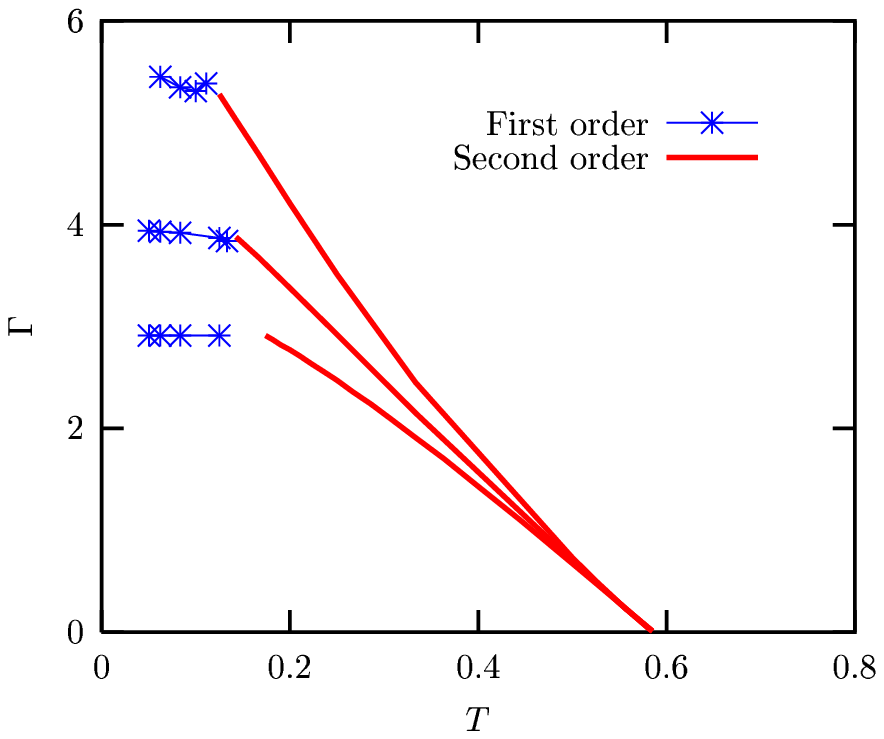}
\includegraphics[width=0.49\columnwidth]{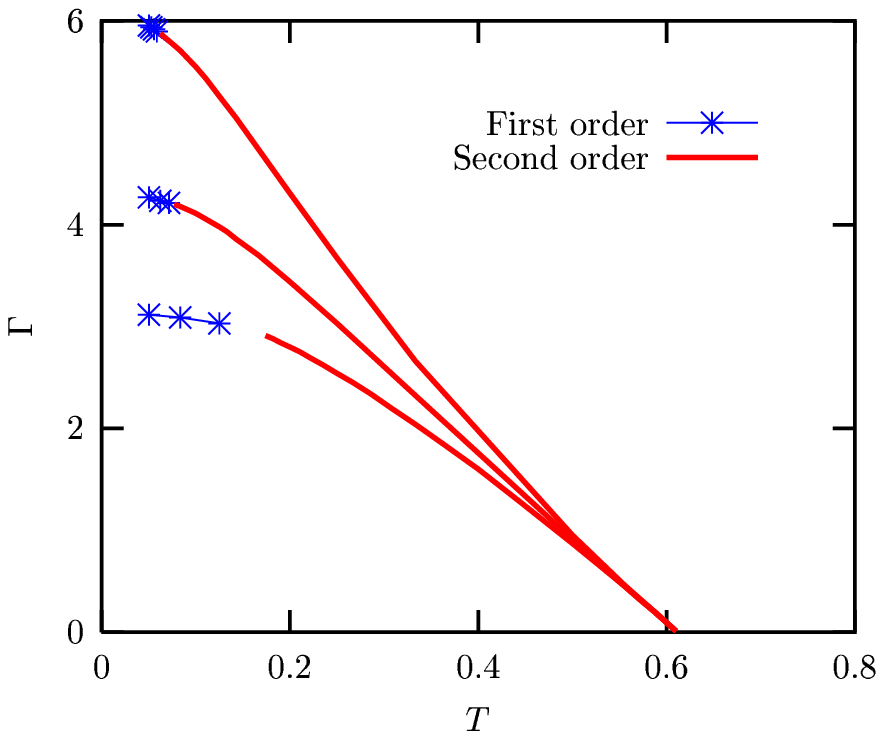}
 \caption{Static (left) and dynamic (right) phase diagrams for the
$p=3$ spin model coupled to an Ohmic bath ($s=1$). The couplings to
the bath are $\alpha =$ 0, 0.25, and 0.5 from bottom to top. The
solid line and line-points represent second and first order
transitions, respectively. (From \figcite{qsg-cugliandolo2}).
\label{fig-qsg-diss}}
\end{figure}

On the right side of Fig.~\ref{fig-qsg-diss} the dynamic phase diagrams
obtained for $p=3$ and three values of the coupling to an Ohmic bath,
$\alpha=0,0.25,0.5$ is shown.  The full line and the line-points
represent second and first order transition, respectively.

The first observation that can be made is that in the limit $\Gamma \to 0$
the transition temperature is independent of the strength of the
coupling to the bath.  This is a consequence of the fact that in the
limit $\Gamma \to 0$ the partition function is essentially determined
by the zero-frequency components of the pseudo-spin which are
decoupled from the bath. This result is however non-trivial from a
dynamical point of view, since it implies that the dynamic transition
of a classical system coupled to a colored classical bath is not
modified by the latter.

The second observation is that
the size of the region in the phase space where the system is in the
ordered state increases with $\alpha$. Coupling to the dissipative
environment thus stabilizes this state. This follows from simple
physical considerations.  The interaction term in the action favors
spin-glass order.  Coupling to the bath favors localization and its
effect is to reduce the effective tunneling frequency. Therefore, in
the presence of the bath, the value of the bare tunneling frequency
needed to destroy the ordered state must increase with $\alpha$.  Even
if the localized state and the glassy state may seem superficially
similar, they are indeed very different.  In the former, the
correlation function $C(t+t_w,t_w)$ approaches a plateau as a function
of $t$ and never decays toward zero while in the latter the
relaxation first approaches a plateau but it eventually leaves it to
reach zero for $t\gg t_w$. The fact that the coupling to the
environment favors the ordered state also reflects itself in the value
taken by the order parameters $C(\tau)$ and $q_{EA}$. As $\alpha$
increases, ${q_d}(\tau)$ reaches a higher plateau level at long
imaginary times.

\subsection{Mean-Field Theory --- Dynamics}

The out-of-equilibrium dynamics in real time of 
the quantum spherical $p$-spin glass model coupled to a dissipative
environment, which was discussed in the last
subsection, was actually studied earlier\scite{qsg-age} 
than the equilibrium properties. The response and correlation function
are defined in analogy to the classical case;
$C(t+t_w,t_w)=
N^{-1}\sum_i (s_i(t+t_w) s_i(t_w) + s_i(t_w) s_i(t+t_w))$
(note that the time evolution is now governed by the quantum dynamics
and $C$ has to be symmetrized in the operators $s_i(t+t_w)$ and 
$s_i(t_w)$) 
and $R(t+t_w,t_w)=N^{-1}\sum_i \delta s_i(t+t_w)/\delta h_i(t_w)$. 

In equilibrium the quantum FDT relates $R(t)$ and $C(t)$:
\begin{equation}
R(t)=\frac{2i}{\hbar} \theta(t)\int \frac{d\omega}{2\pi}
e^{-i\omega t} \tanh(\beta\hbar\omega/2)C(\omega)
\end{equation}
Away from the critical line, $C$ and $R$ decay to zero very fast with
oscillations. Approaching the critical line $T_d(\alpha)$, the decay slows
down and if $T_d>0$ a plateau develops in $C$. At the critical line
the length of the plateau tends to infinity. 

In the glassy phase (below the transition) the system does not reach
equilibrium. For small time differences the dynamics is stationary and
time translational invariance as well as the QFDT holds:\ 
$\lim_{t_w\to\infty}C(t+t_w,t_w)=q+C_{\rm eq}(t)$.  For large times
the dynamics is non-stationary, time translational invariance nor the
QFDT does not hold, and the correlations decay from $q$ to $0$. The
decay of $C$ becomes monotonic in the aging regime, which implies 
$C_{\rm aging}(t+t_w,t_w)=c(h(t_w)/h(t+t_w))$ (see subsection \ref{twotimes}).
One can generalize the QFDT in the same spirit as the classical FDT
was generalized (see subsection \ref{sec-fdt}):
\begin{equation}
R(t+t_w,t_w)=\frac{2i}{\hbar} \theta(t)\int \frac{d\omega}{2\pi}
e^{-i\omega t} \tanh(X(t+t_w,t_w)\beta\hbar\omega/2)C(t+t_w,\omega)
\label{xfdt}
\end{equation}
with $C(t,\omega)=2{\rm Re}\int_0^{t} ds\, \exp[i\omega
(t-s)]C(t,s)$. Again, as in the classical case (see subsection
\ref{sec-fdt}), $T_{\rm eff} \equiv T/X(t+t_w,t_w)$ acts as an
effective temperature in the system. For a model with two 
time-sectors it is proposed 
\begin{eqnarray}
X(t+t_w,t_w) = 
\left\{
\begin{array}{ll}
X_{\rm st} =1
\;\;\; &{\mbox{if}} \;\;\;
t \leq {\cal T}(t_w) 
\nonumber\\
X_{\rm age}(\hbar,T)
\;\;\; &{\mbox{if}} \;\;\; 
t > {\cal T}(t_w)
\end{array}
\right.
\; .
\end{eqnarray}
with $X_{\rm age}$ a non-trivial function of
$\hbar$ and $T$ and ${\cal T}(t_w)$ is a certain time-scale that separates the
stationary and aging time-regimes.
When $t$ and $t_w$ are widely separated, the integration over $\omega$ in 
(\ref{xfdt}) is dominated by $\omega\sim 0$. 
Therefore, the factor 
$\tanh(X_{\rm age}(t+t_w,t_w) \beta\hbar\omega/2)$ can be substituted by 
$X_{\rm age}\beta\hbar\omega/2$ 
({\it even at} $T=0$ if $X_{\rm age}(\hbar,T) =x(\hbar) T$ when $T\sim 0$). 
Hence, 
\begin{equation}
R_{\rm age}(t+t_w,t_w) \sim \theta(t) X_{\rm age} \beta
\partial_{t_w} C_{\rm age}(t+t_w,t_w)
\label{Xclassical}
\end{equation}
and one recovers, in the aging regime, the {\it classical} modified 
FDT.\scite{cuku,qsg-giam}

\begin{figure}
\includegraphics[width=0.49\columnwidth]{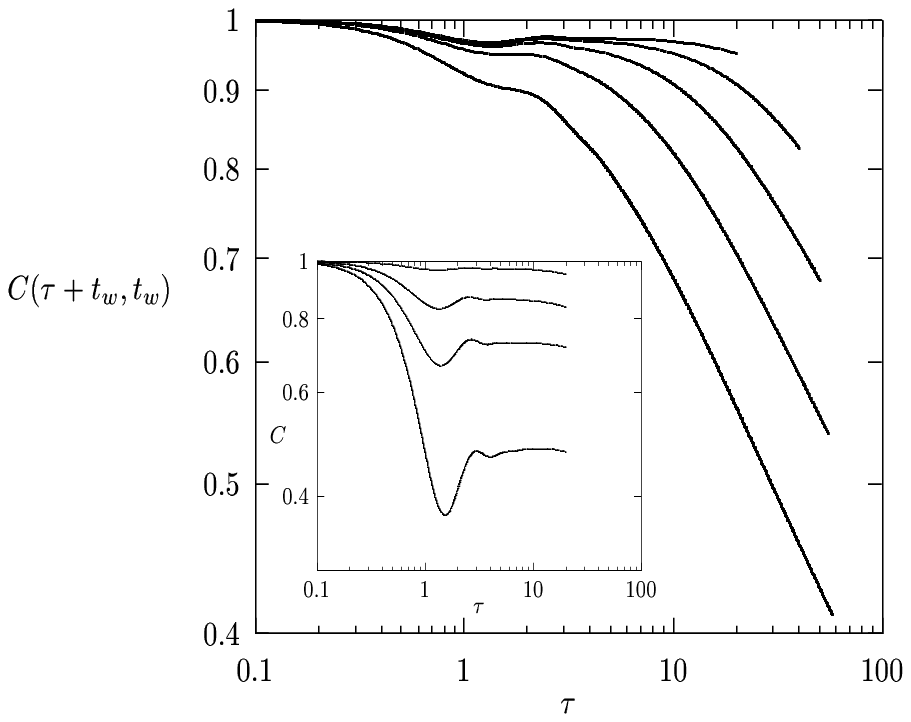}
\includegraphics[width=0.49\columnwidth]{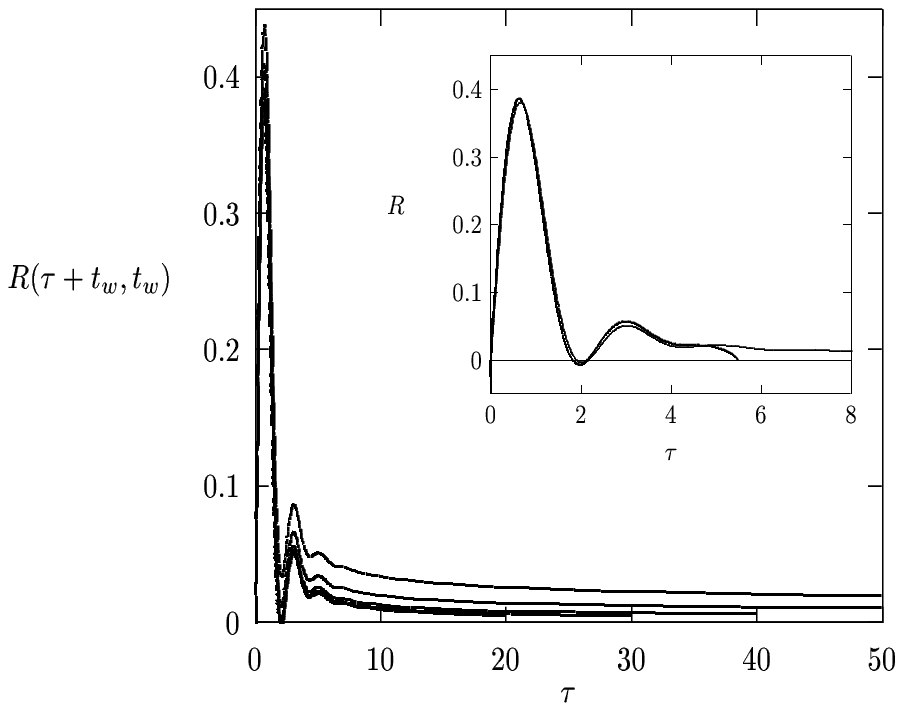}
 \caption{{\bf Left:}The correlation function $C(\tau+t_w,t_w)$ vs
$\tau$ for the $p=3$ quantum spherical $p$-spin SG model.
The waiting times are, from bottom to top,
$t_w=2.5,5,10,20,40$. $q_{EA}\sim 0.97$. In the inset, the same curves
for $t_w=40$ and, from top to bottom, $\tilde{\hbar}=0.1,0.5,1,2$.  
{\bf Right:} The response function for the same model as in the left
part. The waiting-times increase from top to bottom. In the inset,
check of FDT in the stationary regime. The full line is $R(t+t_w,t_w)$ for
$t+t_w=40$ fixed and $t_w\in [0,40]$.  The dots are obtained from
Eq.(\protect{\ref{xfdt}}) with $X_{\rm st}=1$, using the numerical data for
$C_{\rm stat}(t)=C(t+t_w,t_w)-q$ ($q_{EA}\sim 0.97$, see
left part).  In both cases the response is plotted against $t$.
(From \figcite{qsg-age}).
\label{fig-qsg-age}
}
\end{figure}

The self-consistency equations for $C(t+t_w,t_w)$ and $R(t+t_w,t_w)$
were evaluated numerically in \bcite{qsg-age}. An example of the
solution is shown in Fig.\ \ref{fig-qsg-age} for $p=3$. In all figures
the following parameters have been chosen: zero temperature $T=0$, the
width of the coupling distribution $J=1$, the frequency cut-off for
the oscillator bath set to $\omega_c=5$, the mass in the kinetic
energy term $M=1$, and the strength of the quantum fluctuations
$\tilde{\hbar}=\alpha\hbar$ (where $\alpha$ is the spin-bath coupling
strength) is $\tilde{\hbar}=0.1$.

These plots demonstrate the existence of the stationary and aging
regimes.  For $t < {\cal T}(t_w)$ (e.g.  ${\cal T}(40) \sim 5$) time
translational invariance and fluctuation dissipation theorem are
established while beyond ${\cal T}(t_w)$ they break down.  For
$\tilde{\hbar}=0.1$ the plateau in $C$ is at $q\sim 0.97$.  $C$
oscillates around $q$ but is monotonous when it goes below it.  In the
inset the dependence of $q_{EA}$ on $\tilde{\hbar}$ for $T=0$ is
presented. Quantum fluctuations generate a $q_{EA}<1$ such that the
larger $\tilde{\hbar}$ the smaller $q_{EA}$. The addition of thermal
fluctuations has a similar effect, the larger $T$, the smaller
$q_{EA}$.  In order to check the FDT in the stationary regime, in the
inset of the right part of Fig. \ref{fig-qsg-age} a comparison is
shown of $R(t+t_w,t_w)$ from the numerical algorithm for $t+t_w=40$
fixed and $t_w\in[0,40]$ (full line) with $R(t+t_w,t_w)$ from
Eq.(\ref{xfdt}) with $X=1$ using $C_{\rm stat}(t)=C(t+t_w,t_w)-q$,
$q\sim 0.97$ obtained from the algorithm (dots). The accord is very
good if $t \leq {\cal T}(t_w) \sim 5$.  Finally, when one plots
parametrically the integrated response $\chi$ vs. $C$ one finds that
for $C < q \sim 0.97$ the $\chi$ vs $C$ curve approaches a straight
line of {\it finite} slope $1/T_{\rm eff} = X_{\rm age}/T\sim 0.60$.

\subsection{Heisenberg Quantum Spin Glasses}

The spin-1/2 Heisenberg quantum spin glass is defined by the
Hamiltonian (\ref{q-heisen}) where the random exchange interactions
$J_{ij}$ can be ferromagnetic and anti-ferromagnetic. 
The system cannot be studied efficiently with quantum Monte-Carlo methods, 
due to the sign problem arising from the frustration.
Therefore, not much is known about these models in finite dimensions, and also
the mean field theory becomes tractable only in certain limits and
approximations.

\subsubsection{Finite Dimensions}

In \bcite{nonomura-ozeki} and later in \bcite{oitmaa-sushkov} small
clusters of the two-dimensional Heisenberg quantum spin glass
were studied using exact diagonalization. 
The average total spin in the ground state
turned out to scale as $S\propto\sqrt{N}$, where $N$ is the
number of sites. The spin glass order parameter in the ground state
extrapolates to a small but non-vanishing value in the thermodynamic
limit and the spin stiffness does not scale to zero either in the
thermodynamic limit. Ma-Dasgupta-Hu renormalization group 
studies\scite{dasgupta-ma,fisher-xxx} were performed for randomly frustrated
spin-ladders\scite{lin1} and in $d=2$ and $3$\scite{lin2} for various
lattices and spin-values. The general idea of this RG procedure was
already described in subsection \ref{rtim}; large energies (in the
form of exchange interactions) are successively decimated,
ferromagnetic bonds lead to large spin formation and anti-ferromagnetic
bonds to a spin reduction or even elimination in case of equal
effective spins connected by the bond to be decimated. In two and
three dimensions, it was also observed that the final magnetic moment to be
eliminated increased with system size as $\sqrt{N}$, which
corresponds to the aforementioned observation for the ground state spin
in small clusters. In addition the Ma-Dasgupta-Hu RG calculations in
\bcite{lin2} showed that the probability distribution of the low energy
excitations scales as $P(\Delta)\sim\Delta^\omega$ with $\omega=0$
for $2d$ and $3d$ and that the dynamical critical exponent is $z=2$ in
$d=2$ and $z=3/2$ in $d=3$.

\subsubsection{Mean-Field Model}

The first analytical treatment of the mean-field model of the Heisenberg
quantum spin glass was performed in \bcite{qsg-bray-moore} applying the
replica theory. Although the solution was confined to the paramagnetic 
state, the arguments for the existence of a low-temperature spin-glass 
phase were given and the critical temperature was estimated.

Later a Landau theory for quantum rotors on a regular $d$-dimensional
lattice was studied in \bcite{qsg-sachdev-read2}, which is defined by
the Hamiltonian
\begin{equation}
H=\frac{g}{2}\sum_i {\bf \hat L}_i^2-\sum_{\langle ij\rangle}
J_{ij} {\bf \hat n}_i {\bf \hat n}_j\;,
\end{equation}
where ${\bf \hat n}_i$ are $M$-component vectors of unit length 
(${\bf \hat n}_i^2=1$) and represent the orientation of the rotors on the
surface of a sphere in $M$-dimensional rotor space. The operators 
${\bf \hat L}_{i\mu\nu}$ ($\mu<\nu$, $\mu,\nu=1,\ldots,M$) are the
$M(M-1)/2$ components of the angular momentum ${\bf \hat L}_i$ of the
rotor: the first term in $H$ is the kinetic energy of the rotor with
$1/g$ the moment of inertia. The different components of ${\bf \hat
n}_i$ constitute a complete set of commuting observables and the state
of the system can be described by a wave function $\Psi(n_i)$. The
action of ${\bf \hat L}_i$ on $\Psi$ is given by the usual
differential form of the angular momentum 
${\bf \hat L}_{i\mu\nu}=-i(n_{i\mu}\partial/\partial_{i\nu}-
n_{i\nu}\partial/\partial_{i\mu})$. The difference between rotors and
Heisenberg-Dirac quantum spins is that the components of the latter at
the same site do not commute, whereas the components of the 
${\bf \hat n}_i$ do. 

In \bcite{qsg-sachdev-read2} a Landau theory for this model is derived
and it is shown that for a suitable distribution of exchange constants
$J_{ij}$ this model displays spin-glass and quantum paramagnetic
phases and a zero-temperature quantum phase transition between them.
The mean-field phase diagram near the $T=0$ critical point is mapped
out as a function of $T$, the strength of the quantum coupling $g$ and
applied fields. The spin glass phase has replica symmetry
breaking. Moreover, the consequences of fluctuations in finite
dimensions are considered and above $d=8$ the transition turned out to
be controlled by a Gaussian fixed point with mean-field exponents.
Below $d=8$ a runaway RG flow to strong coupling was found.

Recently the mean-field Heisenberg quantum spin glass model was
generalized from the $SU(2)$ spin algebra to an $SU(N)$ symmetry and
solved in the limit $N\to\infty$.\scite{georges-parcollet-sachdev}
Certain universal critical
properties are shown to hold to all orders in $1/N$. A spin-glass
transition was found for all values of the spin $S$ and the phase
diagram as a function of the spin S and temperature $T$ was
described. The quantum critical regime associated with
the quantum transition at spin value $S=0$ and the various regimes in
the spin-glass phase at high spin are analyzed. The specific heat is
shown to vanish linearly with temperature.

The out-of-equilibrium dynamics of the the same model in the same
limit $N\to\infty$, but coupled to a thermal bath, was studied in
\bcite{biroli-parcollet}. It was found that the model 
displays a dynamical phase transition between a paramagnetic and a
glassy phase. In the latter, the system remains out-of-equilibrium and
displays an aging phenomenon, which we characterize using both
analytical and numerical methods. In the aging regime, the quantum
fluctuation-dissipation relation is violated and replaced over a very
long time-range by its classical generalization, as in models involving
simple spin algebras studied previously.

In the context of Heisenberg spin glasses also the work on metallic
spin glasses should be mentioned, which were first considered in
\bcite{qsg-hertz} and later more extensively in \bcite{oppermann},
\bcite{sachdev-read-oppermann} and \bcite{sengupta-georges}. The main
ingredient of a metallic spin glass is an itinerant electron systems
with random (and frustrated) exchange interactions between the
electron spins. Thus in contrast to the spin glass systems discussed
so far the spins are not fixed to particular sites but can diffuse
(quantum mechanically) from site to site. These systems are motivated
by experiments on heavy-fermion compounds such as $Y_{1-x}$U$_x$Pd
$_3$,\scite{metal-exp}
which appear to show a paramagnetic to spin-glass
transition with increasing doping, $x$, in a metallic regime.  To be
concrete the Hamiltonian studied in \bcite{sachdev-read-oppermann} is
\begin{equation}
H=-\sum_{i<j,\alpha} t_{ij} c_{i\alpha} c_{j\alpha}
-\sum_{i<j,\mu} J_{ij} S_{i\mu} S_{j\mu}
+H_{\rm int}\;,
\end{equation}
where $c_{i\alpha}$ annihilates an electron on site $i$ with 
spin $\alpha=\uparrow,\downarrow$, and the spin operator is given 
by $S_{i\mu}=\sum_{\alpha\beta} c_{i\alpha}^+
\sigma_{\alpha\beta}^{\mu}c_{i\beta}/2$, with $\sigma^\mu$ the Pauli
spin matrices. The sites $i,j$ lie on a $d$-dimensional lattice, the
hopping matrix elements $t_{ij}$ are short-ranged and possibly random,
and the $J_{ij}^\mu$ are Gaussian random exchange interactions,
possibly with spin-anisotropies. The remainder $H_{\rm int}$ includes
other possible short-range interactions between the electrons,
and the resulting total Hamiltonian $H$ has a metallic
ground state.

Starting from this Hamiltonian, in \bcite{sachdev-read-oppermann}, an
effective field theory for the vicinity of a zero temperature quantum
transition between a metallic spin glass (``spin density glass'') and
a metallic quantum paramagnet was introduced. Following a mean-field
analysis, a perturbative renormalization-group study was performed and
it was found that critical properties are dominated by static
disorder-induced fluctuations, and that dynamic quantum-mechanical
effects are dangerously irrelevant. A Gaussian fixed point was found
to be stable for a finite range of couplings for spatial
dimensionality $d > 8$, but disorder effects always lead to runaway
flows to strong coupling for $d \leq 8$. Moreover, scaling hypotheses
for a {\em static} strong-coupling critical field theory were
proposed. The non-linear susceptibility has an anomalously weak
singularity at such a critical point.

In \bcite{sengupta-georges} the competition between the Kondo effect
and RKKY interactions near the zero-temperature quantum critical point
of an Ising-like metallic spin-glass was studied. In the `quantum-
critical regime,' non-analytic corrections to the Fermi liquid
behavior were found for the specific heat and uniform static
susceptibility, while the resistivity and NMR relaxation rate have a
non-Fermi liquid dependence on temperature.

\section{Summary and Remaining Problems}

In this review we have tried to provide an overview on the recent
developments in spin glasses. We concentrated on the topics to which
substantial efforts have been devoted in recent years and for which,
in our view, the most significant progress has been achieved:
1) The numerical investigation of the equilibrium thermodynamics of
finite-dimensional spin glass models with short range interactions
using new and powerful methods, such as combinatorial optimization for
ground state calculations (through which excited states can also
be studied) and extended-ensemble Monte Carlo methods 
(e.g., the replica exchange method) for the location and 
characterization of the phase transition. 
The most challenging task remaining here is still the
unification of the two paradigmatic pictures: the droplet picture and
the mean-field scenario. Although being unsolved for several decades now, 
%and still a source of sometimes ideological fighting,
we see promising steps toward the unification of these two approaches. 
2) The experimental and theoretical investigation of
non-equilibrium dynamics and aging phenomena in spin glasses, 
especially the study of fluctuation-dissipation-theorem violations in
the glassy phase and the concept of an effective (non-equilibrium)
temperature. Still a lot of work has to be done,
experimentally in particular, 
in order to put the fascinating theoretical ideas on
a firm and consistent experimental ground. 3) The theoretical
exploration of quantum effects in spin glasses, statically and
dynamically, for mean-field models as well as finite dimensional
models. Here the most demanding challenge appears to be 
handling the {\it real} time quantum dynamics at low temperatures for
realistic (i.e. finite dimensional short-range interacting) models.

Unfortunately, we could not cover in this review many ``glassy'' 
topics related to and inspired by the spin-glass world.
First and most actively pursued in recent years is the theory of the
structural glass transition and the glass ``phase''. A lot of progress
has been made to make contact between spin glasses and structural
glasses, the main difference being that the disorder is self-induced
in the latter during the freezing process, whereas spin glasses live
with frozen, time-independent disorder. 
Recently, it was discussed extensively that this might not be 
a major obstacle in relating the two systems.
From a theoretical point of view, 
a more substantial difference between them 
with concrete experimental consequences is
that spin glasses have a well-defined order parameter.
A diverging correlation length and a divergent susceptibility 
defined in terms of this order parameter signify the transition. 
In the theory of structural glasses, apparently one
cannot depend on these helpful vehicles.

Other related issues are vortex, gauge and Bragg glasses.
All three appear in the context of disordered superconductors. 
The gauge glass model is essentially an $XY$ spin-glass model 
with a random (quenched)
vector potential, originally devoted for describing amorphous granular
superconductors but later also taken as a paradigmatic model for
amorphous high-T$_c$ superconductors. It can be analyzed in the same
way as $XY$ spin glasses and also has the same order parameter. The
major questions concern the existence of a finite-temperature 
phase transition with or without screening effects. The vortex-glass model
is a model of interacting, elastic magnetic flux lines in a random
potential, which freezes at low temperatures into a glassy phase
(which however escapes a clean theoretical description, since an order
parameter is hard to define). If the disorder is weak, topological
defects can be neglected and an elastic description is possible,
starting from the Abrikosov flux line lattice and taking into account
its small elastic deformation via thermal fluctuations and
disorder. For such weakly disordered elastic systems
a Bragg glass phase is predicted
in which the true long-range order of
the Abrikosov flux line lattice is transformed into a quasi long-range
order. This glassy phase manifests itself also via an extremely
sluggish dynamics. Upon increasing the strength of the disorder or 
the density of the lines (via an increased magnetic field), topological
defects will proliferate, the quasi long-range order of the Bragg
glass phase vanishes and the system becomes a vortex glass.

Finally one should mention Bose, Fermi and Coulomb glass models, which
occur in the context of the low-temperature physics of 
quantum-mechanical, disordered, electronic or bosonic systems. 
The origin of the interesting physics is the competition between 
the quenched random
potential and the interactions between particles, 
usually long ranged as in the Coulomb case. The major issues
concern the phase transition between conducting 
(metallic or superconducting) and insulating phases, 
in which the particles are localized. Again the name
``glass'' for the low-temperature phases is justified by the
anomalously slow dynamics present here.

To conclude this tour through the glass zoo we hope to have
demonstrated that a review on spin glasses should not only be useful
for people working in the field of frustrated magnets, but also for
those encountering strong disorder and strong interactions at low
temperatures in other fields of condensed matter physics.

%==============================================================================
\section*{Acknowledgments}
%==============================================================================
\addcontentsline{toc}{section}{Acknowledgments}
N.K.'s work was supported by the grant-in-aid (Program No.14540361)
from Monka-sho, Japan.
H.R. acknowledges financial support from the Deutsche
Forschungsgemeinschaft (DFG) and from the European Community's Human
Potential Programme under contract HPRN-CT-2002-00307, DYGLAGEMEM, and
thanks the KITP at UC Santa Barbara for its hospitality, where 
he composed large parts of this review.
The authors are indebted to H.~E.~Castillo, C.~Chamon, L.~Cugliandolo,
K.~Hukushima, H.~Kawamura, J. Kurchan, C.~M.~Newman, D.~L.~Stein, and
A. P. Young for useful comments.

%==============================================================================

\end{document}